\documentclass[%
reprint, 
amsmath,amssymb,
aps,
prx,
]{revtex4-2}

\usepackage{graphicx}% Include figure files
\usepackage{epsfig} %Works faster than the graphicx package}
\usepackage{dcolumn}% Align table columns on decimal point
\usepackage{bm}% bold math
\usepackage{epstopdf}
\usepackage{multirow}
\usepackage{amsmath}
\usepackage{booktabs}
\usepackage{color}
\usepackage{gensymb}
\usepackage{kantlipsum}  
\usepackage[colorlinks=true, allcolors=blue, breaklinks]{hyperref}
\usepackage{xurl}
\usepackage[normalem]{ulem}
\usepackage{braket}
\usepackage{physics}
\usepackage{url}
\usepackage[utf8]{inputenc}
\usepackage[T1]{fontenc}
\usepackage{mathptmx}
\usepackage{lineno}

\def\fig#1{Fig.~\ref{#1}}

\begin{document}

\title{Interpretable Machine Learning in Physics: A Review}

\author{Sebastian~J.~Wetzel$^{1,2,3}$, Seungwoong Ha$^4$, Raban Iten$^{5}$, Miriam Klopotek$^{6,7}$, Ziming Liu$^{8,9}$}

\affiliation{%
$^1$University of Waterloo, Waterloo, Ontario N2L3G1, Canada
}
\affiliation{%
$^2$Perimeter Institute for Theoretical Physics, Waterloo, Ontario N2L2Y5, Canada
}
\affiliation{%
$^3$Homes Plus Magazine Inc., Waterloo, Ontario N2V2B1, Canada
}
\affiliation{%
$^4$Santa Fe Institute, Santa Fe, NM 87501, USA
}

\affiliation{%
$^5$ETH Zurich, Wolfgang-Pauli-Str. 27, 8093 Zurich, Switzerland.
}
\affiliation{%
$^6$Stuttgart Center for Simulation Science, University of Stuttgart, Universit\"atsstra{\ss}e 32, 70569 Stuttgart, Germany
}

\affiliation{%
$^7$Heidelberg Academy of Science and the Humanities, Karlstraße 4, 69117 Heidelberg, Germany
}
\affiliation{%
$^8$Massachusetts Institute of Technology, Cambridge, MA, 02139, USA
}
\affiliation{%
$^9$The NSF AI Institute for Artificial Intelligence and Fundamental Interactions
}
%\begin{document}

\date{\today}% It is always \today, today

\begin{abstract}
Machine learning is increasingly transforming various scientific fields, enabled by advancements in computational power and access to large data sets from experiments and simulations. As artificial intelligence (AI) continues to grow in capability, these algorithms will enable many scientific discoveries beyond human capabilities. Since the primary goal of science is to understand the world around us,  fully leveraging machine learning in scientific discovery requires models that are interpretable -- allowing experts to comprehend the concepts underlying machine-learned predictions. Successful interpretations increase trust in black-box methods, help reduce errors, allow for the improvement of the underlying models, enhance human-AI collaboration, and ultimately enable fully automated scientific discoveries that remain understandable to human scientists.
This review examines the role of interpretability in machine learning applied to physics. We categorize different aspects of interpretability, discuss machine learning models in terms of both interpretability and performance, and explore the philosophical implications of interpretability in scientific inquiry.
Additionally, we highlight recent advances in interpretable machine learning across many subfields of physics. By bridging boundaries between disciplines -- each with its own unique insights and challenges -- we aim to establish interpretable machine learning as a core research focus in science.

\end{abstract}

\maketitle

\onecolumngrid

\vskip -0.5cm
\begin{figure}[h!]
\centering
    \includegraphics[width=0.65\linewidth]{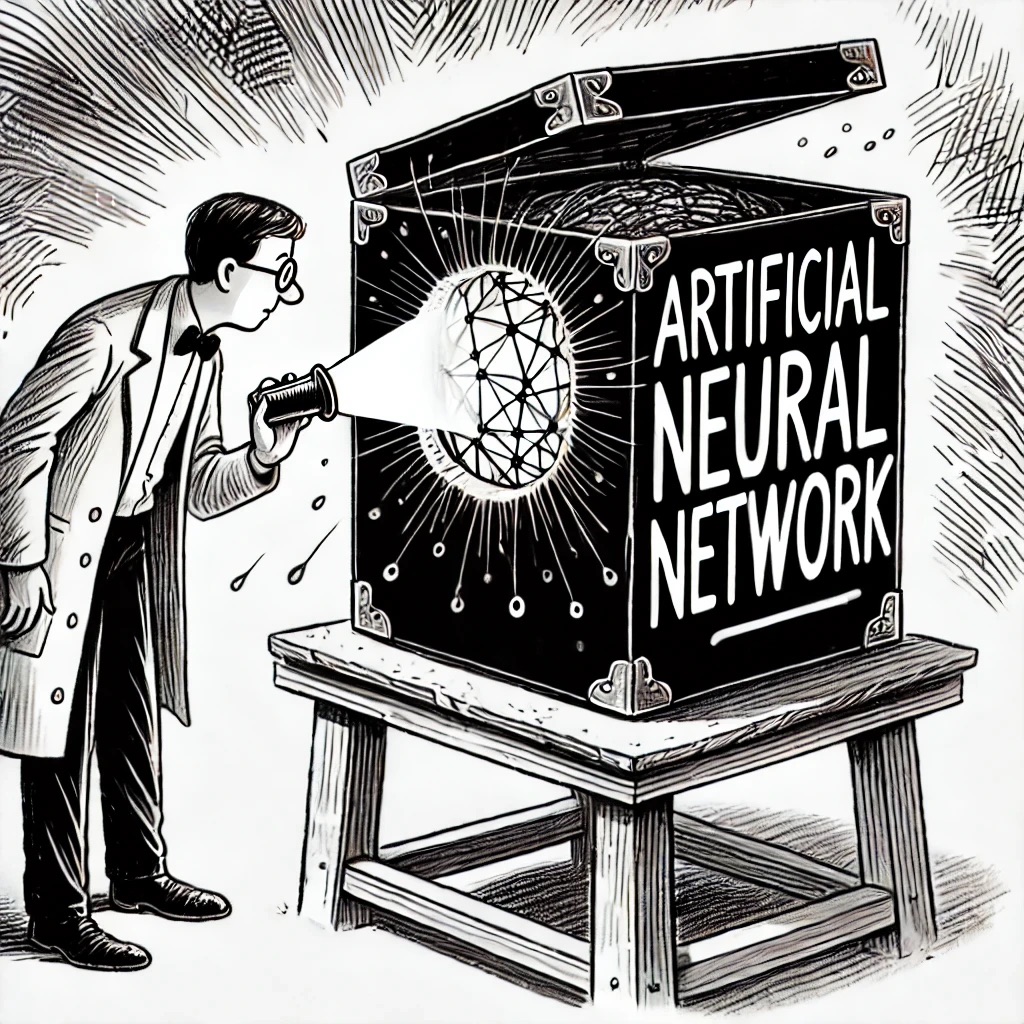}
    \caption{A scientist attempts to peep into the black box of artificial intelligence.}
    \label{fig:open-blackbox}
\end{figure}

\twocolumngrid

\newpage

\tableofcontents

\section{Introduction}
\label{chapter:introduction}

\subsection{Why Interpretable Machine Learning?}
The current revolution in artificial intelligence is profoundly transforming many aspects of science. The revolution is driven by increasing computational power, enormous amounts of data, and powerful machine learning algorithms, most notably artificial neural networks.

Machine learning is a computational paradigm rooted in statistical principles, where algorithms learn patterns directly from data through a process known as training. It serves as a versatile tool for extracting insights, making predictions, and optimizing processes across a spectrum of scientific disciplines including physics. In the context of physics, these machine learning algorithms often need to discern intricate and complex relations within data -- tasks traditionally performed by human scientists. These methods range from simple algorithms like linear regression or principal component analysis to various types of artificial neural networks. While simpler models can only capture relatively simple relations, their learned features are relatively easy for human scientists to interpret. In contrast, more sophisticated models like artificial neural networks, have achieved enormous success in solving complex problems that exceed human capabilities. However, their ``black-box" nature makes it increasingly challenging to understand how they arrive at specific solutions.

Interpretable machine learning is the process of translating machine-learned concepts into a representation and language that can be understood by humans. There are intrinsically interpretable machine learning algorithms whose decision-making processes can be understood and explained by humans. On the other hand, there also exist "black-box" models that produce predictions without revealing their internal logic. In these cases, additional tools must be developed and applied to reveal their inner workings.

\begin{figure}
    \centering
    \includegraphics[width=\columnwidth]{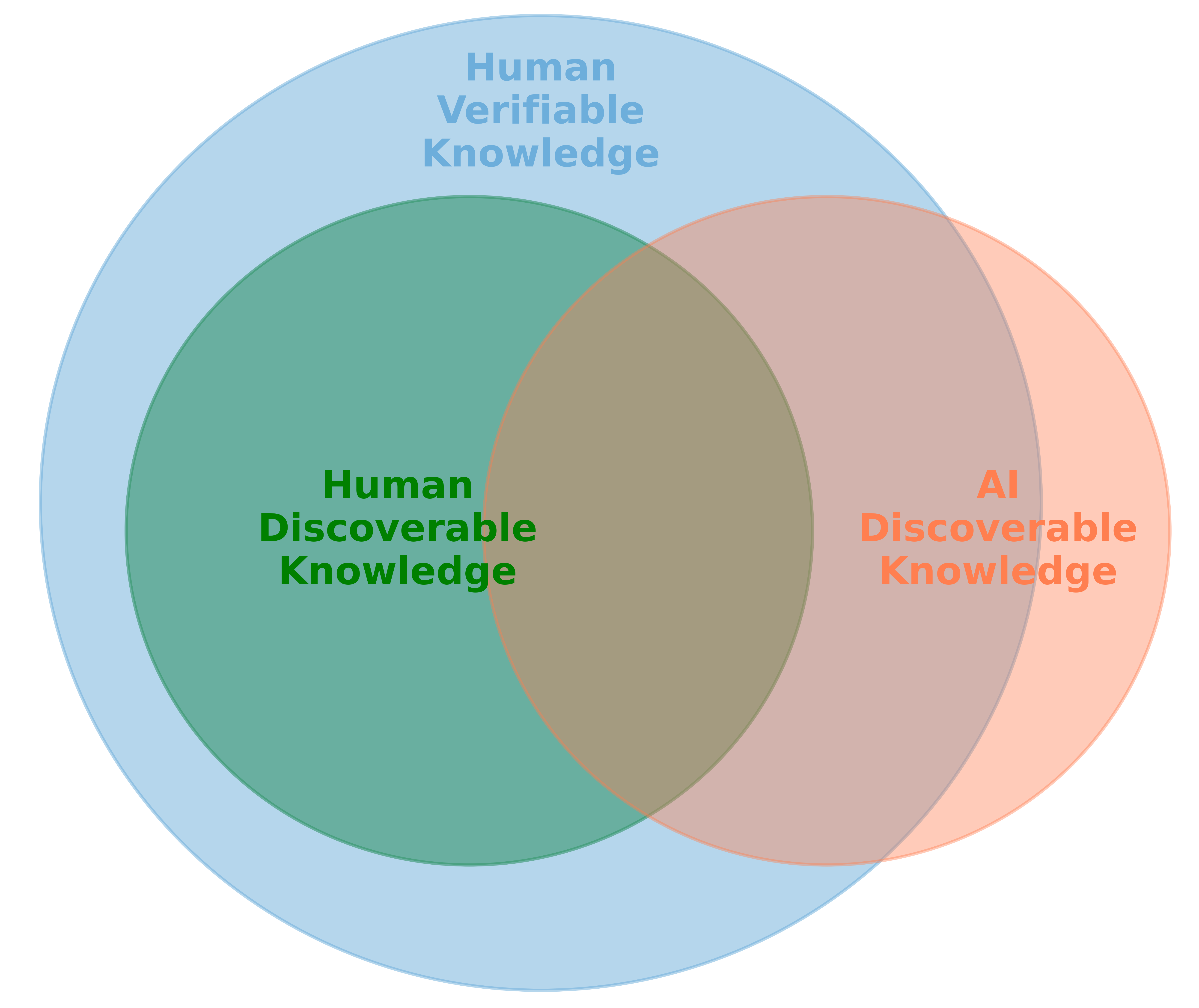}
    \caption{In the near future, AI is expected to drive scientific discoveries beyond the capabilities of human scientists. In science, the primary goal is to facilitate a human understanding of new and unknown concepts. To bridge the gap between AI-generated insights and human understanding, it is crucial to interpret AI systems. By doing so, human scientists can access and integrate knowledge that lies at the intersection of AI-discoverable knowledge and human-verifiable understanding -- unlocking scientific concepts beyond what humans alone can discover.}
    \label{fig:knowledge}
\end{figure}

Interpretable machine learning is critical in science because it bridges the gap between complex models and human understanding, enabling insights that:  

\begin{itemize}
\item \textbf{are reliable, transparent, and aligned with scientific principles.} In scientific applications, interpretability in machine learning is crucial for trust because it enables researchers to understand and validate how models generate predictions, which is essential for scientific rigor. Interpretability allows scientists to ensure that models are making decisions based on meaningful patterns rather than artifacts or biases in the data, thereby reinforcing the reliability and reproducibility of findings. Additionally, interpretable models facilitate accountability, as scientists can trace specific predictions back to underlying data features, supporting transparency and advancing scientific understanding.
\item \textbf{help to debug and improve machine learning models.} 
Interpretable machine learning is vital for finding errors in scientific research because it allows scientists to identify and understand why a model may be making incorrect or unexpected predictions. By analyzing the decision-making process of the model, researchers can pinpoint specific data features or steps that may be causing errors or biases, enabling them to adjust the model or refine the data accordingly.
\item \textbf{facilitate scientific understanding of new and unknown concepts.} Interpretable machine learning enhances scientific understanding by providing insights into the relationships and patterns a model identifies in data, allowing scientists to see which variables or features are driving certain outcomes. This transparency helps researchers draw meaningful conclusions about the underlying processes and phenomena, rather than merely accepting a model’s predictions at face value. By revealing how and why a model reaches its conclusions, interpretable ML can lead to new scientific hypotheses and even uncover new and unknown concepts, see figure~\ref{fig:knowledge}.
\end{itemize}

This review examines various aspects of interpretability in machine learning within the context of physics. We categorize different concepts of interpretability and present an overview of machine learning models, evaluating them in terms of both interpretability and performance. Additionally, we contextualize recent developments across multiple areas of physics within the framework of interpretable machine learning. Our goal is to connect diverse fields, each with its own perspectives and challenges, and to promote interpretable machine learning as a central paradigm in scientific research. We expect that this field will drive groundbreaking AI-driven discoveries, unveiling new physical concepts and insights. 

\subsection{Related Reviews about Machine Learning in Science}

Machine learning is rapidly becoming an integral part of scientific research. Many research communities have outlined the major application areas of machine learning in their fields in various review articles. 

The most prominent overview of machine learning in physics can be found in "Machine learning and the physical sciences\cite{Carleo2019}", while a more focused review tailored towards quantum systems can be found in "Machine learning for quantum matter~\cite{Carrasquilla2020}". A didactical summary of this field can also be found in "Modern applications of machine learning in quantum sciences~\cite{https://doi.org/10.48550/arxiv.2204.04198}". A review that addresses the need for understanding the results of machine learning applied to science is "On scientific understanding with artificial intelligence"~\cite{Krenn2022}. A general review advocating for the need of interpretable machine learning in science can be found in "Explainable Machine Learning for Scientific Insights and Discoveries
\cite{Roscher2020}". Although physicists have yet to establish interpretable AI as a distinct subfield, neighboring disciplines like chemistry and materials science have promoted this paradigm. This is reflected in reviews like "Interpretable and Explainable Machine Learning for Materials Science and Chemistry~\cite{Oviedo2022}" and "Explainable Machine Learning in Materials Science\cite{Zhong2022}". Additionally, "Quantum machine learning for chemistry and physics" \cite{Sajjan2022} outlines the use of quantum ML mostly in a chemistry setting and includes a discussion of statistical-physics-based approaches for understanding learning mechanisms and expressivity. A more recent review "Scientific Discovery in the Age of Artificial Intelligence"~\cite{wang2023scientific} provides a bird's eye view of the use of AI in many scientific disciplines, focusing on how AI learns meaningful representations, generates hypotheses, and aids experiments and simulation.

\subsection{Overview}

%\seb{For Ziming: Pease add an overview figure here and explain how it connects to the review. This is your section.}

\begin{figure}[t]
    \centering
    \includegraphics[width=1.0\linewidth]{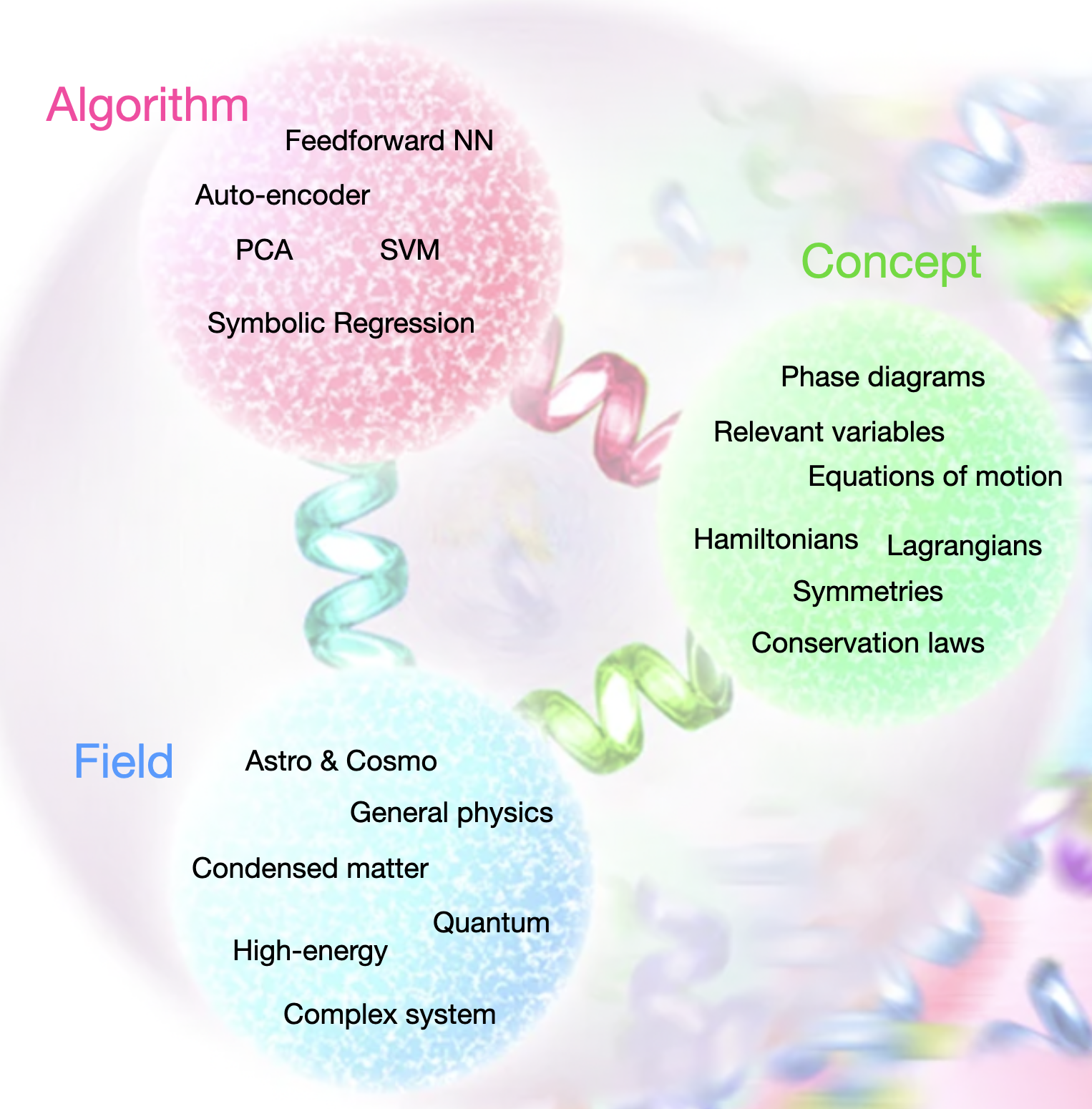}
    \caption{How about "Overview of the scientific scope of the paper: The physics sub-fields, the most frequently reported algorithms, and overarching concepts addressable with or useful for interpretable machine learning in physics". The background image is credited to  Lawrence Berkeley Lab~\cite{quark_picture}.}
    \label{fig:intro_quarks}
\end{figure}

%maybe refer to structure of the review according to the table of contents and refer to the sub elements

This paper focuses on interpretable machine learning for physics, which consists of three components -- \textit{interpretability}, \textit{machine learning}, and \textit{physics}.

{\bf Interpretability} We define what we mean by interpretability in Section~\ref{chapter:interp-notion} and discuss philosophical aspects of interpretability in Section~\ref{chapter:interp-philosophy}. 

{\bf Machine Learning} We review interpretable machine-learning tools in Section~\ref{chapter:algorithms}, posthoc interpretability methods in  Section~\ref{chapter:interpretability_methods}, and symbolic regression in Section~\ref{chapter:symbolic}.

{\bf Physics} Many physical concepts (e.g., phase diagrams) are shared among fields, so we can categorize physics either based on fields or concepts. Categorization using fields: we review quantum physics in Section~\ref{chapter:quantum}, classical condensed matter physics in Section~\ref{chapter:condensed}, high energy physics in Section~\ref{chapter:high_energy}, astrophysics \& cosmology in Section~\ref{chapter:astro}, complex systems in Section~\ref{chapter:complex}, and general physics in Section~\ref{chapter:general}. Since the concepts of conservation laws (symmetry invariants) and phase diagrams are important on their own, we also review how machine learning can be used to discover these concepts in Section~\ref{chapter:conserved} and Section~\ref{chapter:phase_diagrams}.

A visual illustration of the overview is Figure~\ref{fig:intro_quarks}, which suggests that "interpretable machine learning for physics" requires three pillars -- algorithms, fields, and concepts, and they are closely intertwined and constantly interacting.

\section{Notions of Interpretation}\label{chapter:interp-notion}

\subsection{Interpretability Controversy}

Interpretable machine learning is a growing field that has attracted many scientists and sparked much controversy. A common problem is that "interpretable" or "explainable" machine learning can carry many different meanings which we categorize in section~\ref{chapter:interpretation}. 

The article "The Mythos of Model Interpretability: In machine learning, the concept of interpretability is both important and slippery \cite{Lipton2018}" highlights the ambiguity in defining interpretability, noting that academic motivations and techniques are diverse and sometimes conflicting. The article critiques the common assumption that certain models, like linear models, are inherently interpretable, while deep neural networks are not. The article calls for a clearer, more rigorous understanding of what interpretability entails and when it is beneficial.

Similarly, in "Against Interpretability: a Critical Examination of the Interpretability Problem in Machine Learning \cite{Krishnan2019}" the authors criticize the widely accepted "black-box problem" in machine learning, which suggests that many algorithms are opaque and lack interpretability. It argues that terms like "interpretability" are imprecise and do not clearly define the issues often associated with machine learning. 

Against these criticisms, it is possible to interpret machine learning models in various aspects, even complex deep neural networks. The tutorial article "Methods for interpreting and understanding deep neural network \cite{Montavon2018}" provides a detailed explanation of different categories of interpretation and explains methods to interpret artificial neural networks. The book Ref.~\cite{Molnar2024-book} describes the problem of interpretability in a non-technical fashion and and discusses which expectations of interpretable ML can be met for science.

\subsection{Categories of Interpretation}
\label{chapter:interpretation}

Interpreting a machine learning algorithm involves making sense of how it works by \textbf{translating learned concepts into a language humans can understand}. Human understanding can vary, from recognizing everyday things to having specialized knowledge in certain areas. We suggest categorizing interpretations along the following five different attributes below, which is more fine-grained than the categories proposed in \cite{Montavon2018}:

A) Mechanistic vs. Functional: Mechanistic interpretation explains how the mechanisms of the mathematical operations within machine learning algorithms learn and express a solution to a task, while functional interpretation focuses on understanding the whole function, connecting input to output, beyond the interactions of elementary operations. 

B) Local vs. Global: Local interpretation is typically concerned with which parts of a single data point affect the prediction, while global interpretation considers what features are important for a learned concept in general, independent of what data point is looked at. 

C) Verifying vs. Discovering: Sometimes, people studying machine learning have specific ideas about what concepts a algorithm might learn. It's usually easy to check if these ideas match the features the network learned. However, finding a new concept without any initial ideas is usually very difficult.

D) Low-Level vs. High-Level Features: Several machine learning algorithms, including neural networks or decision trees, learn a hierarachy of features that work together to transform an input into output, successively building more complex high-level features from low-level features close to the input. High-level concepts might be encoded in the output neuron in the final layers of neural networks or in latent spaces of autoencoders.  Low-level features close to the input are usually easy to interpret, while high-level features often involve increased complexity concealing their nature.

E) Intrinsic vs. Post-hoc: Some machine learning algorithms like linear regression or PCA are intrinsically interpretable, while others, especially neural networks are considered black-boxes where additional techniques need to be applied to a fully trained model to extract meaningful information.

F) Interpreting ML algorithms vs. Gaining scientific understanding: An interpretation  \textit{of} ML algorithms does not necessarily lead to a gain of understanding of a scientific phenomenon of interest modeled \textit{through} them (see Sec.~\ref{chapter:interp-philosophy}). The bulk of literature in this review reports on how interpretable ML can try to reveal understanding in terms of target physical phenomena, as opposed to purely algorithmic aspects or theories of learning.

\section{Philosophical perspectives}\label{chapter:interp-philosophy}

To provide a wider context on interpretable ML, it is important to note that there are widespread concerns regarding the use of ML in various societal contexts, particularly in high-stakes automated decision-making processes. The core of these concerns stems from the immense difficulty in tracing the steps leading to their decisions, forming explanations of their behavior (i.e., their `reasoning'), and establishing trust in black-box algorithms. 
Research under the umbrella of eXplainable AI (XAI)  emerged to tackle these challenges. However, it is a largely uncoordinated effort that lacks many conceptual foundations, see, e.g., Refs.~\cite{BarredoArrieta2020,Freiesleben2023,Weber2024,Murdoch2019,Gunning2019,Hagras2018,Pez2019,Lakkaraju2019,Miller2019,Bjerring2025}. 

In contrast, a question of central interest in this review is whether, while learning to predict their target quantities concerning a physical phenomenon, ML models capture, discover, or adhere to specific concepts, and how we can understand the latter. This inquiry is about an interpretation ability that can potentially originate a scientific explanation, of both possibilities and matter-of-facts, tied to theories of the natural world (Sec.~\ref{subsec:explanation_interpretation}). Thus, it is not necessarily the same as looking for an explanation of ML behavior on its own (Sec.~\ref{subsec:black_box}). For one thing, it is essential to clarify what we expect from an explanation before attempting to find one \cite{Buchholz2023} in the context of `opening up the black box' (Sec.~\ref{subsec:black_box}). Furthermore, the explanations offered by current XAI may not align with those relevant to science and its underlying philosophy \cite{Pez2019}. The distinction lies in how scientific understanding is approached and (the open question of)  how ML can ultimately relate to it (Sec.~\ref{subsec:understanding_and_ML}).

The aim of XAI research, as stated by Ref.~\cite{Gunning2019}, is ``to make its behavior more intelligible to humans by providing explanations.'' 
There are two types of XAI according to Ref.~\cite{Verhagen2021}, only one of which is related to interpretability as we defined it in Sec.~\ref{chapter:interp-notion}: (1) Data-driven XAI is about understanding and explaining inner mechanisms of algorithms \cite{Anjomshoae2019,Guidotti2018}, which can facilitate ML interpretation relevant for physics if embedded in a truly scientific context, as discussed in this section. (2) The second kind is \textit{not} related to the kind of scientific interpretability in focus here: Goal-driven XAI refers to building goal-driven intelligent systems (like robots) ``explaining their actions and reasons leading to their decisions to lay users'' \cite{Anjomshoae2019}~\footnote{Actually, interpretability of AI systems had been considered a better aim for XAI \cite{Lipton2018,Rudin2019,Rudin2022,Pez2019}.}. See further Refs.~\cite{Langley2017,Gunning2019b,Gunning2019}.

What is considered `interpretable ML' or `XAI' is often difficult to discern, as the same methods (see Sec.~\ref{chapter:interpretability_methods}) are used in both, and the words `interpretation' and `explanation' are used inconsistently \cite{Gilpin2018,Lipton2018,Murdoch2019,Weber2024,Buchholz2023,Miller2019,Renftle2024,Beisbart2022,Bellucci2021,Palacio2021,BarredoArrieta2020,Zednik2019,Krishnan2019}.

%\seb{the following section appears to be really important, but I do not understand what you want to say. can you please reformulate it in a concise way.}\mk{ok}
%
\subsection{Explainability and interpretability} 
\label{subsec:explanation_interpretation}

Explanations play a central role in the formation and use of scientific theories  \cite{Woodward2021,vanFraassen2008-book,Hempel1965}.  There are various forms of explanation~\footnote{There are more forms of explanation than these two, like seeking unification and possibilities \cite{DeRegt2009-book,Grimm2017b-book}. We refer to further literature starting with  Ref.~\cite{Woodward2021}.}: Typically, they encompass the types where causal(-mechanistic) relationships are described, and types when justifications are sought  \cite{Freiesleben2024, Fleisher2022,Ross2023,Woodward1989,Johansson2024,Krishnan2019,Langer2021,Craver2006,Glennan1996,DeRegt2009-book}. Justification relates to verifying a model or hypothesis (as described in Sec.~\ref{chapter:interp-notion}). However, explanations are only effective if we can interpret them.
For example, a numerical (ML) model offers an `explanation' directly via faithful computation  \cite{Zerilli2022,Rudin2019,Guidotti2018}, but this must be interpretable in order to ultimately understand how a particular decision was reached  \cite{Zerilli2022}. Thus, to gain scientific understanding, explanatory information is essential, but explanations only work when they are embedded within a theoretical framework intelligible or comprehensible to the interpreting agent \cite{DeRegt2017-book,Barman2024} (see below for clarification on these words). 
Interpretable systems are by their very nature accessible to our cognition: They enable the formulation of a theory in a `language' with a well-defined structure \cite[p.113]{Hempel1965}. 

The word \textit{interpretability} is frequently used to refer to understanding how an ML model operates~\cite{Murdoch2019,Fleisher2022}. (See also Ref.~\cite{Lutz2023}.)
It is often regarded as a `know it when you see it'  property \cite{Fleisher2022,https://doi.org/10.48550/arxiv.1702.08608}, but what is actually meant by an interpretable model is diverse~\cite{Lipton2018,Krishnan2019,BarredoArrieta2020}:  For example, an interpretable model may be one that a human can contemplate in its entirety at once (\textit{fathomability}~\cite{Lipton2018}), 
or one that allows for targeted manipulations of its components (\textit{intelligibility}~\cite{BarredoArrieta2020,Ratti2018}; but compare \cite[Ch.2]{DeRegt2017-book}). 
Ref.~\cite{https://doi.org/10.48550/arxiv.1710.00794} differentiates interpretable systems, where the internal algorithmic mechanisms are analyzable, and \textit{comprehensible} systems that produce understandable symbols (compare discussion in Ref.~\cite[Ch.5]{DeRegt2017-book}). 
The ability to decompose an algorithm's functioning into smaller parts serving specific purposes \cite{BarredoArrieta2020,https://doi.org/10.48550/arxiv.1702.08608} is certainly beneficial. Intuitively, this can enable one to construct a new model or modify it in a controlled way.
When humans feel they can code a model `by hand' (e.g., possibly with decision trees, decision lists, linear models, and logistic regression \cite{Lakkaraju2019,Beisbart2022}), they perceive it as more interpretable. However,  `felt' understanding does not necessarily equate to true understanding \cite{Barman2024}.
Interpretability might also involve comprehending how the correlations in data are found by the ML algorithm \cite{Desai2022,LpezRubio2019,Ratti2018}, although many algorithms deemed `interpretable' do not deliver this. Beyond  goals such as prediction and explanation, in scientific contexts, interpretability determines whether scientific theories are viable  \cite{DeHaro2018} and is inherently tied to understanding \cite{Faye2014-book} and also discovery \cite{Boge2021} (Sec.~\ref{subsec:understanding_and_ML}). For example, if one can anticipate the outcomes without having to produce a full-blown derivation, then one demonstrates understanding of a theory \cite{DeRegt2005,DeRegt2017-book}.

\subsection{The opacity/black-box problem}
\label{subsec:black_box}

A \textit{black box} is defined as something whose inner mechanisms are unknown to the investigator or observer \cite{Ashby1956,Meskhidze2021}. The term \textit{opacity} of a model typically refers to be the issue of `not knowing how the algorithms work'  or `why you get the output you do'. In essence, this refers to a lack of interpretability. Yet, such formulations do not fully encompass the opacity or black-box phenomenon \cite{Krishnan2019}. It is nuanced and varies depending on the observer's level of sophistication \cite{Zednik2019,Humphreys2008,vonEschenbach2021}. 

A model looks opaque if too many elements of the information processing are epistemically relevant or unknown, making it difficult to follow the computational steps within a reasonable timeframe (that fits our attention span)~\cite{humphreys2004extending,Humphreys2008,Durn2018,Creel2020,Beisbart2021}. This opacity hinders our reasoning about, debugging, or improving the corresponding algorithm \cite{Burrell2016}.  Additionally, opacity arises when the input data is either unknown or only known partially~\cite{Burrell2016}, or when our understanding of it is insufficient~\cite{Boge2023}. 
Moreover, a model may be highly sensitive to the specifics of input parameters; a strong dependency on such `initial conditions' makes it difficult to tell which details really matter when drawing conclusions \cite{Scorzato2024,Desai2022}.  Opacity also occurs when we lack understanding of how the components of the ML model relate to the real-world features it is tracking  \cite{Sullivan2022,Fleisher2022}. This is particularly relevant for applications in science.  If we fail to comprehend how the algorithm generalizes from data using induction, we remain uncertain about what can be known through the model (it is epistemically uncertain)  \cite{Hllermeier2021}\footnote{A particular root for opacity may be a mismatch between complex mathematical optimization processes behind learning algorithms and the interpretation semantics of our normal understanding \cite{Burrell2016}.}.

The perception of the model being `too complex' leads to the idea that simpler models are less opaque, or that interpretability declines as model complexity increases \cite{Sullivan2022,Rz2024,Scorzato2024}.  However, the complexity of a model may be a detached issue, and it depends on what we mean by `simple'~\footnote{
A significant challenge in quantifying the simplicity of a machine learning model is the absence of a universal definition of model complexity \cite{Scorzato2024}, whether in computational terms or otherwise. Some researchers, such as Ref. \cite{Krakauer2023}, suggest shifting the focus from the complexity of a model's basic description to the complexity inherent in the process that generates the model.}. 
To be comprehensible, a model must limit the number of relevant factors it presents, as too many can overwhelm human cognition \cite{LpezRubio2019,Ratti2018}. However, these factors may not be easily identifiable.
For instance, the number and type of presumptions underlying a model might portray its complexity, yet it can be difficult to recognize and evaluate these presumptions \cite{Scorzato2024}\footnote{As noted by Ref.~\cite{Scorzato2024}, the question may arise,  ``[W]hich assumptions are acceptable to assess our model assumptions?''}. Moreover, complexity and opacity do not necessarily hinder the simplicity of learned representations: Deep neural networks have implicit regularization mechanisms that guide them there \cite{Martin2021,Dherin2022,Buchholz2022}, and many cases reported in this review attest to this useful phenomenon, as well. Thus, it is important to recognize that a model's complexity and  opacity (as initially perceived) do not inherently limit the understanding potentially providable \textit{by} it  \cite{Molnar2024-book,Sullivan2022,Sullivan2022b,Boge2021,Zednik2022,Bjerring2025,Meskhidze2021,Bjerring2025}; generating useful conceptions with such a model is a separate issue.

The notion of a `lack of trust' is often conflated with the opacity issue in ML. The belief that `opening up the black box' will resolve both is common~\cite{Kastner2021,vonEschenbach2021,Buchholz2022}.
However, providing a precise explanation of the inner workings is neither necessary or sufficient to induce trust ~\cite{Baron2025,Bordt2022,Kastner2021,Blanco2022}.
In many contexts, striving for reliability is a more appropriate goal~\cite{Grote2024,Scorzato2024,Duede2022,Blanco2022}.

`Opening up the black box' is largely about interpreting the decision-making process within an opaque algorithm, and usually involves additional modeling. Post hoc interpretation techniques  (refer to Sec.~\ref{chapter:interpretability_methods}) ought to reduce the complexity of (mostly) deep learning models \cite{Bordt2024}. These substitute how the original system performs inference. Yet, the original and substitute models have different operational structures \cite{Fleisher2022,Sullivan2022b}, which means the latter can be highly unreliable for generating explanations of the original \cite{Bordt2022,Rudin2019}. The larger problem is that analyzing input-output relations, or their sensitivities, offers only limited insight into how a complex ML model works \cite{Creel2020,Beisbart2022,Bordt2022,Bjerring2025}.
Nonetheless, if the substitute model is subjected to scientific inquiry, then genuine understanding can still be gained even if it is `wrong' \cite{Fleisher2022,Sullivan2022,Sullivan2023,Boge2018,Boge2020b}\footnote{The authors of Ref.  \cite{Bordt2024} suggest XAI may be better understood as `applied statistics.'}. Therefore, perfect model fidelity is not necessary for insight into the black box \cite{Fleisher2022}, just as any good explanation will not be faithful to every aspect of a system \cite{Freiesleben2023}~\footnote{Explanations (of a black box) are relative to what we expect of them: The roles of the means and ends of an explanation of an ML model are often masked and may first become clear when trying to attain one  \cite{Buchholz2023}.}\footnote{Some authors suggest taking inspiration from how scientists study biological \cite{Kstner2024} or other complex systems \cite{Krakauer2023} to interpret the functional organization of trained ML systems.}.

Opening up the black box in these ways can resolve some but certainly not all aspects of opacity \cite{Krishnan2019,Sullivan2022,Fleisher2022}.
Even if a human learns which features of the world a model tracks and the role each unit of the algorithm plays in tracking those features, the features may still appear inscrutable or gerrymandered \cite{Sullivan2022,Fleisher2022}; these will not be immediately understandable \cite{Boge2021}. 
It is often unclear whether the hidden latent variables within the model (see Sec.~\ref{chapter:algorithms}) contain information about the source of the data or if the representations of the internal states are merely artifacts \cite{Humphreys2020}. Unlike classical models, where the intermediate steps of the prediction problem are construed a priori based on scientific intuition and theoretical interpretation \cite{Humphreys2020}, interpretations of (black box) deep learning model may stem entirely from the conceptualization of input and output \cite{Boge2021}~\footnote{Explanations of the ML are contingent upon the explanations of the physical theory which, in turn, rely on the scientific context, usage, and practices ~\cite[p.31]{vanFraassen2008-book}.}.

However, this review reports on many exceptions where physical intuition is employed to design ML algorithms in accordance with the relevant physical theory describing the data or target phenomenon of interest. Thus, incorporating and reflecting basic physical principles is essential for enhancing the interpretability of machine learning in the realm of physics, which may be unique, given the theoretical nature of this field. 

 %The diverse approaches and conclusions presented in this review emphasize the strong relationship between our capacity to interpret machine learning and our understanding of pertinent physical theories within the applicable domain.

\subsection{Scientific understanding and ML}
\label{subsec:understanding_and_ML}

Interpretable ML in physics is motivated differently than explainable ML (XAI) in non-scientific contexts \cite{Pez2019,Zednik2022,Freiesleben2024}. The primary question is how an explanation of ML can contribute to scientific understanding \cite{Rz2022,Rz2024,Boge2020,Sullivan2022,Sullivan2023}, a central aim of the scientific endeavor \cite{DeRegt2009-book,DeRegt2017-book}. While ML algorithms are designed to solve prediction tasks, an `oracle' that provides perfect predictions does not actually offer scientific insight; we would still want to ask `why' or `how' \cite{Krenn2022}. 

Explanations serve a variety of roles in science \cite{vanFraassen2008-book,DeRegt2005,Baumberger2017}, reflecting different modes of  scientific understanding \cite{DeRegt2017-book,Faye2014-book}. There are ways to understand that are not direct consequences of explanation. For example, a scientist can understand by foreseeing the consequences of a theory without performing a full-blown derivation \cite{DeRegt2005,DeRegt2017-book}.  A unificationist understanding advocates that science encapsulates more and
more phenomena into a single explanatory scheme \cite{Friedman1974,Friedman2001,Faye2014-book}.
Another type of understanding, known as modal understanding, allows one to grasp how a phenomenon might arise by exploring a space of possibilities, even using `false' heuristics to represent the world  \cite[Ch.6]{Grimm2017b-book}. According to the `picture theory of science' or \textit{Bildtheorie} \cite{vanFraassen2008-book},  scientiﬁc theorizing involves creating an inner picture or mental construction. Visualizability is a marker of understanding \cite{Cushing1994,Cushing1991,Faye2014-book}.

Scientific understanding is tightly woven with representation. When appropriately chosen or found, a representation generates understanding by resembling something that is already comprehended, such as a pictorial or simple mechanical model \cite{vanFraassen2008-book,DeRegt2017-book}. 
Idealized models thus play a crucial role in science \cite{Fleisher2022,Baumberger2017,potochnik2019idealization,elgin2017true,Sullivan2019,Grimm2021,Grimm2017b-book}. They serve as interpretable representations resembling aspects of the real world, connecting to specific features of it  \cite{vanFraassen2008-book,Frigg2006}.
Even if idealized models do not perfectly represent the target phenomena they aim to depict (they can be  `wrong'), they still provide valuable insights into them \cite{Boge2018,Boge2020b,Boge2021,Boge2023,Sullivan2022,Sullivan2023,Grimm2017b-book}. This perspective can apply to ML models, as well, which can be viewed as `toy models' as long as they facilitate genuine scientific inquiry. How they actually link to the target phenomena of interest needs to be assessed.

The open inquiry about relations between ML models and the real world lies at the heart of interpretable ML in physics, as highlighted in this review.
Utilizing interpretable ML or explainability techniques can thus be seen as part of a broader scientific exploration process \cite{Zednik2022,Duede2023}, which is fundamentally different from validation and justification \cite{Popper1935,Reichenbach1938,Simon1973,DeRegt2017-book}. 

This exploration is inherently constructive. 
ML can sometimes surpass human scientific understanding: When deep neural networks learn to predict based on well-generalizing features of vast datasets, which are poorly understood, they appear to learn concepts that humans lack~\cite{Boge2023}. By trying to interpret these black box algorithms,  new models could be fostered that better promote understanding \cite{Boge2021}.
In this review, we report on how understanding can spring forth from the connections between the ML model and the target phenomena that it is applied to by uncovering and interpreting the learned concepts~\footnote{Unsupervised ML methods are interesting and likewise challenging for discovery \cite{Boge2021,Huntingford2019}: They point to new connections within the target system of interest, yet the underlying physics concepts behind the connections may need to be found first.}.

% \textit{The literature reviewed in this article demonstrates that when interpreting the inner workings of an algorithm, linking it to the target phenomenon in terms of physical concepts relevant to it,  one can discover that ML models mimic these concepts.}

A `concept' may be inherent to living systems that process information about their environments \cite{allen1991concept}, hence may be emergent. Concepts and representations are intricately connected \cite{vanFraassen2008-book} (as are representation and computation \cite{Horsman2014}). As noted in Ref.~\cite[p.232]{allen1991concept}, ``An organism whose internal representations are concept-like should be able to generalize information obtained from a variety of
 perceptual inputs.'' Machine learning methods that aim for generalization across entire sets of models (for example, via ensemble methods, see \ref{chapter:algorithms}) or different categories of tasks (i.e. knowledge transfer), entail a particular type of robustness in their representations \cite{Freiesleben2023b,Gaimann2025,parascandolo2021learning} that can underscore the notion of learning concepts~\footnote{Explanations seem `good' if they are robust, which can be employed for ML approaches \cite{,parascandolo2021learning}.}. Yet, as there is no universal guarantee that these concepts will align with human scientific understanding \cite{Boge2023,Freiesleben2023}, a process of translation will generally be involved.

In the context of scientific understanding, how much theoretical insight concepts render is captured by how `illuminating' they are  \cite[Ch.2]{Grimm2017-book}: For concepts to be effective, they need to have the sort of information required to resolve questions regarding the \textit{why}, \textit{what}, or \textit{how}, in order to identify laws, causes, mechanisms \cite{Salmon1984,Salmon1988,DeRegt2005,Hempel1965}, unifying principles \cite{Friedman1974,Kitcher1981}, dependence relations, and other items relevant for scientific understanding.
In this sense, interpretable ML in physics is a new process to find illuminating conceptions about the physical world; whether humans or ML direct these discoveries is open~\footnote{While the identification and discovery (or validation) of new causal relationships are how fully automated science with ML could be realized, it is not clear whether humans will ultimately need to be involved in the identification process \cite{Humphreys2020}. (See the thematically related Sec.~\ref{subsec:AI_scientist}.)}.

%
%\footnote{Epistemic goals like description, prediction, and understanding, and explanation are affected, generally, in different ways with opaque models \cite{Cichy2019,Durn2018,Beisbart2021,Humphreys2008}.}

% \subsection{Chemistry and Materials Science}
%\subsection{Other fields}
%\label{chapter:related}

%

% Reweighted autoencoded variational Bayes for enhanced sampling (RAVE) \cite{Ribeiro2018}

% Unsupervised Machine Learning for Analysis of Phase Separation in Ternary Lipid Mixture
% \cite{Lpez2019}

% Latent Models of Molecular Dynamics Data: Automatic Order Parameter Generation for Peptide Fibrillization
% \cite{Charest2020}

% Deep molecular dreaming: inverse machine learning for de-novo molecular design and interpretability with surjective representations\cite{Shen2021}

% Interpretable machine learning model for the deformation of multiwalled carbon nanotubes\cite{Yadav2021}

% Decoding Interaction Patterns from the Chemical Sequence of Polymers Using Neural Networks
% \cite{Werner2021}

% Interpretable learning of voltage for electrode design of multivalent metal-ion batteries
% \cite{Zhang2022b}

\begin{figure*}
    \centering
    \includegraphics[width=\textwidth]{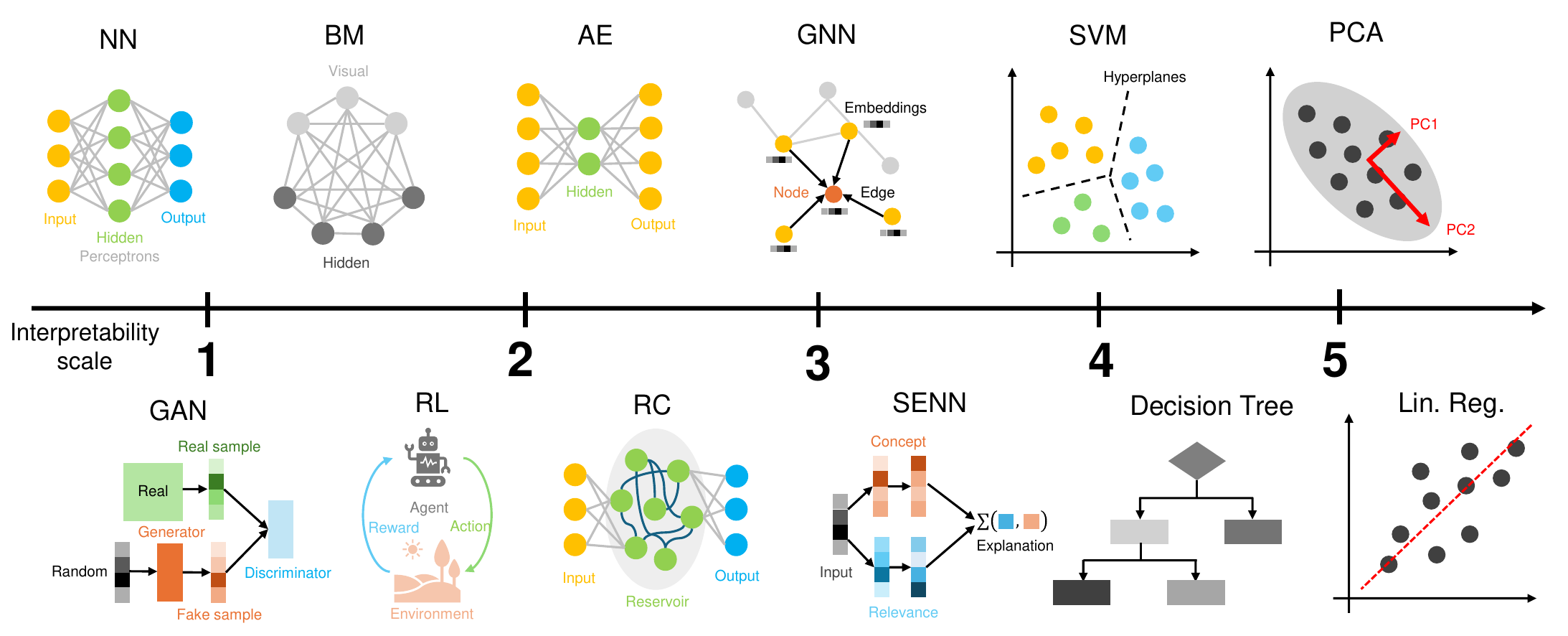}
    \caption{Machine learning methods and their intrinsic interpretability on a 5-point scale. Depicted algorithms and their interpretability scores are NN (Neural network, 1), GAN (Generative adversarial network, 2), BM (Boltzmann machine, 2), RL (Reinforcement learning, 2), AE (Autoencoder, 3), RC (Reservoir computing, 3), GNN (Graph neural network, 3), SENN (Self-explaining neural network, 3), SVM (Support vector machine, 4), Decision tree (4), PCA (Principal Component Analysis, 5), and Lin. Reg. (Linear regression, 5) respectively (from left to right).}
    \label{fig:MLalgo}
\end{figure*}

\section{Machine Learning Algorithms}
\label{chapter:algorithms}

This section briefly introduces popular machine learning methodologies employed in physics and indicates their interpretability on a 5-point scale. Figure \ref{fig:MLalgo} depicts (representative) architectures in each algorithm and its interpretability visually.

\subsection{Principal Component Analysis}
Principal Component Analysis (PCA) is a widely used dimensionality reduction technique in machine learning and data analysis~\cite{Hotelling1933, Pearson1901}. Its primary objective is to transform a high-dimensional dataset into a lower-dimensional subspace while retaining as much of the original data's variance as possible. PCA achieves this by identifying the orthogonal axes, called principal components, along which the data exhibits the most variation. The first principal component captures the highest variance, followed by the second, and so forth. By selecting a subset of these components, one can effectively reduce the dimensionality of the data while preserving its essential structure. 

\textbf{Interpretability: 5/5}: PCA is highly interpretable as it provides a clear mathematical framework for dimensionality reduction.

\subsection{Decision Trees}
Decision Trees (DTs) are a widely used class of supervised learning algorithms that model data using a tree-like structure. Each internal node represents a decision based on a feature, and branches correspond to possible feature values or conditions. The process continues recursively until reaching a leaf node, which represents a final prediction (for classification) or value (for regression)~\cite{Breiman2017}. Decision trees are particularly effective for structured data and problems requiring rule-based decision-making.

Decision trees are easy to interpret but prone to overfitting, especially when trees grow too deep~\cite{Quinlan1996}. To address this limitation, ensemble techniques like Random Forests~\cite{Breiman2001} and Boosted Decision Trees~\cite{Friedman2001} have been developed. These methods improve performance by combining multiple trees to reduce variance and bias, leading to more robust models.

\textbf{Interpretability: 4/5}:
Decision Trees are considered highly interpretable since their structure reflects human decision-making processes and allows for straightforward rule extraction. However, ensemble methods like Boosted Decision Trees sacrifice some interpretability due to the complexity of combining many small trees.

\subsection{Support Vector Machines}

Support Vector Machines (SVMs) are powerful machine learning models used for both classification and regression tasks~\cite{Cortes1995}. At their core, SVMs aim to find an optimal hyperplane in a high-dimensional feature space that best separates different classes or maximizes the margin between them. This hyperplane, also known as the decision boundary, is chosen to have the greatest distance between the nearest data points of each class, allowing for robust generalization to unseen data. SVMs can handle non-linearly separable data through a process called kernel trick, which implicitly maps the input space into a higher-dimensional feature space.

\textbf{Interpretability: 4/5}: SVMs are considered highly interpretable, particularly in their linear form.  Even in non-linear cases using kernel methods, understanding the influence of individual data points remains feasible. However, the interpretability slightly decreases when complex kernels are used, as the transformed feature space may become less intuitive.

\subsection{Neural Networks}

Artificial Neural Networks (ANNs) are computational models inspired by the structure of biological neural networks, designed to tackle complex tasks in machine learning~\cite{Fukushima1980}. Neural networks are computational systems consisting of nodes, called neurons or perceptrons, organized into layers ~\cite{LeCun2015}. In a neural network, each neuron processes information from its inputs, applies a transformation using weights, and passes the result to the next layer. Through a process of training, where weights are adjusted based on the error in predictions, neural networks can learn to approximate complicated functions and make accurate predictions or classifications.

What sets them apart is their abundance of free parameters, allowing them to represent highly non-linear relationships in data. This characteristic grants them formidable modeling capabilities, rendering them applicable to a diverse range of tasks. The surge in computational power, particularly with the advent of Graphics Processing Units (GPUs), has revolutionized ANN training, enabling the handling of huge datasets. Additionally, the availability of robust training libraries like TensorFlow~\cite{tensorflow} and PyTorch~\cite{Ansel2024} has democratized the application of ANNs, making them accessible to a wider community of researchers and practitioners. These advancements collectively propel ANNs to the forefront of modern machine learning and artificial intelligence.

\textbf{Interpretability: 1/5}: Neural Networks, especially deep models, are often considered black-box models due to their complex, layered structures and a large number of parameters. While methods like Layer-wise Relevance Propagation (LRP), SHAP, and saliency maps provide insights into individual predictions, their overall decision-making process remains difficult to interpret. Consequently, despite their high predictive power, their interpretability is low, making them challenging to use in critical applications where transparency is required.

\subsection{Autoencoder Neural Network} \label{sec:autoencoder}
Autoencoders are a class of neural network models designed for unsupervised learning and dimensionality reduction tasks. They consist of an encoder network, which compresses input data into a lower-dimensional representation, and a decoder network, which reconstructs the original input from this compressed representation. By training the model to minimize reconstruction error, autoencoders learn meaningful representations that capture important features of the data. Variational autoencoders (VAEs) are a specific type of autoencoder that extends the concept by introducing a probabilistic framework~\cite{kingma2022autoencodingvariationalbayes}. VAEs not only learn a deterministic mapping from input to hidden representation but also model the probability distribution of the hidden variables.

\textbf{Interpretability: 2/5}: Autoencoders provide moderate interpretability, as their latent space can reveal underlying structures in the data. However, without additional constraints, these representations may be difficult to interpret directly. Techniques such as disentangled representations can improve interpretability by ensuring that each latent dimension corresponds to a distinct feature of the data.

\subsection{Graph Neural Networks}
Graph Neural Networks (GNNs) are a class of deep learning models designed to operate on data structured as graphs, where entities are represented as nodes and relationships between them as edges~\cite{Scarselli_2009_graph}. This architecture enables GNNs to capture complex dependencies and interactions inherent in graph-structured data.

GNNs leverage a message-passing mechanism where each node iteratively aggregates information from its neighbors to update its representation~\cite{kipf2017}. This iterative process allows nodes to capture both local and global structural information, making GNNs particularly effective for learning from relational data and irregular structures.

\textbf{Interpretability: 3/5}: 

While GNNs provide powerful representational capabilities, their interpretability can be challenging due to the complexity of the graph structure and layered message-passing operations. Recent advancements focus on providing explanations for predictions by identifying influential subgraphs and node features. One notable method, GNNExplainer, highlights the most significant features and connections for a given prediction, enhancing transparency~\cite{Ying2019}.

\subsection{Self-explaining Neural Network}
Self-explaining Neural Network (SENN) \cite{alvarez2018towards} is a model that aims to incorporate interpretability into the architecture level by building the network progressively from simple linear classifiers to more complex models with explicit structure and tailored loss function. To achieve this, the model employs an interpretable concept encoder, which transforms input into a small set of interpretable features, and a parametrizer that generates relevance scores, helping to link these features to model predictions. The approach enforces faithfulness and stability through regularization, ensuring that the model's internal representations remain interpretable while maintaining predictive performance. With these, the representation learned by SENN satisfies important features for interpretable representation such as fidelity (preserve information), diversity (concepts do not overlap with each other), and grounding (human-understandable). Recent advances in SENN research have introduced improvements such as contrastive self-explaining networks (C-SENN) \cite{sawada2022c}, which further refine the learning of distinct and meaningful concepts while minimizing the overlap between them, thus enhancing the model’s interpretability across different domains.

\textbf{Interpretability: 3/5}: SENNs significantly improve interpretability over traditional deep neural networks by enforcing structured explanations at the architectural level. However, some level of interpretability trade-off remains, as understanding the meaning of learned concepts may still require domain knowledge or further refinement techniques.

\subsection{Reservoir Computing}
Reservoir Computing (RC) is a framework designed for processing time-dependent and sequential data by exploiting the dynamic properties of recurrent networks. It consists of a fixed, randomly initialized reservoir that projects input data into a high-dimensional dynamic space, while only the readout layer is trained. This simplifies learning and makes RC particularly effective for tasks like time-series prediction, speech recognition, and chaotic system modeling~\cite{Patankar2012, Lukoeviius2009}.

Key architectures include Echo State Networks (ESNs), which maintain the echo state property to ensure past input influence diminishes over time~\cite{Patankar2012}, and Liquid State Machines (LSMs), which simulate spiking neural activity for biologically inspired computation~\cite{Maass2002}. Recent advancements also explore integrating RC with deep learning and quantum computing, enhancing scalability and representational power~\cite{nakajima2021reservoir}.

\textbf{Interpretability: 3/5}: 
RC offers moderate interpretability due to its fixed reservoir structure and trainable output layer. While the internal dynamics can be complex, methods like dynamic systems analysis have been used to gain insights into reservoir behavior~\cite{Lukoeviius2009}.

\subsection{Boltzmann Machines}
Boltzmann Machines (BMs) are a class of stochastic neural networks introduced by Geoffrey Hinton and Terrence Sejnowski in 1985~\cite{Hinton1986}. They consist of symmetrically connected neurons that make binary stochastic decisions, enabling the network to learn complex probability distributions over input data. BMs are particularly useful for unsupervised learning tasks such as combinatorial optimization, feature learning, and density estimation~\cite{Ackley1985}.

A significant variant, the Restricted Boltzmann Machine (RBM), simplifies the architecture by restricting connections between neurons—only allowing connections between two layers: the visible layer (representing the input data) and the hidden layer (representing learned features)~\cite{Smolensky1986}. This bipartite structure eliminates intra-layer connections, making RBMs easier and faster to train compared to standard BMs. RBMs have been foundational in deep learning, particularly in constructing Deep Belief Networks (DBNs), where multiple RBMs are stacked to learn hierarchical representations~\cite{Hinton2006}.

\textbf{Interpretability}: 

\textbf{Interpretability of BMs: 2/5}. Due to their fully connected and stochastic nature, BMs often suffer from limited interpretability. The complex interactions between all units make it challenging to discern how specific inputs influence outputs directly.

\textbf{Interpretability of RBMs: 3/5}. The bipartite structure of RBMs enhances interpretability compared to standard BMs. In classical machine learning tasks, this structure allows for insights into how input features activate hidden units, and visualizing the learned hidden features can offer partial interpretability, especially in applications like image recognition~\cite{Hinton2006}.

In the context of quantum state representation, RBMs have demonstrated notable interpretability. They can effectively model complex quantum systems, capturing entanglement structures and quantum correlations~\cite{Carleo2017}. This capability makes the hidden units interpretable in terms of physical quantum properties, facilitating tasks such as quantum state tomography and variational studies in quantum computing~\cite{Torlai2018}. 

\subsection{Generative models} 

This is an umbrella term for a diversity of different architectures and theoretical approaches. Examples include contrastive learning methods, such as Generative Adversarial Networks (GANs)~\cite{Goodfellow2014}, approximate inference methods, such as Variational Autoencoders (VAEs)~\cite{kingma2022autoencodingvariationalbayes}, newer diffusion-based generative models~\cite{ho2020denoisingdiffusionprobabilisticmodels}, and autoregressive models, such as Transformer-based models (e.g., GPT, XLNet) and sequence-generation architectures like PixelCNN and PixelRNNn~\cite{Vaswani2017}. Many Large Language Models (LLMs), including GPT, are based on autoregressive modeling, generating text token-by-token in a sequential manner. Their common purpose is to emulate the hidden data-generating process that produces the observed training data—a form of reverse engineering that constructs a probabilistic model capable of generating new samples with similar qualitative features as the data. These methods often entail a stronger degree of generalization, making them valuable for a wide range of machine learning applications.

\textbf{Interpretability: 2/5}: Generative models generally have low interpretability since they primarily aim to replicate complex data distributions rather than provide explicit decision-making processes. Some methods, like VAEs (see Section~\ref{sec:autoencoder}), allow for a structured latent space that can be partially understood. LLMs based on the transformer architecture encode pairwise affinities between input features.  GANs—lack inherent interpretability, as their generator and discriminator networks operate as black-box functions. Techniques such as latent space disentanglement and concept attribution can improve interpretability but are not intrinsic to most generative models.

\subsection{Reinforcement Learning} Reinforcement Learning (RL) is a learning framework where an agent interacts with an environment by taking actions to maximize cumulative rewards over time. The agent observes the state of the environment, selects actions based on a policy, and receives feedback through rewards. Over repeated interactions, the agent learns to optimize its policy by balancing exploration (trying new actions) and exploitation (choosing the best-known actions)~\cite{Kaelbling1996}.

Key RL algorithms include Q-learning~\cite{Watkins1992} and Deep Q-Networks (DQN), which combine value-based learning with deep neural networks~\cite{Mnih2015}. More advanced techniques, such as Policy Gradient and Actor-Critic methods, directly optimize policies through gradient-based approaches~\cite{schulman2017trustregionpolicyoptimization}. RL has achieved notable success in various domains, including game-playing, robotics, and autonomous systems.

\textbf{Interpretability: 2/5}: RL models typically have low interpretability due to the complexity of learned policies and stochastic decision-making. Recent efforts aim to improve transparency by visualizing policies and incorporating symbolic reasoning~\cite{puiutta2020explainablereinforcementlearningsurvey}. An additional approach, Projective Simulation (PS), introduces a memory-based framework where the agent's decision-making process is modeled through a network of "clips" representing percepts and actions~\cite{Briegel2012}. This structure provides more insight into the agent’s reasoning by making the sequence of decisions and learned associations more transparent and easier to interpret.

\section{Existing Interpretability Methods}
\label{chapter:interpretability_methods}

This section introduces computational methods to achieve interpretability that are extensively used in machine learning and computer science. Most of these methods aim to understand the rationale behind the model decision, typically a result of a classification or a regression. Understanding the characteristics and limitations of these methods is crucial when one attempts to apply these methods (or their variants) to physics and other domains of science.

\subsection{Feature importance}
Feature importance or attribution is fundamental in machine learning interpretability, serving as a tool to assess which input variables significantly influence a model's predictions. Early foundational work, such as Breiman's introduction of permutation importance in the context of Random Forests \cite{Breiman2001}. This method involves shuffling a feature's values to measure its impact on prediction accuracy, providing a straightforward, model-agnostic approach to gauging feature influence. In the mid-2010s, Layer-wise Relevance Propagation (LRP) \cite{Bach2015} was introduced to interpret features in multi-layered perceptrons. LRP propagates the model's output backward through its layers, distributing relevance scores to the input features. This method is particularly valuable for deep learning models, allowing practitioners to visualize which input features contribute most to a specific prediction. LRP has become a cornerstone in neural network interpretability due to its capacity to handle non-linear relationships effectively. 

In parallel, SHAP (SHapley Additive exPlanations) \cite{lundberg2017unified} leverages concepts from cooperative game theory to assign each feature an importance value based on its contribution across all possible combinations of features. This approach not only ensures consistency but also provides both global and local interpretability, offering deep insights into model behavior across various inputs.

% Added attention/transformer/LLM part
Perhaps the most widely used approach in this area is the attention mechanism \cite{Vaswani2017} and the architecture that uses it. Attention mechanism is a process of computing the relative importance between two features (usually representations in hidden layers), which allows the neural network to learn the importance of given information on its own. Furthermore, it allows the learned importance and the corresponding features to be directly decoupled and visible in an explicit form, which greatly contributes to the interpretability of the model. It was initially used to calculate the relative importance of hidden states of RNNs \cite{bahdanau2014neural}, and since the development of the foundational Transformer structure \cite{Vaswani2017} and subsequent large language models (LLMs), it is now widely used in all domains of machine learning. 

\subsection{Latent representations}

Latent representation learning is central to understanding deep machine learning models, as it maps high-dimensional data into a lower-dimensional space, often revealing hidden patterns. Disentangled representations, in particular, seek to isolate distinct, interpretable factors of variation within this latent space. The importance of these methods lies in their potential to enhance transparency by associating latent dimensions with specific semantic attributes, thereby shedding light on their decision-making processes. For instance, methods like InfoGAN \cite{chen2016infogan} enforce the relationship between certain labels and latent dimensions via mutual information loss, yielding pre-determined and easily understandable latent representations.

One of the most notable works in this direction is $\beta$-VAE \cite{higgins2017beta}, a variation of the variational autoencoder, which uses a regularization term to promote disentanglement by balancing reconstruction accuracy with latent space independence. This seminal paper demonstrated that encouraging orthogonality between latent dimensions leads to more interpretable representations. Later research \cite{chen2018isolating} explored the factors affecting disentanglement in VAEs, isolating key conditions necessary for achieving distinct factorization of data variability. 

More recently, researchers have focused on expanding the concept of interpretability and attempted to provide a diverse range of explanations with given latent representations. For instance, SimplEx \cite{crabbe2021explaining} presents a corpus of training examples that contributes the most to the model prediction and also notifies the most important features that determine the prediction. Another notable example is the human-interpretable representation learning (HRL) framework \cite{Marconato2023}, which integrates causal representation learning with human-centered metrics. Together, these advancements underscore the evolution from purely mathematical latent spaces toward human-aligned, explainable representations—an essential progression for creating trustworthy machine-learning models.

\subsection{Hessian-based methods}

Hessian-based methods in machine learning use second-order derivatives, represented by the Hessian matrix, to analyze how a model’s predictions change concerning its inputs or parameters. The Hessian matrix captures the curvature of the loss function, providing crucial information about the local landscape around a model's parameters. This curvature analysis helps in understanding model sensitivity, stability, and robustness --- key factors in ensuring interpretability. \cite{Murdoch2019}. While it provides a limited range of insights into system behavior compared to other more sophisticated methods, its model-agnostic feature makes it versatile enough to be applied to diverse fields in machine learning.

Initially applied in the 1990s for feature selection and network pruning \cite{Hassibi1992}, Hessians provided early insights into the internal workings of computational models. By the 2010s, researchers leveraged Hessian matrices to explore loss landscapes of deep neural networks \cite{Lakemond2011}, offering a deeper understanding of model behavior and robustness. In recent years, practical implementations of Hessian-based methods have emerged, including toolkits \cite{Dawid2021Hessian} that assess feature contributions and model uncertainty across domains. These tools have extended their impact beyond traditional benchmarks, finding applications in scientific fields like physics, where transparency is crucial for validating complex models.

\section{Symbolic Regression}
\label{chapter:symbolic}

\begin{figure*}
    \centering
    \includegraphics[width=1.0\linewidth]{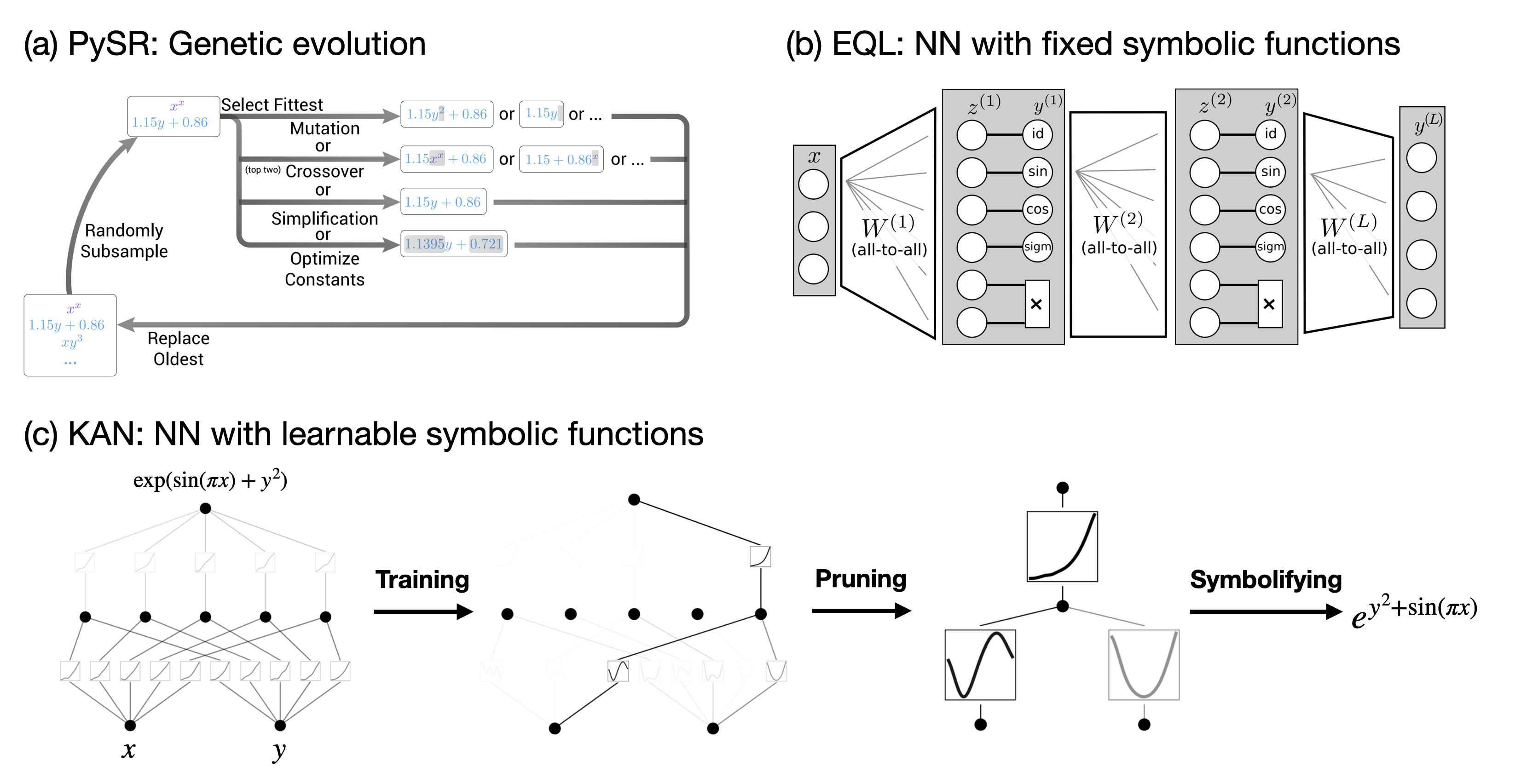}
    \caption{Three representative methods for symbolic regression. (a) PySR uses genetic programming to evolve symbolic formulas, similar to the survival of the fittest principle in Darwinism evolution~\cite{cranmer2023interpretable}. (b) Equation Learner (EQL) replaces activation functions in multi-layer perceptrons with hand-encoded symbolic functions (e.g., sin, exp, multiplications)~\cite{martius2016extrapolation}. (C) Kolmogorov-Arnold Network (KAN) has learnable activation functions parameterized by B-splines, which are matched and converted to symbolic functions after training~\cite{liu2024kan}.}
    \label{fig:symbolic_regression}
\end{figure*}

%[Summary: Symbolic regression methods have two main categories, based on genetic algorithms (GA) and based on neural networks (NN). GA: PySR, Eureqa, RL for GA. NN: AI Feynman, EQL and OccamNet, KAN, transformer that output the formula].

A core challenge for physics is finding symbolic expressions that match data from an unknown function. This is a challenging task in general. %, e.g., it took John Kepler four years to figure out that the orbit of Mars is in fact an ellipse. \raban{I guess that this is not the best example for symbolic regression. The formula of Mars as seen from Earth is quite complex, where it is pretty easy as seen from the sun. We showed in~\cite{Iten2020} that a neural network can switch the representation from the orbits as seen from Earth to the angles as seen from the sun. Hence, I would use another example here and we could mention Kepler's example in the context that symbolic regression may only be the right approach in case one has given a nice representation of the parameters.} \zm{That makes total sense. Any suggested example?}. \zm{I have temporarily deleted this example since I didn't find another significant example in history that is discovered purely from data. I was trying to go for Coulomb's law, but the discovery requires a mixture of theory and experiments. }
More specifically, given data samples structured as $\{x_1,x_2,\cdots,x_n;y\}$, the goal of symbolic regression (SR) is to find a symbolic expression $f$ such that $y=f(x_1,x_2,\cdots,x_n)$. In the narrow sense, ``symbols'' considered by symbolic regression involve basic unary functions (e.g., polynomials, sine, exponential) and binary functions $\{+,-,\times,/\}$, but one can in principle add more functions into the hypothesis space. For example, when studying vibration on a spherical membrane, the library should include Bessel functions, which can characterize vibration eigenmodes. 

Since the space of symbolic formulas is intrinsically discrete, methods are divided to discrete search and continuous search (followed by discrete snapping). Discrete search is exemplified by genetic algorithms, while continuous search usually leverages neural networks. 

\textit{Genetic programming} works with discrete symbols directly. Symbolic formulas can be written down as expression trees (see Figure~\ref{fig:symbolic_regression} (a)) consisting of symbols, analogous to DNA sequences consisting of ATGC bases. Borrowing the idea of how evolution works for biological lives, genetic algorithms introduce mutation and crossover for symbolic formulas and iteratively filter out formulas with the lowest fitness (``survival of the fittest''). After many generations of evolution, a genetic algorithm will provide one formula or many formulas that trade-off between simplicity and accuracy. Genetic algorithms require heuristics to speed up the search. Algorithms using human-designed heuristics include Eureqa~\cite{Schmidt2009}, PySR~\cite{cranmer2023interpretable}, GPLearn and  Operon~\cite{burlacu2020operon}. Instead of manually designing search rules, search policy can also be learned using reinforcement learning~\cite{petersen2021deep,mundhenk2021symbolic}. Ashhab~\cite{ashhab2024using} suggests modifications to SR approaches to make them better suited for the purpose
of finding exact solutions as opposed to good approximations.

\textit{Network-based methods} usually leverage neural networks and use the gradient signal enabled by automatic differentiation to search in the relaxed continuous space. However, to finally obtain symbolic formulas, snapping needs to be performed frequently or at least in the end. Methods include Equation Learner (EQL)~\cite{martius2016extrapolation}, OccamNet~\cite{dugan2020occamnet}, Kolmogorov-Arnold Network (KAN)~\cite{liu2024kan}, Symbolic metamodeling~\cite{alaa2019demystifying}. AI Feynman~\cite{udrescu2020ai,udrescu2020ai2}, end-to-end SR with transformers~\cite{kamienny2022end}, etc. These neural network-based methods differ mostly in how they inject symbolic inductive biases into their models, and how they perform the symbolic snapping which turns neural networks with continuous weights into discrete symbolic formulas. EQL and OccamNet replace standard activation functions in neural networks (e.g., ReLU) with fixed symbolic functions (see Figure~\ref{fig:symbolic_regression} (b)). When strong inductive biases are available such that the target can 
be written down as a sparse linear combination of known symbolic bases, the network is effectively one layer and hence can be solved via sparse linear regression - a notation example is 
SINDy~\cite{brunton2016discovering}. Kolmogorov-Arnold networks have learnable activation functions parameterized as B-splines, which are snapped to symbolic functions after training (see Figure~\ref{fig:symbolic_regression} (c)). AI Feynman aims to recursively simplify the dataset by testing various hypotheses (e.g., ``Does the have translational invariance?'') with neural networks. Transformers treat symbolic regression as a sequence generation task, which has a continuous decision-making process inside the model (albeit uninterpretable to humans) and hence effectively performs some continuous search.

Symbolic search algorithms are typically employed to solve regression problems. However, they can also be employed in neural network interpretation tasks to uncover closed-form concepts encoded in any arbitrary neuron in neural networks~\cite{https://doi.org/10.48550/arxiv.2409.05305,https://doi.org/10.48550/arxiv.2401.04978}.

It is worth mentioning that symbolic regression by itself is a large field, and this paper only focuses on physics-related applications and methods. For a more comprehensive review for this field, please refer to other reviews~\cite{makke2024interpretable,angelis2023artificial}.

%\raban{I would probably put the following reference here (it was listed in "General Physics"). What do you think? Just put it back if you think it fits and I will integrate it.}
%Using machine learning to find exact analytic solutions to analytically posed physics problems
%\cite{ashhab2024using}
%\zm{Thanks for the reference! I have integrated it into this part.}

%\mk{I found this reference for a physics-inspired interpretability algorithm using energy landscape methods, not sure of its relevance for physics, though: \cite{niroomand2023physicsinspired} }

%\newpage

%\zm{Has anyone covered Steve Brunton, and Nathan Kutz's work in their parts (e.g., SINDY)? If not, I should probably mention them in symbolic regression.}

\section{Quantum Systems}
\label{chapter:quantum}

Quantum mechanics is undeniably the best-tested theory of how the world works at atomic and subatomic scales. It is rooted in the abstract: The underlying rules of physics escape our normal intuitions, but, are rigorously describable in a mathematical formalism. 
In quantum systems, both entanglement an symmetries play  key roles. 
Moreover, the time evolution of a (many-body) quantum system is complex, but, often formally describable in well-defined contexts. Quantum systems thus offer highly fruitful grounds for interpretable ML. 
Interpretable quantum ML realized on quantum computing systems is a new and emerging field that has so far received lesser attention. It requires a reconsideration from both a resource (feasibility) and conceptual standpoint, as quantum ML systems operate on fundamentally different physical and computational principles.

\subsection{Entanglement / Quantum Experiment Discovery}

An intuitive first question may be, `How do classical ML algorithms represent quantum systems'?
Entanglement, a concept central to quantum systems, appears behind how they learn representations of small systems:
In Ref.~\cite{https://doi.org/10.48550/arxiv.2306.05694}, "Explainable Representation Learning of Small Quantum States", variational autoencoders were trained on two-qubit density matrices generated by a parameterized quantum circuit, where the entanglement properties are controlled by a single angle. The latent representation directly correlates with the entanglement measure concurrence, and this representation also generalizes to other two-qubit systems. 
(See also Ref.~\cite{Schindler2017} discussed in Sec.~\ref{chapter:phase_diagrams})

In "Entanglement clustering for ground-stateable quantum many-body states" \cite{Matty2021}, the authors use operators relevant for computing entanglement to enable an unsupervised dimensionality reduction and clustering technique to separate ground-stateable wave functions from excited-state many-body wave functions--in a visual, hence interpretable way. Swap operators, which are useful for estimating the entanglement entropy in other numerical contexts \cite{Hastings2010}, are applied onto the wave function data, acting on differently-sized subsystems. Thereafter, the UMAP technique projects this data into two dimensions. Finally, using k-means clustering, ground-stateable functions are visually separated from non-ground-stateable functions. Subclustering of the UMAP projection corresponds to semantically meaningful variations in the used wavefunction datasets.

\textit{Discovering quantum optics experiment setups:}
Experiments on quantum systems have a basic property that makes them unintuitive and it is difficult to foresee the consequences of alterations: There is an intertwining relationship between the setup structure of the experiment and its entanglement properties. Therefore, Ref. ~\cite{FlamShepherd2022} used ML to propose such setups in a generative and exploratory fashion, further asking whether we can then interpret the underlying reasons why these may be suitable.
In the article entitled, "Learning interpretable representations of entanglement in quantum optics experiments using deep generative models", the authors use a variational autoencoder to generate experiments involving even high degrees of entanglement. They uncover intricate but interpretable representations of quantum optics experiments that connect the setup with entanglement properties.
This work suggests that ML can efficiently drive an experimental scientific discovery process while rendering systemic understanding in the making.
A broader follow-up discourse on driving discoveries with ML is found in the related Ref.~\cite{Krenn2022} entitled "On scientific understanding with artificial intelligence". 
  
In "Deep Quantum Graph Dreaming: Deciphering Neural Network Insights into Quantum Experiments" \cite{Jaouni2024}, the authors implement an explainability technique from computer vision, so-called \textit{inception} and \textit{dreaming}, on properties of quantum systems--in the same context of quantum experiment setups.  This allows the users to subsequently `ask' the neural network to imagine a quantum system with a specific property, and how to modify the quantum system to change one property. The authors found that the initial layers are attributed to simple structures, while in deeper ones, quantum entanglement emerges as an identifiable concept.

\subsection{Quantum Condensed Matter}

The authors of "Machine learning of Kondo physics using variational autoencoders and symbolic regression" \cite{Miles2021b} in the context of spectral function reconstruction, i.e. the experimentally measurable functions that contain rich information about a many-body quantum system.  They train what are known as $\beta-$VAEs with varying regularization strengths. To interpret the latent space, they deactivate all but the two remaining neurons that entail predictive power, and tune the values, observing the output. They also plot the projection of the training dataset in latent space and color points according to relevant physical parameters, finding that each latent space variable corresponds to a key physical descriptor for the set of spectral functions used: The VAEs automatically discovered some descriptors like the Kondo temperature, a measure of particle-hole asymmetry, and the presence of competing energy scales. To gain confidence that the VAEs learn meaningful representations, they employed a technique of choosing paths between data points in a two-dimensional interpretable space of the spectral function data and then observe how the topology of the dataset is preserved in the latent space. The authors also use symbolic regression to extract analytic expressions for the physical descriptors in latent space.

\textit{Topological invariants: }
Machine learning has been shown to be able to identify topological invariants in various contexts of solid-state physics. 
The authors of "Deep learning topological invariants of band insulators" \cite{Sun2018} design and train deep neural networks to predict topological invariants for insulators, i.e., the winding number or the Chern number. Despite the complexity of the neural network, one can see that the output of certain intermediate hidden layers resembles either the winding angle for models or the solid angle related to the Berry curvature. This indicates that neural networks essentially capture the mathematical formula of topological invariants. 
In "Unsupervised interpretable learning of topological indices invariant under permutations of atomic bands" \cite{Balabanov2020}, the authors develop a neural network-based protocol to identify topological indices of Bloch Hamiltonians. The neural networks are partly interpretable due to a feature map of the last convolution layer rendering information that is momentum-resolved, which helps in singling out relevant high momentum points, for example. With proper data preprocessing and selection, the protocol ultimately renders the standard classification of band insulators. 
Thematically related is also Ref.~\cite{Zhang2020}, discussed in Sec.~\ref{chapter:phase_diagrams}.

\subsection{Learning a Time Evolution / Dynamic} 

Ref. \cite{https://doi.org/10.48550/arxiv.2111.02385}, entitled "Discovering hydrodynamic equations of many-body quantum systems", uses symbolic regression techniques to extract effective hydrodynamic equations of motion (i.e., partial differential equations) for quantum many-body model systems (XXZ model, 1D Fermi-Hubbard). A key point of novelty in their approach is in discovering the long-wavelength limit of
exact or approximate semiclassical equations directly. Beyond verification in known models, the authors uncovered equations for cases in which an analytical derivation is still outstanding and were able to derive solutions with the intermediate help of ML.

The authors of "Learning quantum dynamics with latent neural ordinary differential equations"~\cite{Choi2022} adopt neural ordinary differential equations (NODE) techniques to learn first-order equations of motion for closed and open quantum system dynamics, rendering a generative method for dynamics with extrapolative qualities. Intriguingly, these extrapolation abilities coincide with fulfillment of the von Neumann and time-local Lindblad master equations for closed and open quantum systems, respectively, which happens without added supervision. The Heisenberg uncertainty principle is recoverable from the learned dynamical representation. 

The method of "Solving The Quantum Many-Body Hamiltonian Learning Problem with Neural Differential Equations"~\cite{https://doi.org/10.48550/arxiv.2408.08639} learns the quantum many-body Hamiltonian as an operator behind quantum trajectory data; the aim is to construct propagators in a way that facilitates interpretation. The authors present a combined approach of neural ordinary differential equations and a model \textit{ansatz} for the Hamiltonian. By providing a basic structure of the Hamiltonian (an inductive bias), robustness, accuracy, and generalization were boosted, but also interpretability.  On the technical level, curriculum learning  emphasized the  \textit{ansatz} of the form of the operator for the model's generalization power as opposed to the expressiveness of the black-box neural network, highlighting its importance for enhanced interpretability.

Similarly, the authors of Ref.~\cite{Yao2023}, "Emulating quantum dynamics with neural networks via knowledge distillation",  present a training framework involving curriculum learning to predict the time evolution of quantum wave packets propagating through a potential landscape, with increasing complexity of the landscape down the training curriculum. Upon employing feature attribution techniques (direct gradients) for post-hoc interpretation, the Cauchy-Riemann relations were re-discovered, a mathematical underpinning for complex differentiable functions. Moreover, new facts about the emulated quantum system were extractable, including detected symmetries and the relative importance of contributing physical processes. 

The approach of Ref. \cite{Cemin2024}, "Inferring interpretable dynamical generators of local quantum observables from projective measurements through machine learning", allows one to infer the dynamical generator governing the evolution of local observables in a many-body sub-system--from a finite set of local measurements at randomly selected times (noisy data). An `optimal' time-independent Markovian description of the evolution is extracted. The interpretability of the method lies in the ability to read out the Hamiltonian and the dissipative dynamical processes in the open systems in terms of jump operators.  

In "Learning the dynamics of Markovian open quantum systems from experimental data" \cite{https://doi.org/10.48550/arxiv.2410.17942}, the authors present a Bayesian learning scheme involving Monte Carlo sampling to extract dynamical models from the underlying data: The method outputs a ranked list of interpretable master equation models of different excitation processes and corresponding rates. 

\subsection{Neural Network Quantum States}

Machine learning models like Boltzmann Machines can efficiently represent quantum many-body states via parametrizing families of trial functions~\cite{Carleo2017,Carleo2018,Pastori2019}, and are optimized using variational Monte Carlo methods. 
It was shown that they are formally connected to tensor networks \cite{Chen2018}, i.e., mathematical structures highly effective at representing quantum many-body systems in the realm of analytical theory. This has inspired some researchers to recast tensor-network-based ML algorithms as inherently interpretable for quantum systems.
For one thing, in Ref.~\cite{Carleo2018}, the authors describe that by changing the network depth from one to two, one moves from a description of a classical spin system to one for describing quantum mechanical states. 

The authors of "Correlation-enhanced neural networks as interpretable variational quantum states"~\cite{Valenti2022} introduce a neural-network variational ansatz that can be customized to a system of interest via the correlations that are expected to dominate. They add an extension to restricted Boltzmann machines by introducing coupling terms in the internal energy function that reflect physical intuition. From the technical side, this leads to a significant increase in the precision and flexibility of the neural nets, while attaining simplicity in the number of optimizable parameters. 

The authors of "Approximately-symmetric neural networks for quantum spin liquids" \cite{https://doi.org/10.48550/arxiv.2405.17541} propose a family of ansatz functions that are approximately symmetric, introducing an inductive bias that aids the search for ground states while maintaining the
flexibility to capture complex quantum states (e.g. spin liquids) that are not exactly symmetric. They provide a partial interpretation of the internal learning process of different physical features in different sections of the network: They detect a remniscent of a quasi-adiabatic continuation \textit{ansatz} of Hastings and Wen (see corresponding Ref. for details);  by a two-step training procedure, starting with only activating the symmetric sub-network and then subsequently the non-symmetric part, they test this hypothesis.

The authors of "Physics-informed Transformers for Electronic Quantum States" ~\cite{https://doi.org/10.48550/arxiv.2412.12248} aim to enhance the interpretability of transformer-based representations of quantum many-body states by modifying the framework to first construct a numerical second-quantized basis that leverages prior physical knowledge, serving as a reference state approximating roughly the true ground state. In this basis, a transformer is then used to parametrize and autoregressively sample the corrections to the reference state. This approach allows for a recovery of the natural order of the different basic states according to their energy levels. 

\subsection{Quantum-Inspired Interpretable ML}
\label{subsec:quantum-inspired_ML}

An emerging discussion is to interpret or create intrinsically interpretable machine learning from the vantage point of quantum mechanical systems and the mathematical structures they embody. 
Ref.~\cite{Planat2024} analyzed the topological structure of ChatGPT--a black-box large language model--drawing on concepts from topological quantum computing. They draw associations directly between natural language processing and the topological structure of anyons from $SU(2)_k$ theories, arguing that the modular tensor structure could offer insights into the algorithms at a systemic level (i.e. robustness, adaptability, etc.) 

\textit{Tensor neural networks:}
The authors of Ref.~\cite{Ran2023}, "Tensor Networks for Interpretable and Efficient Quantum-Inspired Machine Learning", argue that tensor neural networks offer interpretability, as physically-motivated properties like the von Neumann entropy and mutual information are extractable \cite{https://doi.org/10.48550/arxiv.2412.15826} (which are related to entanglement). These can be used to uncover the information content of individual features in data and their contextual dependencies, which may be useful for anomaly detection, for example, \cite{Bai2022,https://doi.org/10.48550/arxiv.2401.00867}. The authors of Ref.~\cite{Bai2024}  claim one can interpret better the representation and generalization power of the ML methods themselves, specifically, as functions of scaling laws of the entanglement entropy \cite{Pastori2019}.

\textit{Other}: On a different line of thinking, Ref.~\cite{https://doi.org/10.48550/arxiv.2406.17583} presents a framework for interpretable ML based on category theory (a mathematical theory behind some classes of quantum field theories). They claim that quantum, deterministic, and probabilistic models can all be treated. Related is also Ref.~\cite{https://doi.org/10.48550/arxiv.2401.08585}.

\subsection{Quantum Information Devices}

This subsection taps into interpretable ML for device characterization, calibration, and control in the context of quantum computing and information. The main message is that physics `knowledge infusion', i.e, providing sufficient inductive biases about quantum mechanical dynamics and other dynamical laws at relevant scales of a measurement device, can majorly improve the ML model's performance but also is utility--for example, in order to be able to know how to improve the device setup or increase the effectivity of quantum control. 

The article "Quantum-Tailored Machine-Learning Characterization of a Superconducting Qubit" \cite{Genois2021} considers characterizing the dynamics of a quantum device and learning device parameters with ML. The aim is to design ML models exploiting domain knowledge of quantum trajectories and dispersive measurements in circuit QED to enhance interpretability, as well as performance--to be able to associate the model with an explicit physical description in order to assess it. A recurrent neural network was modified to better incorporate quantum mechanical effects in its loss function and to respect an underlying stochastic differential equation known to govern the device dynamics. The approach allowed for an efficient calculation of quantum device efficiency (of the measurement chain) and 
extracting theoretical knowledge: the Hamiltonian and Lindblad operators of the experimental system. 
In "Experimental gray box quantum system identification and control" \cite{Youssry2024}, the authors employ `gray box' models that incorporate knowledge of quantum mechanics to first construct a physical model and subsequently design an optimal quantum control protocol. The interpretable mathematical structure of the trained model offers a basis for intuition for practical improvements.  (See also the related Ref.~\cite{Youssry2020}).

In "Automation of quantum dot measurement analysis via explainable machine learning" \cite{Schug2025}, the authors employ explainable boosting machines \cite{Lou2013}, a method that enables an ML model to become interpretable directly (rather than relying on surrogate explanations (i.e., LIME and SHapley) for the purpose of image classification. These images represent measurements acquired during the tuning process of quantum dot devices. Features therein need to be properly analyzed to guide subsequent tuning steps, which is where ML interpretability plays a practical role.

\subsection{Quantum Machine Learning / Quantum Circuits}

In comparison to classical ML algorithms, much less literature has arisen addressing the problem of interpretability in the context of quantum ML algorithms \cite{https://doi.org/10.48550/arxiv.2211.01441,https://doi.org/10.48550/arxiv.2211.04343,
https://doi.org/10.48550/arxiv.2301.09138,
Pira2024,https://doi.org/10.48550/arxiv.2412.14753,Tian2024,Mercaldo2022}. Parametrized quantum circuits are the most promising approach to quantum ML to date \cite{GilFuster2024}. 
A general challenge is interpreting the relationship between input variables (e.g., Hamiltonian parameters) and output quantum states \cite{Mercaldo2022,Pira2024,https://doi.org/10.48550/arxiv.2301.09138,https://doi.org/10.48550/arxiv.2211.01441,Burge2023}.
The authors of  "Toward Transparent and Controllable Quantum Generative Models" \cite{Tian2024} suggest using model inversion, i.e., tracing generated quantum states back to their latent variables, to enhance both interpretability and controllability of quantum generative models (implemented on quantum neural networks). 
One of the most popular methods of generating post-hoc explanations involves Shapley values.
Indeed, the question of explainability of a quantum algorithm can go down to explaining the quantum circuit itself: The authors of "Explaining Quantum Circuits with Shapley Values: Towards Explainable Quantum Machine Learning"~\cite{https://doi.org/10.48550/arxiv.2301.09138} use Shapley values to evaluate the importance of each quantum gate to a given prediction. 
Refs.~\cite{https://doi.org/10.48550/arxiv.2211.01441,Burge2023,https://doi.org/10.48550/arxiv.2412.14639} consider how to calculate Shapley values in quantum ML systems; as measuring a quantum system
destroys the information, it is difficult to translate classical concepts of Shapley values for post-hoc explanations in a way that is computationally efficient.  

In general, quantum ML systems are fundamentally different regarding computational complexity and sources of uncertainty; how precisely or feasibly we can interpret them with post-hoc methods depends on basic limitations.
The authors of Refs. \cite{https://doi.org/10.48550/arxiv.2412.14753,Burge2023,https://doi.org/10.48550/arxiv.2211.01441}  discuss some of these: Firstly, the no-cloning theorem limits the intrinsic algorithmic capabilities, but not necessarily what we can explain about them. 
The phase space of a quantum circuit expands exponentially with the number of qubits, complicating efforts to execute explainability methods in polynomial time: How intermediate states can be represented is strongly limited, i.e., by storing them on classical devices with limited memory; yet, newer techniques for quantum time-efficient approximate representations may be underway.  
Function evaluation, i.e., measurements on quantum circuits, introduces probabilistic errors that impact the convergence of any explanation method. 
Finite shot noise constrains evaluations in polynomial rather than exponential time. 
Moreover, the precision needed to estimate the model limits the precision of calculating the gradients,  a general challenge in quantum machine learning. Hence any explainability method that requires calculating such gradients is subject to this constraint and will be unsuitable when gradients become exponentially small. Yet, quantum ML methods that are linear to a high degree (hence intrinsically interpretable) are superior from a technical standpoint. Nonlinearity is attributed to the parametrization of the unitary gates. The data encoding strategy on the input gates plays a key role for model expressivity \cite{Schuld2021,https://doi.org/10.48550/arxiv.2312.15124}, an aspect to keep in mind when construing interpretable algorithms.

The authors of "Opportunities and limitations of explaining quantum machine learning"  \cite{https://doi.org/10.48550/arxiv.2412.14753} propose two contrasting explainability methods for parametrized quantum circuits, which highlight the distinction between quantum-efficient versus classical explanation techniques. The first is a black-box explanation technique that tests a set of hypotheses, utilizing the Fourier picture of parametrized quantum circuits. Ideally, the total quantum complexity should be the total number of calls to the quantum model times the complexity of evaluating the model, and other computational efficiency issues should not come into play in a major way. The method successfully retained more useful information compared to a classical counterpart. The second method estimates the quantum state in classical memory, adopting the philosophy of `divide and explain' behind the layerwise relevance propagation XAI technique introduced in Ref.~\cite{Montavon2018}.  It offers two-step explanations that take the structure of the intermediate quantum feature maps into account.
The method proposed in "On the interpretability of quantum neural networks"~\cite{Pira2024} is a generalization of a classical explanation technique (LIME, a local interpretable model-agnostic explanation) to quantum nets. They emphasize how quantum explanation may present distinct features, like regions of data that escape from a classical `explanation' when the quantum net becomes inherently indecisive about its output.
Ref.~\cite{Mercaldo2022} applied Gradient-weighted Class Activation Mapping to a quantum neural network. Ref.~\cite{https://doi.org/10.48550/arxiv.2211.01441} entitled, "eXplainable AI for Quantum Machine Learning" further considers the Integrated Gradients method \cite{pmlr-v70-sundararajan17a}, a model-free interpretation method, for quantum ML. 

Ref.~\cite{Tian2024} entitled, "Toward Transparent and Controllable Quantum Generative Models", proposes an inherently interpretable quantum ML model by mapping input data into a human-understandable concept space that represents a basis for algorithmic decisions. The authors employ a quantum variational autoencoder to generate the concept space, although this could also be implemented with classical neural networks.

\textit{Equivariant quantum neural networks:}
In "Theory for Equivariant Quantum Neural Networks" \cite{Nguyen2024}, the authors provide a general framework for constructing quantum neural networks that are equivariant, i.e. the network respects specified symmetry groups. These architectures entail a grade of interpretability at the group-theoretic level, relevant to the targeted physical systems of interest in the data. The authors interpret each network layer as a form of generalized Fourier-space action. (Each layer thus operates as a group Fourier transform that acts on Fourier components and subsequently transforms them back). They claim that this enables the possibility of using different group representations as hyperparameters--an interesting prospect.

\textit{Quantum versus classical ML:}
A broader aspect of interpreting quantum ML is understanding clearly why and when a quantum advantage comes into play \cite{Anschuetz2023}, i.e., where quantum models are more expressive compared to classical ones. For example, the authors of "Interpretable Quantum Advantage in Neural Sequence Learning"~\cite{Anschuetz2023} consider sequence-to-sequence learning tasks and show that a quantum advantage is due to quantum contextuality. 
 The authors of Ref.~\cite{GilFuster2024}, entitled "Understanding quantum machine learning also requires rethinking generalization", discuss how corresponding notions of complexity need to be reformulated for quantum ML.

\textit{Quantum projective simulation and active inference:}
To another architectural end of quantum ML, the authors of "Towards interpretable quantum machine learning via single-photon quantum walks" \cite{Flamini2024} present a variational method to find a physically-realizable \cite{Franceschetto2024} quantum mechanical equivalent of the projective simulation method \cite{Briegel2012}, a reinforcement learning model that entails episodic memory (a deliberation process). 
In the quantum variant of projective simulation, they employ a single-photon-based quantum computer:  Episodic memory comes from a quantum walk of single photons in a lattice of tunable Mach-Zehnder interferometers; the training procedure utilizes variational algorithms. 
On a generic level, Refs.~\cite{https://doi.org/10.48550/arxiv.2402.10192,https://doi.org/10.48550/arxiv.1910.06985} discuss how the projective simulation captures the essential components of chains-of-thought;  deliberation processes are  random walks of a  particle on a graph with vertices representing concepts or thoughts.

%\textit{Other}

%Recently, the authors of "Quantum many-body physics calculations with large language models" \cite{Pan2025} asked ChatGPT to derive the analytical expressions for Hartree-Fock Hamiltonians and associated self-consistency equations in various test cases, also asking the machine to explain its own answers. 

\section{Classical Condensed Matter}
\label{chapter:condensed}

Condensed matter systems are inherently complex, as they are composed of many interacting bodies, perceptible both in structural properties as well as the dynamical behavior of, e.g., liquids, glasses, solids, gels, etc. Regarding equilibrium properties, all observable statistical properties are ultimately generated out of a principle of a minimization of a free energy, and ML has aided the search for the explicit form of this generating function.
Out of equilibrium (Secs.~\ref{subsec:stat_mech_nonequi} and~\ref{subsec:glassy_fluids}), there are many deep-seated questions where ML can help, for example, on the nature of irreversibility as well as how structure intertwines with dynamics to render the overall complex behavior.

Regarding the equilibrium theory of inhomogeneous fluids, the paper
"Analytical classical density functionals from an equation learning network"~\cite{Lin2020} uses an equation learner (EQL)~\cite{martius2016extrapolation} to extract symbolic expressions for classical density functionals, i.e. the functional generating equilibrium particle density distributions when minimized. The analytical expressions for most density functionals, i.e. for realistic fluids, are not known exactly. It is important to interpret the functionals generated by ML~\cite{https://doi.org/10.48550/arxiv.2406.07345} in order to be able to advance existing approximations and theory. Ref. \cite{ShangChun2019} uses  a convolutional neural network architecture as an inductive bias aligning with the known theoretical structure of the functionals, avoiding a purely `black box' approach. The approach of Ref.~\cite{Sammller2023} uses neural networks to construct density functionals via direct one-body correlation functions. The authors claim the approach can avoid a complete `black box' when the resulting correlation functional is inspected via functional calculus rules, which can include those stemming from generalized Noether invariances \cite{Sammller2024}.

\subsection{Statistical Mechanics Out of Equilibrium / Driven Systems} 
\label{subsec:stat_mech_nonequi}

Many-body systems that are out of equilibrium states generally attain their dynamics through both dissipative and coherent driving processes in interacting with their environment;  they showcase the property of irreversibility, which gives rise to a direction time. Thus there are interesting questions for interpretable ML revolving around how they might contain information on the arrow of time, which has a thermodynamic origin in the second law, as well as abilities to separate coherent from incoherent motion. 

In "Machine learning the thermodynamic arrow of time"~\cite{Seif2020}, the authors train a convolutional neural network, as well as a logistic regression classifier, to infer the direction of time on microscopic trajectory data of nonequilibrium systems like a Brownian particle in a moving potential, a spin chain in a magnetic field, or with time-dependent couplings. Such microscopic trajectories generally display strong fluctuations under nonequilibrium conditions; hence, guessing the arrow of time is challenging.
To interpret the network's representations of forward and backward trajectories visually, they apply an \textit{inception} and \textit{dreaming} technique from computer vision, also modifying it to act on gradients. (Similar techniques were also applied in Refs.~\cite{Schindler2017,Jaouni2024}.) For the spin systems, the trajectories are first projected onto the (interpretable) magnetization and nearest-neighbor correlation functions and averaged over small time intervals. How the weights of the convolutional networks correspond to these quantities is analyzed. Information about the magnetization is encoded most strongly therein for the case of the time-dependent magnetic field, and that of the correlation function in the case of the time-dependent couplings.  The learned values of the weights agree with analytical results that reproduce the correct likelihood value. 
The authors infer that the important features that the networks discover are the time derivative of the Hamiltonian as well as the related dissipated work.

In "Measuring Irreversibility from Learned Representations of Biological Patterns"~\cite{Li2024}, VAEs were combined with thermodynamic inference to render lower-bound estimates on irreversibility in spatiotemporally evolving systems; the latent space representations are used directly in the estimation. A factorizing VAE (FVAE) loss function was utilized~\cite{Kim2018-disentangling}. The disentangled latent space variables are identifiable with scale and complexity of the spatiotemporal patterns.

\textit{Reducing (many-body) out-of-equilibrium dynamics to drift--diffusion equations of motion:} The authors of "Machine learning stochastic differential equations for the evolution of order parameters of classical many-body systems in and out of equilibrium"~\cite{Carnazza2024} infer governing equations of motion of order parameters in the interpretable form of a drift-diffusion equation. (They used an integral version as a technical starting point). 
They observed the near-equilibrium dynamics of the kinetic Ising model upon relaxing from a temperature quench; the learned drift forces allow for the reconstruction of an interpretable, effective potential for the order parameter that shows the characteristic double-well shape below the critical temperature. For the contact process, a high non-equilibrium model disobeying detailed balance, this reconstructed effective potential signals the phase transition into an absorbing state. In the latter case, the diffusion explicitly depends on the evolution of the drift forces. Both models were  quality-checked in their abilities to recover critical scaling behavior via the drift forces. Related approaches are discussed in Sec.~\ref{subsec:complex_systems_supervised}, i.e., Ref.~\cite{Gao2024}, which explores flocking behavior that resembles that of known active matter (Vicsek) models, and Ref.~\cite{Bae2024}, which explores the stochastic van der Pol oscillator, a spiking neuron model, and a Brownian Carnot engine. See also related ML approaches in Refs.~\cite{Dietrich2023,Garca2022,fernandez2023}. (Ref.~\cite{fernandez2023}  uses variational autoencoders to extract a minimal number of parameters rendering the stochastic process well-approximated, applied to anomalous diffusion datasets.)

\textit{Intelligent computing with active matter:} Active matter systems are inherently out of equilibrium. In Ref. \cite{Gaimann2025}, the authors study and interpret the information processing in active-matter model systems used as a reservoir computer. The physical substrate is intrinsically interpretable in contrast to neural networks that are normally used. They uncover a nonequilibrium dynamical regime of the active matter that renders robustly optimal performance readable in the system's intrinsic damping dynamics, tested across different tasks and variations of physical control parameters. Physical quantifiers known from supercooled fluids (next section) and visual inspection are used to indicate and interpret optimal behavior, illuminating information transfer and memory erasure, as well as a way to interpret different prediction tasks. See also an earlier Ref.~\cite{Lymburn2021}. 
Controllable active matter agents are also used for reinforcement learning; their learning and signaling strategies have physical roots~\cite{https://doi.org/10.48550/arxiv.2307.00994,https://doi.org/10.48550/arxiv.2501.08632}.

\subsection{Supercooled and Glass-Forming Fluids}
\label{subsec:glassy_fluids}

The interplay between structure and dynamics is a perplexing problem in glassy systems. Cooling a liquid rapidly or compressing it past a certain point slows down the observed dynamics by many orders of magnitude, although the overall structure remains similar; it is both structurally disordered and dynamically heterogeneous. Despite being an object of study for over 150 years, the exact mechanism of the glass transition has not yet been understood. %Machine learning has become a tool for prediction models of dynamics based solely on structural information. As stated in the recent review, "Roadmap on machine learning glassy dynamics" \cite{Jung2025}, interpretable ML has become part of a mid-term vision.   

In "Dynamics of supercooled liquids from static averaged quantities using machine learning"~\cite{Ciarella2023}, the authors aim to use ML to model the dynamics of supercooled fluids in a physics-inspired way via a Generalized Langevin Equation, which incorporates memory effects and is thus non-Markovian in nature. They first hinge on mode coupling theory to tractably solve for the memory function alone, which avoids the need of a black box ML method, and this memory function can be represented as a linear sum of stretched exponential functions that ultimately aids interpretability. Employing an evolutionary learning strategy, they find that the memory kernel is well-approximated by two such stretch exponentials. Thus they conclude that two predominant relaxation modes drive the nonequilibrium dynamics.

The authors of "Heterogeneous Activation, Local Structure, and Softness in Supercooled Colloidal Liquids"~\cite{Ma2019} find a means to explain nonexponential relaxation seen in glasses by exploring more closely (interpreting) a  physics-inspired, ML-defined  `softness' parameter (introduced first in Ref.~\cite{Schoenholz2016}). The latter `somehow' characterizes the local susceptibility of particles to structural changes but is a phenomenological black box. They examine this parameter in bidisperse liquids both in terms of local microscopic structure, and local dynamical relaxation processes: Particles dubbed `soft' tend to have fewer nearby large particles; those with the same `softness' had local structural environments similar enough to give rise to exponential relaxation with a single activation time. These display multiple thermal activations characterized by the same decay time. 

\textit{Dynamic propensity:}
Many pieces of work have focused on predicting the so-called dynamic propensity parameter, 
which is indicative of heterogeneous environments and dynamics. 
In "Unveiling the predictive power of static structure in glassy systems"~\cite{Bapst2020}, the authors applied a linear \textit{post hoc} interpretation technique to visualize the propensity parameter predicted by a graph neural network that learns to organize local structures in the fluids in an unsupervised way.
Graph neural networks were tested against more interpretable linear models like linear regression and support vector machines in the article, "Averaging Local Structure to Predict the Dynamic Propensity in Supercooled Liquids"~\cite{Boattini2021}. Here, a large set ($10^3$) of pre-calculated structural descriptors are used on the input: The authors borrowed an operational feature of the GNNs, i.e., particle shell-averaging, and made handcrafted descriptors on the same principle, asking if this would lead to similar predictions. The linear regression model was highly competitive in performance, and combined with the full interpretability of the input descriptors, the authors could question which structural features are most relevant for predicting the dynamics. %For example, ``[I]s radial (or density) information sufficient to accurately predict the dynamics, or do we also need angular information?''
The linearity of the model also meant the number of necessary input descriptors could be reduced drastically. 
The authors of "Dynamic heterogeneity at the experimental glass transition predicted by transferable machine learning"~\cite{Jung2024} use neural networks and add an attention layer to enhance transferability of predictions to different temperatures and times; again, coarse-grained state descriptors are calculated \textit{a priori} such as local particle densities, potential energy, Voronoi parameters, and variation in potential energy. The authors interpret the attention layer \textit{post hoc}, showing that it can represent different temperatures via different (calculatable) dynamical correlation lengths. The multilayer-perception architecture includes a bottleneck layer: Although they have difficulty in interpreting it directly in terms of physical quantities, they at least conclude that essentially only two variables are needed. 
%

%"Predicting the propensity for thermally activated β events in metallic glasses via interpretable machine learning" Ref.~\cite{Wang2020} [doesn't actually interpret anything]

In "What do deep neural networks find in disordered structures of glasses?"~\cite{Oyama2023}, a ML approach is proposed to extract characteristic local meso-structures associated with glassy behavior (i.e. liquid-like vs. solid-like) from particle configurations in glassy systems.
First,  a neural network classifies configurations of liquids and glasses correctly. Then, the authors quantify the grounds of the decisions made by the network using Gradient-weighted Class Activation Mapping, i.e., interpretability methods.  Finally, they measure two distinct local multibody structural indicators to interpret the scores in physical terms. 
A very similar approach was taken in Ref.~\cite{Liu2024}, 

The authors of "Classifying the age of a glass based on structural properties: A machine learning approach"~\cite{Janzen2024} uses a set of known order parameters as input to predict long-time (aging) dynamics in deeply supercooled liquids. They employ principal component analysis and a Shapley additive explanation techniques to the multilayer perceptron to reveal the main, structural features that most strongly encode the age of the fluid.

\section{High Energy Physics}
\label{chapter:high_energy}

High energy physics, or particle physics, aims to uncover the fundamental components and interactions that make up the universe. By studying elementary particles like quarks, leptons, and gauge bosons, and their interactions via the four fundamental forces—gravitational, electromagnetic, strong nuclear, and weak nuclear—scientists aim to formulate a comprehensive understanding of the underlying principles that govern all matter and energy. This involves using advanced particle accelerators, such as the Large Hadron Collider, to produce high-energy collisions that recreate conditions similar to those just after the Big Bang. These experiments allow physicists to probe the predictions of the Standard Model, search for new particles and forces, and explore phenomena like symmetry breaking or dark matter. It explores the confinement of quarks and gluons to hadrons and thus provides the foundations of nuclear physics. On the side of experiments, machine learning can help to find patterns in a huge amount of data created by particle colliders. On the theory side, machine learning allows theoretical physicists to formulate concepts and equations describing particles.

\subsection{Theory}

In "Machine learning of explicit order parameters: From the Ising model to SU (2) lattice gauge theory"~\cite{Wetzel2017b} construct a neural network architecture that iteratively reduces the neural network capacity until a performance drop is observed, see \fig{fig:PhaseDiagramNN}. The remaining artificial neural network can only learn a small number of correlations that are present in the data, which can be extracted through polynomial regression. The authors extracted the formulas of several physical observables, including the magnetization and energy of the Ising model and the Polyakov loop order parameter of SU(2) lattice gauge theory.

The authors of "Towards novel insights in lattice field theory with explainable machine learning~\cite{Blcher2020}" apply artificial neural networks to predict action parameters. During this process, they used layer-wise relevance propagation (LRP) to identify the most important observables depending on the location in the phase diagram in a scalar Yukawa model. It is then possible to reconstruct all order parameters from the learned filter weights.

In "Discover the Gell-Mann–Okubo formula with machine learning"~\cite{Zhang2022}, symbolic regression guided by physical constraints, such as dimensionality and symmetry, is employed to re-discover the Gell-Mann–Okubo formula. This formula provides a sum rule for the masses of hadrons within a specific multiplet.

\subsection{Experiment}

The authors of "Mapping machine-learned physics into a human-readable space"~\cite{Faucett2021} devise a method to translate a black-box machine-learned classifier operating on a high-dimensional input space into a small set of human-interpretable observables. The method iteratively selects observables from a large space of high-level discriminants by finding those with the highest decision similarity relative to the classifier. This technique is demonstrated in the example of a convolutional neural network addressing jet classification in collider physics.

In attempting to search for new physics, anomaly detection can be conducted by trained convolutional autoencoders. Yet, the black-box nature of these algorithms makes it difficult to identify the nature of a such potentially new discovery. For this reason, the authors of "Creating simple, interpretable anomaly detectors for new physics in jet substructure"~\cite{Bradshaw2022} map out the physical observables learned by the autoencoder by the method introduced in the previous article~\cite{Faucett2021}.

The authors of "Resurrecting $b\bar{b}h$ with kinematic shapes"~\cite{Grojean2021} address the problem of separating signals and background from Large Hadron Collider (LHC) data. This is achieved by employing game theory concepts like Shapley values and Boosted Decision Trees.

In "Explainable AI for ML jet taggers using expert variables and layerwise relevance propagation"~\cite{Agarwal2021}, the authors train an artificial neural network to determine the substructure of jets. In order to interpret the neural network they augment the input data by meaningful physical composite variables and apply layerwise relevance propagation.

In "An Exploration of Learnt Representations of W Jets"~\cite{https://doi.org/10.48550/arxiv.2109.10919}, a variational autoencoder is trained to encode collider physics data (specifically boosted W jets) into a latent representation. One can observe that the VAE learns a meaningful latent representation of the data manifold, with semantically meaningful and interpretable latent space directions. 

The authors of "Explaining machine-learned particle-flow reconstruction"~\cite{https://doi.org/10.48550/arxiv.2111.12840} address the problem of reconstructing a collision in particle colliders by combining information from different subdetectors. For this purpose, the authors replace conventional algorithms with a graph neural network. In order to gain insight into the model's decision-making process, the authors employ layerwise-relevance propagation.

"A detailed study of interpretability of deep neural network based top taggers"~\cite{Khot2023} treats a number of interpretability techniques, borrowing them from the computer science community and applying them to artificial neural networks. The latter are tasked with identifying jets at the Large Hadron Collider. Inspired by their findings, the authors design a "Particle Flow Interaction Network" and demonstrate its superior top-tagging performance.

The authors of "Interpretability of an Interaction Network for identifying $H\rightarrow b\bar{b}$ jets"~\cite{Roy2022} employ an Interaction Network (IN) model designed to separate boosted $H\rightarrow b\bar{b}$ jets from a quantum chromo dynamics (QCD) background. They employ different interpretability methods developed by the computer science community to figure out how their classifier network makes its decisions, and use this knowledge to optimize the model.

In "Interpretable machine learning methods applied to jet background subtraction in heavy-ion collisions"~\cite{Mengel2023} a deep neural network is trained to subtract the background from measurements of jets in relativistic heavy-ion collisions. The authors interpret this neural network by employing symbolic regression. They show that a neural network learns to approximate an existing algorithm that is based on particle multiplicity.

The article "Interpretable deep learning models for the inference and classification of LHC data"~\cite{Ngairangbam2024} overcomes a central bottleneck in the 
shower deconstruction methodology that is pivotal in distinguishing signal and background in particle collider experiments. A central problem in this algorithm is the combinatorial growth associated with increasing jet constituents. They employ a neural network that efficiently learns the optimal policy for this task, achieving a linear scaling relationship with the number of jet constituents.

PELICAN~\cite{Bogatskiy2024} is a specialized neural network designed for experimental particle physics that implements permutation equivariance and Lorentz invariance/covariance. The PELICAN architecture reduces complexity, which leads to improved interpretability and enhanced performance. It is tested on classification and regression tasks, including Lorentz-boosted top-quark tagging and reconstruction, W-boson identification, and jet classification.

\section{Astrophysics \&  Cosmology}
\label{chapter:astro}

Astrophysics and cosmology are fundamental fields of physics that explore the nature, origins, and evolution of the universe. These two fields focus on different scales: while astrophysics deals with the properties and behaviors of celestial objects and phenomena (e.g., stars, planets, galaxies, black holes), cosmology is concerned with the large-scale properties of the universe as a whole. Nevertheless, these two fields are closely related and share similar techniques, hence we review them together. 

Due to increasingly larger datasets collected with telescopes, satellites, and observatories, astrophysics and cosmology have entered an area of "big data" where many tasks are desperate to be automated by machine learning. Researchers have tried various interpretability methods to extract physical insights from seemingly black-box machine learning models. Depending on the granularity of interpretations, there are two main categories: identifying important features and discovering symbolic rules. 

\subsection{Identifying important features}

To make predictions from machine learning models more trusted, we should at least answer: which input features are picked up to make the prediction? For example, given the waveforms from LIGO, what features are used by the model to determine whether they are gravitational wave signals or not~\cite{safarzadeh2022interpreting}? To this end, we need methods to assign importance scores to features, including but not limited to Shapley values, saliency map, and other importance scores. 

The Shapley value is an average of all the marginal contributions to all possible coalitions~\cite{winter2002shapley}, used in Ref.~\cite{machado2021shaping} to understand gas shapes in dark matter halos. They explore a machine-learning approach for modeling the dependence of gas shapes on dark matter and baryonic properties. They show that XGBoost models can accurately predict gas shape profiles in dark matter halos. They explored model interpretability with SHAP, a method based on Shapley values, that identifies the most predictive properties at different halo radii. They find that baryonic properties best predict gas shapes in halo cores, whereas dark matter shapes are the main predictors in the halo outskirts. SHAP has also been used to understand molecular abundances in star-forming regions~\cite{heyl2023understanding}. 

A saliency map is an image (array) that highlights the most relevant regions (features) for machine learning models.  ExoMiner~\cite{valizadegan2022exominer} used an Occlusion-based saliency map to explain a classifier that validates
301 newly discovered exoplanets. They introduced a simple explainability framework that provides experts with feedback on why ExoMiner classifies a transit signal into a specific class label. To explain how a galaxy image is classified into different types (bar, bulge, spiral arms), Ref.~\cite{bhambra2022explaining} uses SmoothGrad saliency maps to visualize important input features. Besides analyzing input features, Ref.~\cite{safarzadeh2022interpreting} applies saliency techniques to hidden neurons inside neural networks to understand how models internally discriminate gravitational waves from noises. Interestingly, their analysis suggests that different parts of the network appear to specialize in local versus global features--corresponding to the Hanford and the Livingston branches, respectively. On the other hand, Ref.~\cite{jacobs2022exploring} studied the failure modes of saliency maps for neural networks used to detect gravitational lenses. Using a tool called sensitivity probe, they found that the networks are highly sensitive to color, the simulated PSF used in training, and occlusion of light from a lensed source, but are insensitive to Einstein radius, and performance degrades smoothly with source and lens magnitudes. This highlights that neural networks may be leveraging non-physical spurious features to perform the classification, calling for attention to understanding the biases and weaknesses of deep learning models.

%\mk{
A technique of feature omission can render a (though weaker) degree of interpretability of deep neural network models: By omitting certain subsets of feature variables, one can pinpoint which of them are relevant predictors of the target variables, i.e., are essential for the prediction task. This was applied in Ref.~\cite{Kruijssen2020} on a deep neural network model that used simulation information about galactic globular clusters associated with five different progenitors to quantify properties (stellar masses and accretion redshifts) of satellite galaxies from which the Milky Way was assembled. The authors inferred that the Milky Way's stellar mass grew mostly via in-situ star formation and then offered a highly detailed reconstruction of the accretion events in the Milky Way's history. Feature omission and a similar deep neural network model were repeated in Ref.~\cite{TrujilloGomez2023}, while focusing on classifying the origin of globular clusters in external galaxies based on observable properties. Somewhat related is Ref.~\cite{Gilda2021}, which introduced MIRKWOOD: a tool comprising an ensemble of supervised ML-based models that map galaxy fluxes to their galactic physical properties (stellar masses, etc.). To enhance interpretability, they employed Shapley values, i.e., a sensitivity measure of specific features in how much each contributes to the output.
%}

While previous methods can identify important features that are provided by researchers, sometimes we hope to learn new features that might be unknown. Autoencoders can meet this goal by having bottlenecks in the middle of the network. The bottleneck layer restrains the amount of information that can go through, hence learned variables at the bottleneck should be important and probably physically relevant. For example, 
Ref.~\cite{lucie2024explaining} used an interpretable variational encoder (IVE) to connect the evolutionary history of dark matter halos with their density profiles.  Ref.~\cite{ntampaka2022importance} attempts to understand galaxy cluster cosmology, where they find that the terse layer (bottleneck) is strongly dependent on the amplitude of matter fluctuations $\sigma_8$.

Besides neural networks, decision trees are also used in astrophysics and cosmology due to their interpretability. Ref.~\cite{lucie2019interpretable} uses gradient-boosted trees to study important factors in dark matter halo formation. They used a metric known as feature importance to measure the relevance of each input feature in training the algorithm to predict the correct target variable. Their interpretation approach suggests that there is only a little role for the tidal shear field in establishing final halo masses.

\subsection{Discovering symbolic rules}
While identifying important features might suffice in many cases, it would be more informative, yet more challenging, to discover symbolic rules from neural networks. Symbolic regression is a tool for this. In "Learning symbolic Physics with Graph Networks"~\cite{cranmer2019learning} and the follow-up work "Discovering Symbolic Models from Deep Learning with Inductive Biases"~\cite{cranmer2020discovering}, they encourage a graph neural network to sparse latent representations in a supervised setting, and then they apply symbolic regression to components of the learned model to extract explicit physical relations. Remarkably, their approach finds a new symbolic formula in a non-trivial cosmology example--a detailed dark matter simulation--which can predict the concentration of dark matter from the mass distribution of nearby cosmic structures. Based on the same framework with additional improvements, they trained a graph neural network to simulate the dynamics of our Solar System's Sun, planets, and large moons from 30 years of trajectory data. They then used symbolic regression to correctly infer an analytical expression for the force law implicitly learned by the neural network, which their results showed is equivalent to Newton's law of gravitation. 

Besides symbolic laws that are continuous functions, symbolic rules can also be extracted from machine learning models. In Ref.~\cite{pasquato2024interpretable}, extreme gradient boosting classifiers (XGBoost) are trained on globular cluster simulations that can, in principle, predict intermediate-mass black-hole (IMBH) host candidates based on observable features. They used the anchors method to extract symbolic rules that can explain their classifier. They also independently train a natively interpretable model using certifiably optimal rule lists.

\section{Complex Systems}
\label{chapter:complex}

A complex system is an umbrella term for a myriad of systems in different domains, comprising many interconnected components that interact in ways that can lead to non-trivial emergent behavior \cite{Mitchell2009}. One of the main goals of complexity science is to identify the underlying dynamics of the components in a given system and build an analytic (or at least computational) model of it, which falls into the category of inverse problem \cite{Tarantola2005}. This is a well-known, notoriously challenging task due to its multifaceted complexity. First, the interaction between components in complex systems is non-linear, non-Markovian, and sometimes even non-conservative, complicating straightforward and intuitive modeling approaches. Also, inferring and understanding the relationship between several thousand components and variables in a network structure often imposes a qualitatively different kind of difficulty than other domains in physics. Furthermore, in a complex system, a time series of state variables (the only observable we have in most cases) comes from the summation of an unknown number of unknown interactions. This makes the nature of the inverse problem fundamentally ill-defined, which is what causes many optimization-based approaches to fail.

Indeed, various attempts have been made to address this infamous problem with the help of machine learning. Especially, the domain of complexity science greatly benefits from the interpretable machine learning approaches, since one of the primary focuses in this field is to go beyond simply modeling interactions and fully understand how those interactions give rise to emergent behavior. Many of them either analyze the data in an unsupervised manner to find meaningful patterns or train the predictive model with the data and then further investigate the model. Many of the interpreted results and findings in these researches can be further distilled into more interpretable forms via other techniques like symbolic regression. In this section, we introduce some of the noteworthy works in both approaches.

\subsection{Unsupervised approach}

Unsupervised machine learning techniques such as clustering and dimensionality reduction have been used consistently  to give researchers an interpretable representation of puzzling patterns and phenomena in complex systems. Especially, these methods excel at providing unsupervised yet interpretable classification of data points when there exist multiple different emergent behaviors that typically cannot be simply inferred from raw data. 

For instance, in "Analyzing collective motion with machine learning and topology" \cite{Bhaskar2019}, the phenotypes of emergent collective motion from the flock model are successfully classified and model parameters are recovered by using $k$-medoids clustering on its time series of system order parameters, without the necessity of the ground-truth label. The authors of "Physically-interpretable classification of biological network dynamics for complex collective motions" \cite{Fujii2020} approached a similar problem with graph dynamic mode decomposition (GDMD), which is a dimensionality reduction algorithm that approximates the modes and eigenvalues of the Koopman operator on a graph. "Unsupervised manifold learning of collective behavior" \cite{Titus2021}, as its name suggests, addresses a similar problem by adopting a similarity measure and diffusion map to construct a meaningful embedding from fish schooling data. In "Analysis of the collective behavior of boids" \cite{Inomata2020}, principal component analysis (PCA) is applied to the flock model simulation data for the automatic classification of a diverse range of collective motion. Similarly, in "Dynamics of collective motion across time and species" \cite{Papadopoulou2023}, PCA is employed to construct a so-called 'swarm space', where each group trajectory of the swarm is embedded to visualize their collective characteristics and facilitates easy comparison between collective motions of two different fish species.

Some of the work adopted more information-theoretical approaches to extract relevant variables and structures in complex systems. The authors of "Information theory for data-driven model reduction in physics and biology" \cite{Schmitt2024} employed the concept of information bottleneck, which is a framework that compresses an input while preserving only the most relevant information for predicting an output. They linked relevant variables to the eigenfunctions of the transfer operator and demonstrated its application in extracting slow collective variables from videos of atmospheric flows and cyanobacteria colonies. In "The Distributed Information Bottleneck reveals the explanatory structure of complex systems" \cite{murphy2022distributed}, the Distributed Information Bottleneck, a modification of the Information Bottleneck framework that distributes bottlenecks across input components is proposed, enabling interpretable deep learning in science by deconstructing relationships into meaningful approximations, demonstrated in Boolean circuits and sheared glass systems.

\begin{figure*}
    \centering
    \includegraphics[width=1.0\linewidth]{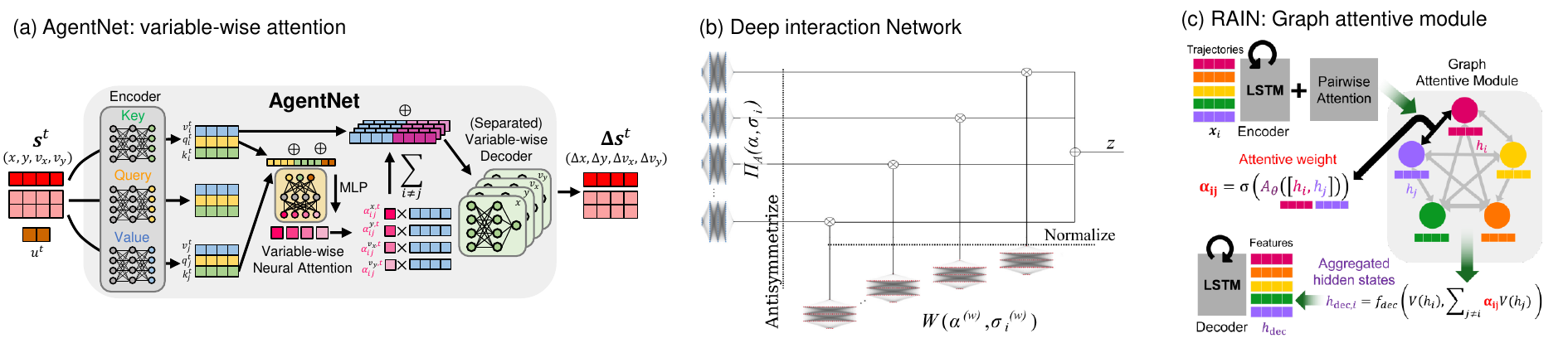}
\caption{{\bf Attention mechanism to capture microscopic interaction dynamics of agents in complex systems.} (Figure taken from (a) Ha and Jeong et al., ~\cite{Ha2021B}, (b) Heras et al., \cite{Heras2019} and (c) Ha and Jeong et al., \cite{ha2022learning}).
(a) The figure details the neural network-based architecture (AgentNet) by which the attention mechanism is designed to infer the strength of interactions between the state variables of an agent from a time series in a complex system. Here, variable-wise attention values are naturally distilled during the supervised learning process aimed at predicting time series and provide the human researcher with an interpretable, visual intuition of the system's decision rules. (b) The structure of a deep attention network to predict the next state of individuals in a zebrafish school. The structure shows how the inputs of the pair-interaction and aggregation subnetworks are integrated to produce a single logit $z$ for the focal fish turning right after $1$ second. (c) The structure of relative attentive interaction network (RAIN). Here, the model first encodes each agent’s trajectory with an LSTM encoder and applies pairwise attention (PA) to the hidden states for constructing a pair of embeddings for each agent pair, then the graph attentive module extracts the interaction strength from a pair of embeddings in the form of an attention weight using multi-layered perceptrons, different from the typical inner product.}
\label{fig:complex_systems_attention_1}
\end{figure*}

\subsection{Supervised approach}
\label{subsec:complex_systems_supervised}

There are several different ways to enhance interpretability and facilitate better human understanding of the complex system via supervised learning. One of those ways is understanding the causal relationship between multiple components, which has been thoroughly explored by researchers in diverse domains. The seminal work in this direction is "Neural Granger Causality" \cite{Tank2021}, where authors first employed neural network as a non-linear model of Granger causality and showed great performances in capturing causal relationships from non-linear time series generated by climate dynamics, gene regulation, and human joint motion. This concept has been extended in the context of other complex systems as well, as in "Learning interaction rules from multi-animal trajectories via augmented behavioral models" \cite{Fujii2021}. In this work, the authors constructed a theory-based neural architecture in which independent components learn the sign and amplitude of non-linear Granger causality in an interpretable form.

Also, machine learning can contribute to human interpretability by inferring the characteristics of the network that underlies the complex system. In "Finding key players in complex networks through deep reinforcement learning" \cite{Fan2020}, authors employed graph reinforcement learning to detect "key players" on the network, which denotes influential and vital nodes for certain network functionality. Recently, in "Reconstructing the evolution history of networked complex systems" \cite{Wang2024}, the output of a graph neural network is directly trained to predict the generation order of the given edges, recovering the valuable information of graph evolution process which is unknown in most cases.

Another notable direction is optimizing only the unknown components of the governing equation of motion for each individual. This gray-box method combines the expert's domain knowledge and machine learning's capability of high-dimensional optimization, and making a good distinction between which parts of a complex system to model and which parts to approximate with a neural network is critical to success. For example, in "Inferring the Langevin Equation with Uncertainty via Bayesian Neural Networks" \cite{Bae2024}, the non-homogeneous drift force and diffusion matrix of the given Langevin system are inferred solely from noisy trajectory by isolating the training module for force and diffusion components with Bayesian neural network. The authors of "Learning interpretable dynamics of stochastic complex systems from experimental data" \cite{Gao2024} propose a similar but more integrated approach, where each terms of Langevin dynamics are modeled via neural network and further distilled into symbolic form, providing enhanced interpretability which is enough to build a fully analytic stochastic differential equation of the system.

\subsection{Attention as a relational strength}

Recently, the concept of attention \cite{Vaswani2017} has been widely used in the machine learning literature, due to its powerful performance and interpretable nature. Typically, many studies in complex system analysis attempted to treat the attention weight assigned to a certain component as a proxy of an interaction strength from that component. The idea behind this view is simple; attention weight represents the importance of information coming from a particular source, and a weight value of zero means that the information coming from that source is irrelevant to the final prediction. In "Deep attention networks reveal the rules of collective motion in zebrafish" \cite{Heras2019}, authors demonstrate that deep attention networks can identify the interaction among a flock of zebrafish by providing the interaction strength between fishes as a form of attention weight. The authors of "Deep learning-assisted comparative analysis of animal trajectories with DeepHL" \cite{Maekawa2020} trained a model to classify animal trajectories, and let attention highlight specific segments of those trajectories that contributed to the classification the most. The significance of this work comes from the fact that machine-highlighted segments helped human experts narrow down the significant points on trajectories and facilitated researchers to formulate new hypotheses. 

This concept has since been explored further, leading to the development of a unified framework that can encompass all varieties of complex system data. In "Unraveling hidden interactions in complex systems with deep learning" \cite{Ha2021B}, the authors introduce a novel interpretable framework called AgentNet that aims to predict trajectories of agents in various types of complex systems and provide a visual map of variable-wise interaction forms and strengths. This model is generally applicable to all sorts of complex system data as long as each agent's state is governed by pairwise interaction rules with others and demonstrates its versatility by simultaneously showing good prediction performance and unprecedented interpretability of interaction (see Fig. \ref{fig:complex_systems_attention_1}). The same authors further tried to generalize the model and in "Learning Heterogeneous Interaction Strengths by Trajectory Prediction with Graph Neural Network" \cite{ha2022learning}, they propose a framework called relational attentive inference network (RAIN) to infer prefixed, i.e. not according to the decision rule but determined from the beginning, continuous interaction strength between components without any prior knowledge of the dynamics.

\section{General Physics}
\label{chapter:general}

\subsection{Extracting symbolic expressions from experimental data}

Various methods exist to deriving equations from experimental data. Automatizing the process of finding the physical laws underlying experimental data was considered in an early study in~\cite{Crutchfield1987} in 1987.  In this section, we compare and discuss several of these methods, highlighting their advantages and disadvantages. A common limitation among these approaches is the need to pre-identify the relevant parameters that appear in the symbolic expression. To address this challenge, deep learning has emerged as a promising solution, leading to a growing body of research that integrates symbolic regression with deep learning techniques. This hybrid approach combines the parameter identification capabilities of deep learning with the interpretability of symbolic equations (see Section~\ref{sec:general_physics_autoencoder}).

Symbolic regression, introduced in Chapter~\ref{chapter:symbolic}, is a powerful method used to derive equations from given data, a task that is essential in many scientific domains, particularly in physics. One classical example in physics is the determination of the equation of motion from observational data.

In this section, we focus on the case where the parameters in the equation are known and, hence, symbolic regression can be effectively applied. A pioneering work in this area is based on the conceptual framework that most physical laws are grounded in mathematical symmetries and invariants~\cite{Schmidt2009, Schmidt2011}. This perspective implies that discovering many natural laws is closely tied to identifying conserved quantities and invariant equations~\cite{Anderson393, noether, noether_2}. This foundational understanding has driven the development of methods that focus on leveraging these known parameters to uncover underlying physical laws (see also Section~\ref{chapter:conserved}).

Indeed, let us outline the method from Schmidt et al.~\cite{Schmidt2009} through a pendulum example. Consider a pendulum, a nonlinear oscillator described by its angular position \(\theta\). The data set \(\mathcal{S} = \{(t^i, \theta(t^i), \omega(t^i))\}_i\) comprises triples that include time \(t^i\), position \(\theta(t^i)\), and angular velocity \(\omega(t^i)\). Since the conserved quantities are not known a priori, they must be learned in an unsupervised manner. As mentioned in Section~\ref{chapter:conserved}, symbolic regression can be used to find conserved quantities, such as the Lagrangian $L$. The search for mathematical expressions in symbolic regression can be made more efficient by using a corpus of closed-form mathematical models from Wikipedia to derive the posterior probability of each expression based on basic probabilistic principles and explicit approximations as discussed in~\cite{guimera_bayesian_machine_2020}. From the Lagrangian, the equation of motion is then obtained via the Euler-Lagrange equation.

\begin{align}
\frac{\textnormal{d} }{\textnormal{d} t} \left( \frac{\textnormal{d} L }{\textnormal{d} \omega} \right) - \frac{\textnormal{d}L }{\textnormal{d} \theta} = \frac{l}{g} \ddot{\theta} + \sin \theta = 0 \, ,
\end{align}

which leads to \(\ddot{\theta}= -\frac{g}{l} \, \sin(\theta)\),  where \(g\) is the acceleration due to gravity, and \(l\) is the length of the pendulum.

While symbolic regression is a powerful tool for deriving mathematical expressions given relevant variables, its runtime can quickly become prohibitive when dealing with complex physical equations. In~\cite{Udrescu2020}, it was demonstrated that a physics-inspired form of symbolic regression could successfully identify all 100 equations from the Feynman Lectures on Physics, whereas previous publicly available software could only solve 71. For a more challenging physics-based test set, the success rate was improved from 15\% to 90\%. Additionally, certain complex equations that eluded discovery using the methods in~\cite{Udrescu2020} were later identified by leveraging partial derivatives to incrementally construct the formula based on the number of parameters (see~\cite{patrick_semester_thesis} and Remark 6.1 in~\cite{Iten2023_book}). However, fully automating this refined approach remains a challenge for future research.

In cases where we have an a priori guess about the terms that may appear in the equation of motion, sparse regression techniques, such as Sparse Identification of Nonlinear Dynamical Systems (SINDy), provide an efficient tool for finding underlying governing equations of complex dynamical systems directly from data \cite{brunton_discovering_2016}. The fundamental idea behind SINDy is to approximate the dynamics of a system as a sparse linear combination of a library of candidate functions. This approach leverages the fact that many physical systems can be described by a few dominant terms, even when a large number of possible terms are considered.

For a system like a simple pendulum, the state vector $\boldsymbol{X}$ typically includes the angular position $\theta$ and angular velocity $\dot{\theta}$. To discover the equations of motion, we first construct a candidate function library $\boldsymbol{\Theta(X)}$, which includes various possible functions of the state variables. This library may consist of terms such as $1$, $\theta$, $\dot{\theta}$, $\theta^2$, $\theta\dot{\theta}$, $\dot{\theta}^2$, $\sin(\theta)$, and $\cos(\theta)$.

The dynamics of the system are then expressed as:

\[
\dot{\boldsymbol{X}} = \boldsymbol{\Theta(X)} \boldsymbol{\Xi}
\]

where $\dot{\boldsymbol{X}}$ represents the time derivatives of the state variables, and $\boldsymbol{\Xi}$ is a coefficient matrix that we seek to determine. The key insight of sparse regression is that most of the elements in $\boldsymbol{\Xi}$ are zero, meaning that only a few terms in the library $\boldsymbol{\Theta(X)}$ are necessary to accurately describe the system's dynamics.

To find $\boldsymbol{\Xi}$, we solve a sparse regression problem, often using methods like LASSO (Least Absolute Shrinkage and Selection Operator)~\cite{Tibshirani_2018, Hastie2009}. A more optimized version for the given kind of problem is provided in~\cite{brunton_discovering_2016}. This optimization process minimizes the difference between the observed time derivatives $\dot{\boldsymbol{X}}$ and those predicted by the model, while simultaneously encouraging sparsity in $\boldsymbol{\Xi}$.

Once $\boldsymbol{\Xi}$ is determined, the non-zero coefficients indicate which terms from $\boldsymbol{\Theta(X)}$ are significant. For the pendulum, this typically reveals the expected equation of motion, such as:

\[
\ddot{\theta} + \frac{g}{l} \sin(\theta) = 0
\]

This method is effective when there is a good a priori understanding of the terms that may appear in the equations of motion and has been successfully applied to complex physical systems such as water flowing through a pipe (defined by the Navier-Stokes equations). Without such knowledge, the number of possible combinations in the candidate function library can become excessively large, leading to a matrix that is too big and computationally challenging to manage. Therefore, the success of this approach hinges on having a well-informed guess about the relevant terms, ensuring that the problem remains tractable and the resulting model remains interpretable.

In recent years, there has been substantial advancement in the extraction of dynamical equations from experimental data, which is recognized as an NP-hard problem~\cite{Cubitt2012}. Various methods, such as those in~\cite{Daniels2015, Reinbold2021, Lu2022}, have made significant strides in this area. More recently, there has been a growing interest in integrating physical prior knowledge into deep learning frameworks, leading to the development of physics-informed deep neural networks (PINNs)~\cite{Raissi2019}. These networks leverage a priori knowledge that the underlying physical processes are governed by partial differential equations (PDEs), which has enabled the data-driven discovery of PDEs in various applications~\cite{Raissi_2018,Raissi2018_hidden,Raissi2019,Zhang_2019}.

Another recent study~\cite{goessmann2020tensornetworkapproacheslearning} explores an innovative approach to discovering non-linear dynamical laws from observational data using tensor networks. Tensor networks are mathematical structures often used in quantum physics to efficiently represent high-dimensional data with complex correlations. In the context of learning governing equations for dynamical systems, tensor networks provide a scalable and expressive framework that can incorporate physical constraints such as locality and symmetry.

\subsection{Autoencoder}
\label{sec:general_physics_autoencoder}

Finding physical equations is only possible if the relevant variables are known and accurately measured. While it may seem straightforward to work with variables like mass, velocity, and acceleration—well-established in physics textbooks—identifying these variables from raw data or sensory inputs can be equally, if not more, challenging.

For instance, consider the process of discovering the kinematical laws of falling objects. Once you have identified the correct variables (such as position over time and mass) and have a mathematical background, deriving the equations that govern motion, as Galileo did, becomes a more straightforward task. However, determining which variables to measure and how to set up the experiment to capture the essential aspects of the phenomenon can be highly complex. The challenge lies not just in the analysis but in the very act of identifying the right parameters to observe, a step that is critical for uncovering the underlying physical laws.

An autoencoder, as introduced in~\ref{chapter:algorithms}, can learn to compress experimental data in an unsupervised manner by providing the data as input-output pairs of the form \((o, o)\). While it might initially seem that the autoencoder is simply learning a trivial identity function, the task becomes non-trivial when the latent layer is constrained to store only a limited number of parameters. In this case, the autoencoder must learn which parameters are essential to reconstruct the measurement data effectively.

\begin{figure*}
    \centering
    \includegraphics[width=0.8\textwidth]{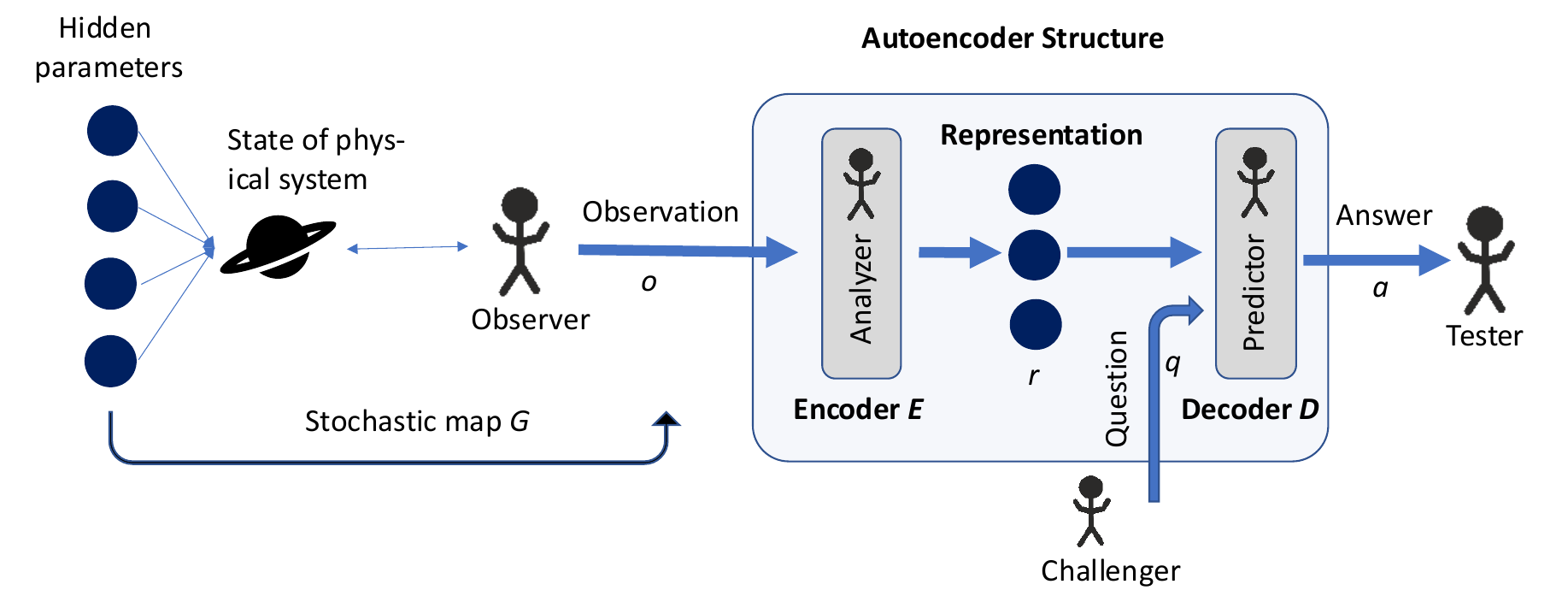}
\caption{{\bf Adapted autoencoder structure for discovering physical parameters.} (Figure taken and modified from~\cite{Iten2023_book}).
 We assume that a small set of hidden parameters fully determines the state of a physical system. The mapping \textit{G} from these hidden parameters to the measurement outputs of an observing agent is allowed to be stochastic. An autoencoder structure can then be employed to recover the relevant hidden parameters. Specifically, the observing agent provides the observation $o$ to an analyzing agent (described by an encoding map $E$), whose task is to recover the relevant hidden parameters $r$ necessary to respond to questions posed by a challenging agent. A predicting agent, which answers a question $q$ based on the representation $r$, is modeled by a decoder mapping $D$. Finally, a testing agent compares the answer $a$ given by the predicting agent with the correct answer obtained through direct measurements of the environment. Note that, in practice, the roles of the analyzer, predictor, and challenger could be performed by a single physicist.}
\label{fig:autoencoder_physical_reasoning}
\end{figure*}

If we slightly adapt the autoencoder structure by adding an additional input (the ``question input'') provided to the decoder, the process mirrors a simplified form of reasoning used by physicists, as illustrated in figure~\ref{fig:autoencoder_physical_reasoning}. This modified architecture can be employed to uncover the key parameters of various physical systems, including quantum systems, in an unsupervised manner, as demonstrated in~\cite{Iten2020}. This approach, along with other methods for extracting physical concepts from experimental data using minimal prior knowledge, is explored in detail in a recently published book~\cite{Iten2023_book}. Moreover, intriguing applications, such as identifying the parameters necessary to describe the dynamics of fire flames~\cite{Chen2022} or learning physical relationships in an electric grid using variational autoencoders~\cite{Multaheb2022}, demonstrate that these methods can uncover key physical parameters. Since autoencoders struggle with cyclic data, e.g., data lying on circles or spheres (as encountered in quantum tomography~\cite{Iten2020}), the work in~\cite{https://doi.org/10.48550/arxiv.2205.05279} suggests preprocessing the experimental data using topological data analysis to infer topological features like Betti numbers, which can then inform the design of the variational autoencoder's architecture and loss function. 

Although the neural network architecture is effective at finding relevant parameters in many examples, interpreting these parameters remains challenging. If you have a hypothesis about the parameters, one approach is to plot the latent variables against your guessed parameters by generating experimental data for different values of these variables. However, this method only works if the parameters are 'naturally' separated in the latent representation.

One way to achieve this separation is by using $\beta$-variational autoencoders ($\beta$-VAEs)~\cite{Higgins2017, Burgess2018}, which encourage the latent variables to be statistically independent with respect to the distribution of the collected experimental data. The impact of the question input has been further explored in~\cite{Lenaerts2022}. This approach has been applied to various physical systems~\cite{solera2023, fernandez2023, Iten2020}, for example, in reduced-order modeling of fluid flows~\cite{solera2023}. However, this approach may not always result in the most natural parameters. An alternative approach was explored in~\cite{PoulsenNautrup2022}, offering a different solution based on an operational principle to separate the parameters, specifically by minimizing the communication between agents, each of whom must answer physical questions.

In addition, one may want to search for parameters allowing for a simple time evolution, or more concretely, for a simple update rule for small time intervals. Such a structure was used in~\cite{Iten2020,Iten2023_book}, to predict the angle of Mars and the sun as seen from the Earth (the data that Copernicus used in the 16th century to find that the solar system is heliocentric). It could be shown that the network switched to the angles as seen from the sun in the latent representation, hence, allowing for the interpretation that the sun is in the middle of the solar system. Similarly, in~\cite{Lu2020}, convolutional neural networks are used to process spatiotemporal systems and extract the parameters from the latent representation. The investigated systems include the chaotic Kuramoto-Sivashinsky equation and the nonlinear Schrödinger equation, identifying essential parameters like viscosity damping and nonlinearity. Notably, the method handled multi-parameter systems, such as the 2D convection-diffusion equation, by separating diffusion and drift velocity components.

Furthermore, recent work~\cite{Qu2021, Murata_Fukami_Fukagata_2020} showcases a nonlinear mode decomposition method using a convolutional neural network autoencoder (MD-CNN-AE) to analyze fluid dynamics, particularly applied to flow around a circular cylinder. This approach significantly improves the reconstruction accuracy over traditional methods like Proper Orthogonal Decomposition (POD) by leveraging the nonlinearity of activation functions. These studies demonstrate the capability of MD-CNN-AE to extract and visualize complex flow features with fewer modes, providing a more interpretable and efficient dimensionality reduction tool in fluid dynamics.

Additionally, combining Proper Orthogonal Decomposition (POD) with Polynomial Chaos Expansion (PCE) has proven to be a robust alternative for multidimensional non-linear fields \cite{Mouradi2021}. The POD-PCE method offers explicit input-output formulations and physical interpretability, addressing the limitations of traditional neural networks. 

\subsection{Combining Autoencoders and methods to extract symbolic expressions}
\label{sec:general_physics_autoencoder_and_symbolic_expressions}

The ultimate goal in the study of physical systems is to identify both the relevant parameters and the underlying physical laws, expressed as mathematical equations, from experimental data. When these parameters are not provided, two main approaches are generally considered:

\begin{itemize}
    \item \textbf{Sequential Approach:} First, search for the relevant parameters and then apply symbolic regression or another method to derive the equations of motion.
    \item \textbf{Parallel Approach:} Simultaneously search for both the parameters and the equations of motion.
\end{itemize}

In addition to these approaches, several papers have explored hybrid methods. These methods incorporate prior knowledge about the system's time evolution directly into the machine learning architecture. This integration not only aids in identifying the relevant parameters but also simplifies both the interpretation of the parameters and the subsequent extraction of symbolic expressions. Furthermore, symbolic expressions have been shown to be directly extractable from video data rather than numerical data~\cite{Chari2023}. While the following discussed methods primarily work with numerical data, many of them could be adapted to process video data using approaches similar to those in~\cite{Chari2023}, albeit with increased computational effort.

A recent study~\cite{https://doi.org/10.48550/arxiv.2401.04978} contributes a tool for the sequential approach by investigating how to extract symbolic expressions from a neural network, such as the encoder of an autoencoder, designed to handle multiple input variables and a single output variable.

In the last few years, significant progress has been made in simultaneously searching for parameters and the corresponding equations. Generally, this method balances the trade-off between using more prior knowledge about the system, resulting in more interpretable results, and using less prior knowledge, which may yield results that are harder to interpret. The prior knowledge can be of a mathematical or physical nature. The following presents methods in increasing order of prior knowledge usage and interpretability.

\textbf{Prior Knowledge of a Simple Mathematical Expression for the Decoder:} A step in this direction was taken in~\cite{Kim2021} (see also Chapter~\ref{chapter:symbolic}). The authors employed Equation Learner (EQL) networks, which are similar to those in~\cite{martius2016extrapolation, Sahoo_learning_2018}. EQL networks utilize activation functions that represent simple mathematical operations, such as multiplication or trigonometric functions like sine and cosine. To ensure the discovery of "simple" mathematical expressions, sparsity is enforced in the network weights, meaning that the more weights are set to zero, the simpler the resulting mathematical expression becomes. In~\cite{kim2019integration}, EQL networks were applied to problems such as the analysis of simple kinematic systems and harmonic oscillators. These networks identified straightforward mathematical expressions for the time evolution of these systems.

\textbf{Mathematical Prior Knowledge for the time evolution:} Koopman operator theory, introduced in 1931~\cite{Koopman1931, Koopman1932}, is a promising approach for linearizing nonlinear dynamics~\cite{Mezi2004, Mezi2005}. Recent advances in this field include both theoretical developments~\cite{Mezic_spectrum_2019, Budii2012, Mezi2013} and numerical methods like dynamic mode decomposition~\cite{SCHMID2010, ROWLEY2009, LBrunton2015}. The increasing availability of measurement data has also enhanced the application of data-driven Koopman theory. Obtaining eigenfunctions of the Koopman operator in practical applications has proven challenging. In~\cite{Lusch2018}, a deep learning-based approach was considered, utilizing essentially the same network architecture used to find simple update rules in~\cite{Iten2020, Iten2023_book}, but with a twist—using complex numbers in the representation. Here, the complex and real parts are each represented by a latent neuron. Their setup allows us to interpret the found representation and obtain the eigenfunctions of the Koopman operator.

\textbf{Prior Knowledge on Object and Interaction Structure:} Graph neural networks (GNNs)~\cite{Scarselli_2009_graph, Li_gated_2017, Gilmer_neural_2017} are particularly useful for handling graph structures in data, such as systems that can be separated in subsystems that interact with each other. Recently, GNNs have been used to extract equations from different parts of the network~\cite{Cranmer_discovering_2020, Pablo_2022}. The authors of~\cite{Cranmer_discovering_2020} applied their techniques to a cosmology example, discovering a new analytic formula that predicts the concentration of dark matter based on the mass distribution of nearby cosmic structures. Similarly, in~\cite{Pablo_2022}, a GNN was trained to simulate the dynamics of the solar system’s Sun, planets, and large moons using thirty years of trajectory data. Symbolic regression was then employed to extract an analytical expression for the force law implicitly learned by the GNN, which closely resembled the true physical law governing the system.

 \textbf{Prior knowledge of various aspects of physics:} In~\cite{Desai2021}, Parsimonious Neural Networks (PNNs) combine neural networks with evolutionary optimization to extract interpretable physical laws directly from data, even capturing fundamental dynamics like Newton’s second law and identifying novel melting temperature relationships in materials. In~\cite{Li2023}, the interpretable meta neural ordinary differential equation (iMODE) method is introduced to rapidly learn generalizable dynamics from trajectories of multiple dynamical systems with varying physical parameters, enabling modeling of unseen systems and inverse inference of physical parameters.

\subsection{Reinforcement learning}

Reinforcement learning (RL) can be an invaluable tool for gaining insights into physical systems, particularly through specialized approaches like projective simulation, introduced in \cite{Briegel2012}. Projective simulation is especially well-suited for RL problems where understanding the agent's decision-making process is crucial. This method allows for the discovery of experimental tool combinations, or "gadgets," that prove useful beyond the agent’s immediate task. Specifically, projective simulation can be employed to design experimental setups aimed at achieving particular goals, as well as to identify gadgets that extend their utility beyond the original experimental design~\cite{Melnikov2018}. The application of such methods to creating setups for quantum experiments is discussed in Chapter~\ref{chapter:quantum} of this review.

On the other hand, reinforcement learning can also play a crucial role in disentangling parameters within latent representations, particularly through state representation learning. Unlike traditional representation learning, where relevant features are extracted from a fixed dataset, state representation learning involves interactive settings~\cite{Lesort2018}. Here, a reinforcement learning agent interacts with an environment, influencing the system's state through its actions. The objective is to develop a low-dimensional representation of the system that evolves over time in response to the agent's actions. This interactive approach offers new opportunities for disentangling representations, as the agent can experiment with various actions and observe their outcomes~\cite{Bengio_2017_independently}. For example, in~\cite{Thomas_2018_disentangling}, the representation is disentangled using an \textit{independence prior}, which encourages independently controllable features of the environment to be stored in separate parameters (see also~\cite{Jonschkowski2015} for a related approach).

As shown in~\cite{FrancoisLavet2019, Jonschkowski2015}, such methodologies can lead to natural representations in certain contexts, like creating an abstract map of a maze for navigation tasks. However, these representations are tied to the actions and policies available to the agent, which are determined by the agent's capabilities and the experimental setup. In physics, these actions are dependent on the experimental apparatus and may be considered less fundamental than the goals or objectives of the agent. A related approach~\cite{PoulsenNautrup2022} involves disentangling representations based on different questions. In~\cite{Jaderberg_2017_reinforcement}, multiple reinforcement learning agents with varying goals share a common representation that is optimized during training. It was observed that learning auxiliary tasks could enhance the agent's performance in achieving the primary objective. A further development in this area was presented in~\cite{Veeriah_discovery_2019}, where the auxiliary tasks are not predefined by humans but are learned autonomously during the training process.

\subsection{Restricted Boltzmann Machines}
Restricted Boltzmann Machines (RBMs) are efficient models for learning complex probability distributions, making them particularly useful for representing systems such as spin configurations, quantum wavefunctions, and statistical ensembles. Moreover, the results obtained from RBMs are often interpretable, providing insights into the underlying physics of these systems. It has been shown that RBMs trained on spin configurations of the Ising model at various temperatures generate a flow of model parameters, including temperature, that approaches the critical value \(T_c\), highlighting a unique feature extraction process reminiscent of renormalization group concepts~\cite{Iso2018}. RBMs have also been used to study thermodynamics and feature extraction, demonstrating their ability to identify relevant physical quantities in datasets and offer insights into their thermodynamic behavior~\cite{Funai2020}. Furthermore, RBMs have served as variational wavefunctions for quantum many-body systems, successfully representing complex quantum states and enabling efficient computation of ground-state properties~\cite{Carleo2017}.

\subsection{AI Scientist}
\label{subsec:AI_scientist}

Recent advancements in AI have enabled systems to autonomously discover scientific principles and generate interpretable models. The "AI Physicist"~\cite{Wu2019} uses strategies such as divide-and-conquer and Occam's razor to identify distinct physical laws from segmented data, producing concise models of complex systems. Meanwhile, "AI-Descartes"~\cite{Cornelio2023} combines logical reasoning with symbolic regression to derive physical laws from minimal data, showcasing a robust integration of data-driven methods with theoretical insights.
Recently, the "AI Scientist"~\cite{lu2024aiscientis} leverages large language models to automate the entire research workflow, from hypothesis generation to paper writing, demonstrating its ability to produce publishable research autonomously.

\subsection{Statistical-Field-Theory- and Renormalization-Group-Inspired Methods}

The article "Learning Interacting Theories from Data" \cite{Merger2023}  considers the generic problem of extracting an effective model of a physically interacting system, in a readable and useful way, from raw data. The authors first employ a (black-box) unsupervised generative model trained on measurements of a physical model system. Then, with a technique involving utilizing invertible networks, a classical action is extracted from the trained black-box (neural network) in a layer-by-layer fashion, represented in an interpretable diagrammatic language. This unveils underlying physical reasons for what the ML algorithm learned to generate a physical model system of interest, in terms of a transparent effective theory. The method was also tested on the MNIST dataset, generating effective theories that approximate the dataset up to the third cumulant.
Technically related is Ref.~\cite{Fischer2022}.

\section{Conserved Quantities and Symmetry Invariants}
\label{chapter:conserved}

Conserved quantities, symmetry invariants, and symmetries are fundamental concepts in physics that shed light on the underlying principles governing the behavior of physical systems.

Conserved quantities are properties of a system that remain constant over time, even as the system undergoes various transformations or interactions. A physical property, denoted as \(Q\), is said to be conserved if
\[ \frac{dQ}{dt} = 0 \]
for all states of the system. Examples include energy, momentum, angular momentum, and electric charge. These quantities play a crucial role in predicting and understanding the behavior of physical systems, as they provide insights into the fundamental laws that govern their dynamics.

Symmetry invariants are mathematical expressions or relationships that remain unchanged under specific transformations. A quantity, denoted as \(I\), is said to be a symmetry invariant under a symmetry transformation \(T\) if
\[ I(T\psi) = I(\psi) \]
for all states \(\psi\) in the system. They serve as powerful tools for identifying and characterizing symmetries within a system. In physics, symmetries refer to operations or transformations that leave the fundamental laws of physics invariant. For example, translational symmetry implies that the laws of physics remain the same regardless of where an experiment is conducted in space, while rotational symmetry suggests that the laws remain invariant under rotations.

The profound connection between symmetries and conserved quantities is encapsulated in Noether's theorem which states that for every continuous symmetry in a physical system, there exists a corresponding conserved quantity. This theorem has had a profound impact on modern physics, forming the basis for many fundamental principles, including the conservation of energy, momentum, and angular momentum. It provides a deep and elegant understanding of how symmetries underlie the behavior of physical systems, making it one of the cornerstones of theoretical physics.

Machine Learning can help in finding symmetry invariants, symmetry transformations, and conserved quantities. These tools might be especially helpful in complex systems where it is difficult to find a solution by hand or where it is impossible to represent the solution in a closed form since machine learning tools can provide accurate approximations guaranteed by the universal approximation theorem. 

Recent articles propose machine learning algorithms that can be grouped into four sections: 
\begin{itemize}
\item Non-neural network-based ML methods for finding symmetry invariants and conserved quantities.
\item Neural network-based ML methods for finding symmetry invariants and conserved quantities.
\item Neural networks that implicitly learn conserved quantities to improve their predictions.
\item ML methods to infer symmetry transformations.
\end{itemize}

\begin{figure*}
    \centering
    \includegraphics[width=\textwidth]{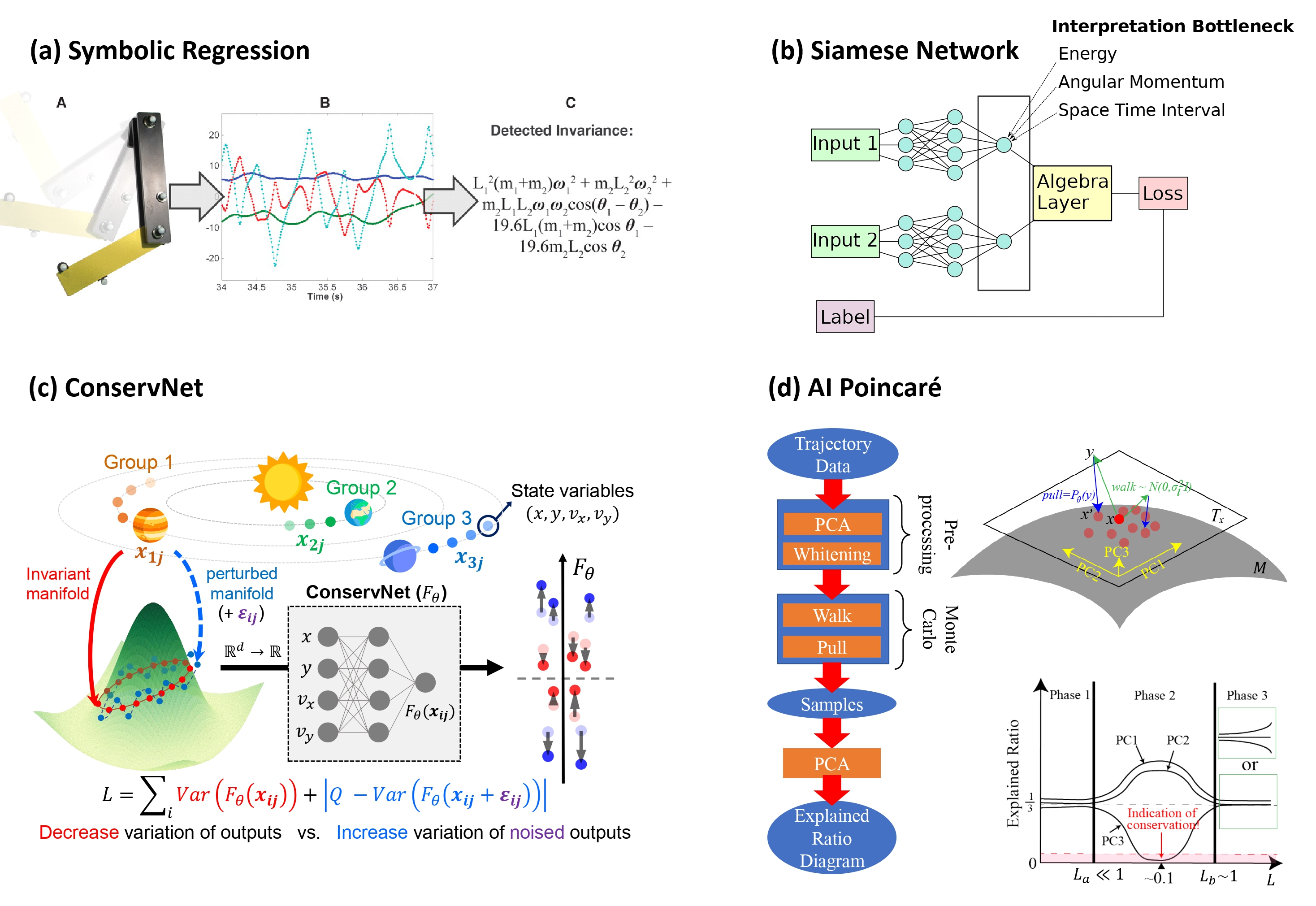}
    \caption{Symmetry invariants and conserved quantities revealed by machine learning. (a) Symbolic Regression can be trained to learn conserved quantities \cite{Schmidt2009}; (b) Siamese Network encode symmetry invariants and conserved quantities in low dimensional latent spaces \cite{Wetzel2020}; (c) ConservNet \cite{Ha2021} that uses a contrastive learning technique to avoid trivial solutions and (d) AI Poincare with Conservation Law Networks~\cite{Liu2022} force neural networks to learn conserved quantities by enforcing conservation on physical paths but allow violations on unphysical distortions., similar to denoising objectives in diffusion models~\cite{stanczuk2022your}.}
    \label{fig:ConservedQuantityNN}
\end{figure*}

\subsection{Non-neural network-based ML methods for finding symmetry invariants and conserved quantities}

Before the hype around artificial neural networks started in 2012, Schmidt and Lipson proposed a symbolic regression-based method for finding conserved quantities in "Distilling Free-Form Natural Laws from Experimental Data"~\cite{Schmidt2009}. Without any prior knowledge of physics, kinematics, or geometry, the algorithm is able to discover Hamiltonians, Lagrangians, and other laws of geometric and momentum conservation in simple systems in symbolic form.

The article "Discovering Conservation Laws from Data for Control"~\cite{Kaiser2018} describes a data-driven architecture for discovering conserved quantities based on Koopman theory. The authors demonstrate their framework with the example of three-dimensional rigid body equations, where they simultaneously discover the total energy and angular momentum.

The central proposal in "Conservation laws and spin system modeling through principal component analysis"~\cite{Yevick2021} is a PCA-based method that can identify conserved quantities from recorded system trajectories as long as they can be written as polynomials.

In "Extracting Dynamical Models from Data"~\cite{https://doi.org/10.48550/arxiv.2110.06917}, the author proposes a machine learning-based strategy that learns the underlying model from data
of a dynamical system. It combines ideas from the fields of Lie symmetries as
applied to differential equations (DEs), numerical integration (such as RungeKutta), and machine learning (ML). During this process, the algorithm constructs quantities that can be interpreted as constants of motion.

The article "Discovering conservation laws using optimal transport and manifold learning
\cite{Lu2023}" reformulates the task of finding conserved quantities as a manifold learning problem. The tools from optimal transport theory and manifold learning provide a direct geometric approach to identifying conservation laws that is both robust and interpretable without requiring an explicit knowledge of the underlying model.

While the previous approaches discuss how to learn conserved quantities in classical systems, the manuscript "Learning conservation laws in unknown quantum dynamics  \cite{https://doi.org/10.48550/arxiv.2309.00774}" addresses this problem for quantum systems. The algorithm combines the classical shadow formalism for estimating expectation values of observables and ideas form singular value decomposition to discover all such conservation laws in unknown quantum dynamics.

In "A Unified Framework to Enforce, Discover, and Promote Symmetry in Machine Learning
" \cite{https://doi.org/10.48550/arxiv.2311.00212}, the authors provide a unifying framework for incorporating symmetry into machine learning models in three ways: 1. enforcing known symmetry when training a model; 2. discovering unknown symmetries of a given model or data set; and 3. promoting symmetry during training. This framework is improved in "Towards Robust Data-Driven Automated Recovery of Symbolic Conservation Laws from Limited Data
" \cite{https://doi.org/10.48550/arxiv.2403.04889}.

The authors of "Machine Learning Conservation Laws of Dynamical systems" \cite{https://doi.org/10.48550/arxiv.2405.20857} propose a method based on an “indeterminate” form of kernel ridge regression where the labels still have to be found, to discover conserved quantities.

\subsection{Neural network-based ML methods for finding symmetry invariants and conserved quantities}

While artificial neural networks are typically seen as the most expressive machine learning methods, they are also among the most elusive to interpret. In this section, we review the developments regarding revealing conserved quantities and symmetry invariants of physical systems with neural network-based methods. These methods are fairly similar in the sense that many submodules are compatible and interchangeable between different algorithms. An overview of different architectures can be found in \fig{fig:ConservedQuantityNN}. They show the initial formulation of the phase detection algorithm via Siamese neural networks \cite{Wetzel2020}, the improvement that can be gained through generating negative data from deforming contours \cite{Ha2021} the combination of such methods with symbolic regression algorithms and the development of loss functions that allow the neural network to learn multiple independent conserved quantities \cite{Liu2022}, similar to how diffusion models estimate the manifold dimensionality~\cite{stanczuk2022your}. 

The article "Discovering symmetry invariants and conserved quantities by interpreting siamese neural networks"~\cite{Wetzel2020} introduces a neural network-based method for finding symmetry invariants and conserved quantities from a data set containing multiple observations of the same system. The neural network is trained to identify functions that evaluate the same value on observations of the same object. It operates in tandem with an identical neural network that forces it to take on different values when evaluated on observations of other objects to penalize the learning of trivial functions.

The algorithm developed in "Discovering invariants via machine learning by deforming trajectories"~\cite{Ha2021} operates on trajectory data. In a similar spirit to the article in the previous paragraph, a neural network is trained to evaluate the same number along a trajectory. However, the negative data that penalizes the neural network from learning trivial functions is generated by perturbing the trajectory itself. It has been shown that this method is more robust and less susceptible to noise compared to the previous Siamese network-based method.

The authors of the articles discussed in the following paragraphs have contributed several important ideas to find conserved quantities with the help of neural networks \cite{Liu2021,Liu2022,https://doi.org/10.48550/arxiv.2305.19525}. Their first proposal is formulated in "Machine Learning Conservation Laws from Trajectories"~\cite{Liu2021}. Similar to the article in the previous section it operates on trajectory data and learns conserved quantities based on artificially perturbing the underlying trajectories. Their method uses the artificial neural network to predict the mapping of the perturbed trajectory back to the original trajectory within a walk-pull Monte Carlo module. This phase can be used to reveal conserved quantities. Their second method is described in "Machine learning conservation laws from differential equations"~\cite{Liu2022}. where they develop a machine learning algorithm that discovers conservation laws from differential equations, both numerically (parametrized as neural networks) and symbolically with the help of their own symbolic regression algorithm AI Feynman -- A physics-inspired method for symbolic regression~\cite{Udrescu2020}. It is also important to note that this algorithm provides a unified approach to learning multiple independent conserved quantities. Finally, this approach is extended and applied to fluid mechanics and atmospheric chemistry in "Discovering New Interpretable Conservation Laws as Sparse Invariants"~\cite{https://doi.org/10.48550/arxiv.2305.19525}, where their machine-discovered conserved quantity was later verified and generalized by domain experts in atmospheric chemistry~\cite{blokhuis2023data}.

The article "Machine learning independent conservation laws through neural deflation"~\cite{https://doi.org/10.48550/arxiv.2303.15958} extends the unified framework to learn multiple independent conserved quantities from the previous paragraph. The authors formalize the mathematical requirements of what it means to have truly independent conserved quantities. Further, they show that their algorithm is able to find the full set of Poisson commuting conserved quantities in various systems.

The authors of "Learning Hamiltonian Neural Koopman operator and Simultaneously Sustaining and discovering conservation laws
" \cite{Zhang2024} formulate the Hamiltonian neural Koopman operator (HNKO) which builds upon integrating mathematical physics into learning the Koopman operator, which allows for the automatic preservation and discovery of conservation laws.

\subsection{Neural networks that implicitly learn conserved quantities to improve their predictions}

The following articles describe how to improve neural networks that are tasked with learning the dynamics of a physical system. All these articles follow a similar spirit of implicitly learning conservation laws which provide an inductive bias for preventing the prediction from violation of physical laws.
The article "Hamiltonian Neural Networks"~\cite{https://doi.org/10.48550/arxiv.1906.01563} describes the Hamiltonian neural network that learns to conserve a quantity
that is analogous to total energy. Similarly, in "Lagrangian neural networks"~\cite{cranmer2020lagrangian} the Lagrangian network learns the Lagrangian inducing a strong physical prior to the learned dynamics.Finally, "Noether Networks: meta-learning useful conserved quantities
"~\cite{alet2021noether} take inspiration from Noether’s theorem to reduce the
problem of finding inductive biases to meta-learning useful conserved quantities. The purpose of learning these conservation laws is to improve prediction quality on learned trajectories even beyond applications in physics for example position prediction in videos.

\subsection{ML methods to infer symmetry transformations}
In this section, we discuss methods that use machine learning to find symmetry transformations that leave a specific system or quantity invariant.

The authors of "Learning a local symmetry with neural networks"~\cite{Decelle2019} explore the capacity of neural networks to detect a symmetry with complex local and non-local patterns. They show how to design a neural network-based framework that is able to learn $\mathbb{Z}_2$ gauge symmetry and how to find compressed latent representations of the gauge orbits.

The paper "Interpretable conservation law estimation by deriving the symmetries of dynamics from trained deep neural networks"~\cite{Mototake2021} is another take on finding conserved quantities. However, the approach taken is based on identifying symmetry transformations with a neural network and then inferring conserved quantities with Noether's Theorem.

In "Detecting symmetries with neural networks"~\cite{Krippendorf2020} the authors discuss how neural networks can be used to identify symmetries and orbits of the symmetry in the input space. They discuss examples based on rotation groups SO(n) and the unitary group SU(2). This work is aimed at the classification of complete intersections of Calabi-Yau manifolds where it is crucial to identify discrete symmetries on the input space.

The authors of the following articles have made several important contributions toward the identification of symmetry transformations with machine learning
\cite{Forestano2023,Forestano2023b,https://doi.org/10.48550/arxiv.2309.07860}.

In their first work "Deep learning symmetries and their Lie groups, algebras, and subalgebras from first principles"~\cite{Forestano2023} they design a deep-learning algorithm for the discovery of continuous symmetry groups. With this approach, it is possible to recover the complete subgroup structure of the rotation groups SO(n), and of the Lorentz group SO(1,3). This method can be extended to derive sparse representations as shown in "Discovering sparse representations of Lie groups with machine learning"~\cite{Forestano2023b}. A deep dive into the structure of ML-learned symmetries can be found in "Identifying the Group-Theoretic Structure of Machine-Learned Symmetries"~\cite{https://doi.org/10.48550/arxiv.2309.07860}.

The article "Finding discrete symmetry groups via Machine Learning"~\cite{https://doi.org/10.48550/arxiv.2307.13457} describes a neural network-based approach capable of automatically discovering discrete symmetry groups in physical systems. This method identifies the finite set of parameter transformations that preserve the system's physical properties. 

In case symmetries are hidden or only exist in some latent space, the transformation from the original space to the latent space is necessary to discover the symmetry. This line of work features "Latent Space Symmetry Discovery"~\cite{yang2023latent} and "Machine Learning Hidden Symmetries"~\cite{liu2022machine}, where they both use neural networks to learn transformations between the original space and the latent space, but they differ in how symmetries are discovered (via losses defined on group operators vs. Lie operators).

\begin{figure*}
    \centering
    \includegraphics[width=\textwidth]{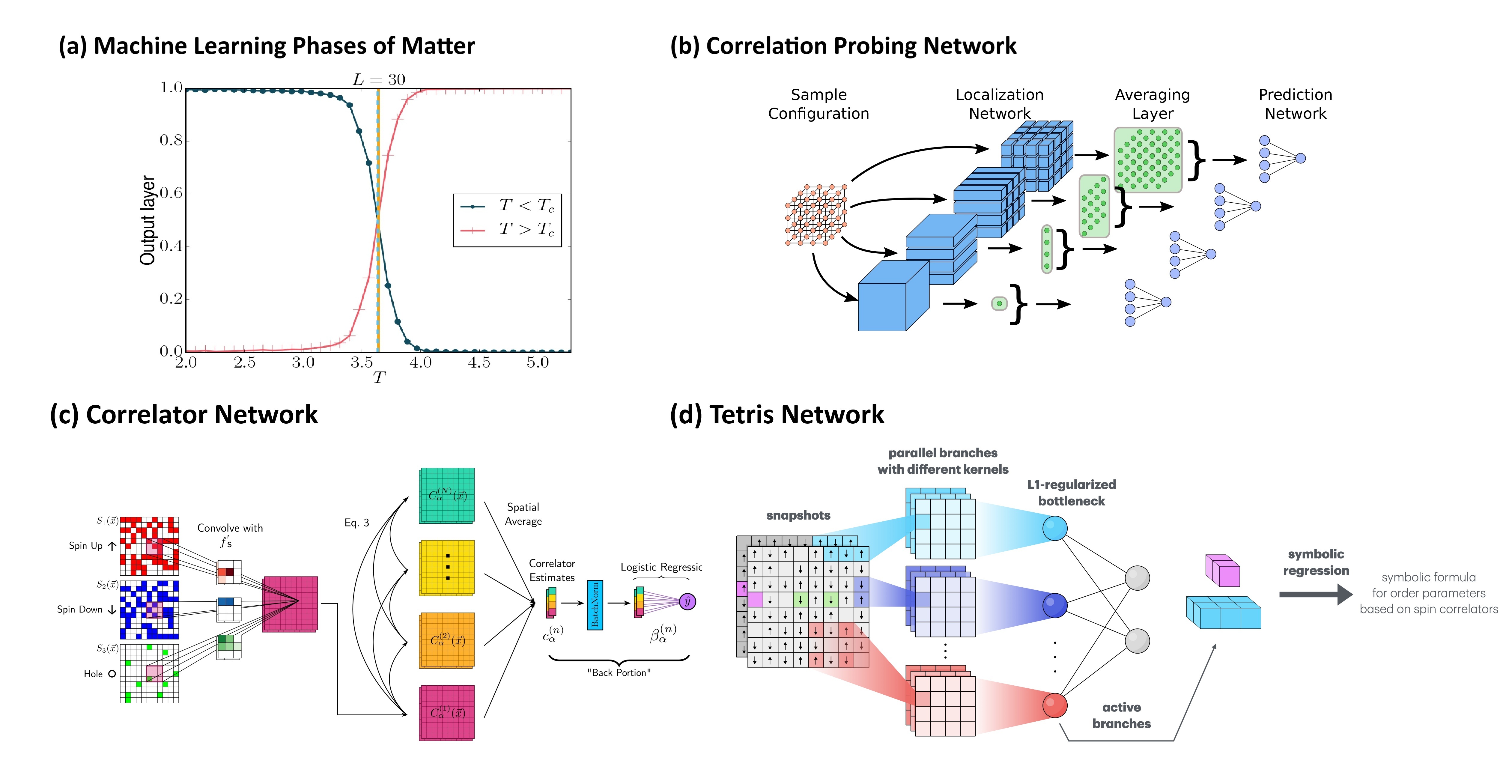}
    \caption{Phase Diagram Signatures from Neural Networks. (a) A neural network trained on Monte-Carlo configurations to distinguish phases of matter. The neural network learns in low and high-temperature regimes and is tasked to predict the phase at intermediate temperatures. This network accurately predicts the phase transition respecting finite size effects \cite{Carrasquilla2017}. In the current review, we discuss what physical quantities such a machine learning framework learns. Promising strategies for interpreting neural networks tasked with phase recognition include (b) Correlation Probing Neural Network \cite{Wetzel2017b}, (c) Correlator Neural Network \cite{Miles2021}, (d) Tetris Convolutional Neural Network \cite{https://doi.org/10.48550/arxiv.2411.02237}}
    \label{fig:PhaseDiagramNN}
\end{figure*}

\section{Phase Diagrams}
\label{chapter:phase_diagrams}

In many disciplines of physics, a central problem is comprehending the fundamental behavior of matter. Often this behavior can be described universally by distinct phases of matter. These include phases like solids, characterized by ordered structures or liquids with mobile yet closely arranged particles. Phase diagrams describe the location of specific phases emerging in physical systems based on external control parameters like temperature or pressure.

Many phase transitions can be universally classified by Ehrenfest's classification by the order of discontinuity of a derivative of the free energy with respect to a control parameter. Landau theory is an attempt to formulate a general theory of continuous phase transitions with the help of the associated concept of the order parameter as a descriptor of the essential character of the transition.

An order parameter contains unique signatures corresponding to a certain phase. It quantifies the level of order during a phase transition, typically varying from zero in one phase to a non-zero value in the other. At the critical point, the susceptibility of the order parameter tends to diverge. An instance of an order parameter is the total magnetization in a ferromagnetic system during a phase transition. In transitions between liquid and gas, the order parameter is the difference in densities. The emergence of order parameters in theory stems from symmetry breaking.

Using machine learning to map out phase diagrams involved collecting sample configurations produced for a range of different control parameters. This data can be collected through experimental observations or numerical calculations such as from Monte-Carlo simulations or tensor networks. A machine learning algorithm can then be trained to assign known phases to certain configurations (supervised learning) or to cluster the configurations within some latent space (unsupervised learning). Typically, the phase transitions can be identified when a phase assignment or the locations of configurations in the latent space undergo a rapid change. 

The original article outlining how to use supervised machine learning to calculate phase transitions is "Machine learning phases of matter~\cite{Carrasquilla2017}". It describes how to train a neural network to distinguish phases of Monte-Carlo simulated configurations of simple spin models. A hybrid method between supervised and unsupervised learning "Learning phase transitions by confusion~\cite{vanNieuwenburg2017}" suggests systematically mislabelling neural network training data to identify a phase separation that produces the least mislabelled training data set. Fully unsupervised methods have been developed in "Discovering phase transitions with unsupervised learning~\cite{Wang2016}" based on principal component analysis and in "Unsupervised learning of phase transitions: from principal component analysis to variational autoencoders~\cite{Wetzel2017}" based on autoencoder neural networks.

\subsubsection{Supervised Artificial Neural Networks}

In this section, we discuss several articles that interpret supervised artificial neural networks that are applied to map out phase diagrams in a diverse range of physical systems. The purpose of this interpretation is twofold: (1) a priori it is not clear that the machine learning algorithm extracts the same features as humans do. Similar to the different physical quantities that are used to identify phase transitions in Landau theory or in Ehrenfests classification, the machine learning algorithm could rely on a completely separate framework to locate these transitions. (2) Since various signatures of phases are inherently linked to the nature of the underlying phase, extracting physical quantities like the order parameter from a fully trained machine learning could give insight into the properties of the underlying phase. Two very successful interpretations have been performed in Ref.~\cite{Wetzel2017b}, where the authors extracted the order parameter in the confinement-deconfinement phase transition in SU(2) lattice gauge theory, and Ref.~\cite{Miles2021}, where the neural network is used to reveal the defining correlations of certain phases in condensed matter systems. The corresponding architectures can be found in \fig{fig:PhaseDiagramNN}.

In "Machine learning of explicit order parameters: From the Ising model to SU (2) lattice gauge theory"~\cite{Wetzel2017b} the authors construct a neural network architecture that iteratively reduces the neural network capacity until a performance drop is observed, see \fig{fig:PhaseDiagramNN}. The remaining artificial neural network can only learn a small number of correlations that are present in the data, which can be extracted through polynomial regression. The authors extracted the formulas of several physical observables, including the magnetization and energy of the Ising model and the Polyakov loop order parameter of SU(2) lattice gauge theory.

The authors of "Smallest neural network to learn the Ising criticality"~\cite{Kim2018} follow a similar idea and find that an artificial neural network with two hidden neurons is sufficient to learn the 2d Ising model phase transition. By inspecting the two neuron architecture they verify that the neural network learns the magnetization. They extend this approach in \cite{Kim2022} "Minimal Neural Network to Learn the Metal-insulator Transition in the Dynamical Mean-field Theory", where they show that patterns encoded in the weight matrix of the neural network correspond to an order parameter that can precisely detect the transition point.

In "Probing many-body localization with neural networks"~\cite{Schindler2017}, a simple neural network trained on entanglement spectra can accurately identify the transition between many-body localized and thermalizing regimes in a disordered Heisenberg spin-1/2 chain. By analyzing the network using the dreaming technique, it is possible to show that the neural network correctly learns the power-law structure of the entanglement spectra in the many-body localized regime.

The article "Parameter diagnostics of phases and phase transition learning by neural networks"~\cite{Suchsland2018} discusses the application of both fully-connected and convolutional neural networks for the two-dimensional Ising- and XY models. By examining the weight matrices and the convolutional filter kernels that result from the fully trained neural networks, they find that the neural network learns the magnetization and domain walls in the Ising model while also extracting differences in spin orientations in the XY-model indicating the neural networks tendency to learn vortex structure.

Artificial neural networks are applied to recognize phases in topological models in the article "Machine Learning Topological Invariants with Neural Networks"~\cite{Zhang2018}. By opening up the neural network, the authors show that the network does learn the discrete version of the winding number formula.

Phase diagrams of dynamical systems are explored in "Learning phase transitions from dynamics"~\cite{vanNieuwenburg2018}. For this purpose, recurrent neural networks are employed for classifying phases of two distinct models of one-dimensional disordered and interacting spin chains. The authors further examine a periodically-driven model featuring an inherently dynamical time-crystalline phase. The results highlight the possibility of extracting a dynamical order parameter from the neural networks.

In "Phase transition encoded in neural network"~\cite{Kashiwa2019} the authors applied neural networks to calculate the phase transition of Ising and Potts models. In agreement with the previous works, by inspecting the weights, they also show that the neural network encodes the magnetization and the energy.

The interpretability of various types of machine learning algorithms in the context of phase recognition is examined in "Interpretable machine learning for inferring the phase boundaries in a nonequilibrium system"~\cite{Casert2019}. The authors apply techniques from statistics and computer science to study the explained variance ratio in principal component analysis and highlight the learning behavior of saliency maps in neural networks.

A game-theoretic neural-network-based framework for the discovery of phase transitions is described in "Automated discovery of characteristic features of phase transitions in many-body localization"~\cite{Huembeli2019}. This approach is used to identify a new order parameter for the disorder-driven many-body localization transition.

The interpretation of neural networks applied to topological phase transitions has been studied in "Interpreting machine learning of topological quantum phase transitions"~\cite{Zhang2020}. The authors examine topological quantum phase transitions in the following systems: Chern insulator, $\mathcal{Z}_2$ topological insulator, and $\mathcal{Z}_2$ quantum spin liquid, using a shallow fully connected feed-forward ANN. To identify the topological phases, the neural networks learn physically meaningful features, such as topological invariants and deconfinement of loops. 

The authors of "Learning what a machine learns in a many-body localization transition"~\cite{Kausar2020} employ a convolutional neural network to explore the distinct phases in random spin systems. In these systems, the neural network depends on different physical quantities depending on the filter size: the smallest nontrivial kernel width establishes the level spacing as the signature to distinguish the many-body localized phase from the thermal phase. In the case of a large kernel width, the neural network detects phases from the raw energy spectrum using a previously unknown signature.

The study "Phase detection with neural networks: interpreting the black box"~\cite{Dawid2020phase} shows that an interpretability method called influence functions can interpret neural networks trained to predict phases of matter in a spinless Fermi–Hubbard model at half-filling. The authors reveal that the network learns order parameters for the phase transition.

The article "Learning Order Parameters from Videos of Skyrmion Dynamical Phases with Neural Networks\cite{Wang2021}" discusses the use of neural networks for classifying dynamical skyrmion phases. Further, a parameter visualization scheme is proposed that enables the interpretation of what neural networks have learned.

%Finding the deconfinement temperature in lattice Yang-Mills theories from outside the scaling window with machine learning\cite{Boyda2021} also Polyakov Loop

As shown in "Observing a topological phase transition with deep neural networks from experimental images of ultracold atoms"~\cite{Zhao2022}, it is possible to analyze topological phase transitions using neural networks applied to experimental data. A convolutional neural network is trained to distinguish experimental data obtained in a symmetry-protected topological system of spin-orbit-coupled fermions. The authors find two topological phase transitions. By visualizing the filters one can see that the CNN uses the same information to make the classification in the system as the conventional analysis, namely, spin imbalance.

The manuscript "Randomized-Gauge Test for Machine Learning of Ising Model Order Parameter"~\cite{Morishita2022} demonstrates that neural networks can extract the order parameter or the energy of the random-gauge Ising model when applied to the phase classification task. The authors also discuss how and where the information of random gauge is encoded in neural networks and attempt to reconstruct the gauge from the neural network parameters.

A successful method to extract meaningful correlations from physical systems in both experimental and theoretical contexts is the Correlator convolutional neural network, see \fig{fig:PhaseDiagramNN}, which was introduced in "Correlator convolutional neural networks as an interpretable architecture for image-like quantum matter data"~\cite{Miles2021}. The key idea is to build a neural network architecture that discovers a hierarchy of features in the data that are directly interpretable in terms of physical observables. By employing $L_1$ (i.e., lasso) feature selection, it is possible to reduce the number of meaningful features to the few key observables that characterize physical phases. At first, this network was applied to simulated snapshots produced by two candidate theories approximating the doped Fermi-Hubbard model.
In a follow-up article, "Machine learning discovery of new phases in programmable quantum simulator snapshots"~\cite{Miles2023}, the authors apply the correlator neural network to experimental data generated using a programmable Rydberg atom array quantum simulator. Further, in a similar work, "Fluctuation based interpretable analysis scheme for quantum many-body snapshots"~\cite{Schlmer2023}, the authors apply the correlator neural network together with confusion learning to study thermodynamic properties of the 2D Heisenberg model. It is possible to identify the full counting statistics of nearest neighbor spin correlations as the most important quantity for the decision process of the neural network. This type of neural network was also of help when analyzing the thermalization behavior of an interacting quantum system that undergoes a nonequilibrium phase transition from an ergodic to a many-body localized phase in "Analyzing Nonequilibrium Quantum States through Snapshots with Artificial Neural Networks"~\cite{Bohrdt2021}. The correlator network can be combined with transformers which are able to capture non-local dependencies and are thus effective for classifying quantum phases of matter. In Ref.~\cite{https://doi.org/10.48550/arxiv.2407.21502}, a correlator transformer is used to classify phases of matter, which is intrinsically interpretable.

In "Characterizing out-of-distribution generalization of neural networks: application to the disordered Su-Schrieffer-Heeger model"~\cite{https://doi.org/10.48550/arxiv.2406.10012}, convolutional neural networks are trained to predict phases in the SSH model. Initially, their model failed to predict the phase diagram. However, interpreting the neural network and understanding the data with class activation mapping (CAM) and PCA helped the authors identify their own sources of error to build a more robust machine-learning pipeline to recognize phases in the SSH model.

Recently, TetrisCNN~\cite{https://doi.org/10.48550/arxiv.2411.02237} was introduced to combine ideas from neural network interpretation \cite{Wetzel2017b,Miles2021} and symbolic regression \cite{cranmer2023interpretable} into a single pipeline capable of classifying phases of matter together with revealing order parameters in symbolic form.

\subsubsection{Unsupervised Artificial Neural Networks} 

Unsupervised Artificial Neural Networks in this section are mostly based on autoencoders, see \fig{fig:Autoencoder}. The interpretation of autoencoders faces the same difficulties as standard feed-forward neural network architectures for supervised learning. However, as we discuss in this paragraph, autoencoders tend to store physically meaningful information about the underlying system in their latent space. If autoencoders are trained to encode configurations of systems undergoing a phase transition, the latent parameters try to approximate a suitable order parameter. This latent space order parameter can then be employed similarly to traditional order parameters in order to examine phases and phase transitions. In this section, we focus on articles that made contributions regarding the interpretability of autoencoders applied to phase recognition.

In the article "Unsupervised learning of phase transitions: from principal component analysis to variational autoencoders"~\cite{Wetzel2017} train different versions of autoencoders on Monte-Carlo-sampled configurations of the ferromagnetic and antiferromagnetic two-dimensional Ising model in 2d and the three-dimensional XY-model. It was discovered that the latent space variables approximate the underlying order parameters, the magnetization, and the staggered magnetization. Based on this observation, the author proposed this procedure as a method for unsupervised learning of phase diagrams.

In "Connecting phase transition theory with unsupervised learning"~\cite{https://doi.org/10.48550/arxiv.1712.05704} the authors apply autoencoders to spin glasses in addition to Ising type lattices. It is also found that the latent variables reproduce the ferromagnetic and antiferromagnetic order parameters in the Ising model. Further, it is possible to formulate an approximate order parameter for the spin glass model. Additionally, it is shown that the cross-entropy loss of a successfully trained autoencoder could be used to estimate the physical entropy.

The article "Unsupervised machine learning account of magnetic transitions in the Hubbard model"~\cite{Chng2018} describes the application of several unsupervised machine learning techniques, including autoencoders, random trees embedding, and T-SNE, to Monte Carlo simulated configurations of the Ising and Fermi-Hubbard models. Results from a convolutional autoencoder for the three-dimensional Ising model can be shown to produce the magnetization and the susceptibility as a function of temperature with a high degree of accuracy. The output of the T-SNE algorithm applied to the Hubbard model shows nearly perfect agreement with the antiferromagnetic structure factor in the weak-coupling regime. T-SNE also predicts a transition to the canted antiferromagnetic phase for the three-dimensional model when a strong magnetic field is present. 

Another application of autoencoders applied to Monte-Carlo simulations of spin models can be found in "The critical temperature of the 2D-Ising model through deep learning autoencoders"~\cite{Alexandrou2020}. Similarly to previous work, their testbed consists of the ferromagnetic and antiferromagnetic Ising models. The major contribution of this article is the finite size analysis of the latent variable order parameter. 

In "Deep learning on the 2-dimensional Ising model to extract the crossover region with a variational autoencoder"~\cite{Walker2020} the two-dimensional Ising model on a square lattice with non-vanishing external field is investigated with a variational autoencoder for the purpose of extracting the crossover region between the ferromagnetic and paramagnetic phases. The latent variable provides suitable metrics for tracking the order and disorder in the Ising configurations that extend to the extraction of a crossover region in a way that is consistent with expectations. 

"Unsupervised machine learning of topological phase transitions from experimental data"~\cite{Kaming2021} describes the application of unsupervised machine learning, most notably autoencoders, for anomaly detection and influence functions, in order to map the topological phase diagram of the Haldane model from ultracold atom experiments.

With the examination of high entropy alloys with variational autoencoders in "Neural network-based order parameter for phase transitions and its applications in high-entropy alloys"~\cite{Yin2021} the authors show that it is possible to learn more complex order parameters beyond spin models. A VAE-based order parameter can reproduce the interplay of several order parameters in multiple refractory high-entropy alloys. Using this order parameter, it is possible to aid in alloy design by mimicking the natural mixing process of elements.

% Variational autoencoder analysis of Ising model statistical distributions and phase transitions\cite{Yevick2022}
% autoencoder is trained on a thermal equilibrium distribution of Ising spin realizations. Synthetic spin realizations are then obtained by decoding sets of randomly assigned latent variable values and interpreting the output as the likelihood of a certain spin orientation. The resulting state distribution in energy-magnetization space then qualitatively resembles that of the training samples. However, this paper demonstrates that because such techniques suppress correlations among spins, the computed energies are unphysically large for low-dimensional latent variable spaces. The features of the learned distributions as a function of temperature, however, qualitatively indicate the presence of phase transitions.

% The article "Application of the variational autoencoder to detect the critical points of the anisotropic Ising model\cite{Baul2023}" generalizes the previous applications of variational autoencoders to the two-dimensional Ising model to a system with anisotropy. The authors reproduce the phase diagram for a wide range of anisotropic couplings and temperatures via a variational autoencoder without the explicit construction of an order parameter. 

The article "Semi-supervised learning of order parameter in 2D Ising and XY models using Conditional Variational Autoencoders"~\cite{https://doi.org/10.48550/arxiv.2306.16822} enhances the autoencoder framework by employing a conditional variational autoencoder and training it in a semi-supervised manner. The framework is applied to Monte-Carlo-generated configurations of the two-dimensional ferromagnetic Ising model and the two-dimensional XY model. It is shown that the conditional variational autoencoder latent variables reproduce the order parameter more faithfully than a traditional variational autoencoder.

The authors of "Group-equivariant autoencoder for identifying spontaneously broken symmetries"~\cite{Agrawal2023} describe the application of group-equivariant autoencoder to locate phase boundaries by determining which symmetries of the Hamiltonian have been spontaneously broken at each temperature. Within their framework, the group-equivariant autoencoder learns an order parameter invariant to these “never-broken” symmetries. The learned order parameter is equivariant to the remaining symmetries of the system. By examining the group representation by which the learned order parameter transforms, it is possible to extract information about the associated spontaneous symmetry breaking. This procedure is applied to the ferromagnetic and antiferromagnetic two-dimensional Ising models.

\subsubsection{PCA}
The principal component analysis projects data onto its principal components via an orthogonal transformation~\fig{fig:MLalgo}. While traditional PCA performs the projection in the input space, kernel PCA is able to do this operation in a kernel space. PCA is typically very interpretable since it is easy to just read out the values of the principal components. When applied to the problem of finding phase transitions and phase diagrams, these principal components tend to correspond to the order parameters of the underlying physical system.

The proposal to employ PCA for discovering phase transitions was first proposed in "Discovering phase transitions with unsupervised learning"~\cite{Wang2016}. The author examined Monte-Carlo generated configurations of the Ising model and conserved order parameter Ising model. Using clustering analysis, it is possible to identify distinct phases in the feature space. This approach successfully
reveals the nature of the underlying order parameter and structure factor.

The article "Discovering phases,  phase transitions,  and crossovers through unsupervised machine learning: A critical examination"~\cite{Hu2017} is a deep dive into various aspects of unsupervised learning of phase transitions, mainly focusing on PCA, but also including autoencoders. The authors employ these algorithms in several classical spin models: the square- and triangular-lattice Ising models, the Blume-Capel model, a highly degenerate biquadratic-exchange spin-1 Ising (BSI) model, and the two-dimensional XY model. Further, the authors also study the limitations of PCA and point out models in which this algorithm does not succeed in discovering phase transitions.

PCA is used to identify structural changes at the melting transition in classical molecular dynamics simulations in "Identifying structural changes with unsupervised machine learning methods"~\cite{Walker2018}. In addition to an estimation of the melting point transition, PCA reveals a pattern criterion that is conceptually similar to how humans describe the melting transition. 

The authors of "Deriving the order parameters of a spin-glass model using principal component analysis"~\cite{Kiwata2019} investigate various spin models with principal component analysis. Similar to previous results in most simple spin models, the first principal component is found to be equivalent to the magnetization in the ordered phase, which is the order parameter. In order to study spin glasses, they apply PCA to datasets corresponding to the Sherrington-Kirkpatrick model and show that PCA is able to find traces of a spin glass order parameter. 

In "Recognition of polymer configurations by unsupervised learning"~\cite{Xu2019}, unsupervised learning in the form of PCA and diffusion maps together with a hybrid neural network is employed to examine phase transitions of polymer configurations. PCA and diffusion maps manage to extract low-dimensional representations of polymer states, distinguishing coiled and collapsed states while revealing key physical insights.

Hidden spin order is examined via PCA in "Unsupervised machine learning of quenched gauge symmetries: A proof-of-concept demonstration"~\cite{LozanoGmez2022}.
In this work, the authors study two random spin systems with a hidden ferromagnetic order that can be exposed by applying a Mattis gauge transformation. It is possible to detect the hidden order, quantify the corresponding gauge variables, and map the original random models onto simpler gauge-transformed ferromagnetic ones.

In "Learning new physical descriptors from reduced-order analysis of bubble dynamics in boiling heat transfer"~\cite{Rokoni2022} PCA is used to extract new physical descriptors of boiling heat transfer from experimental images.

The following articles belong to a two-part series, one focusing on traditional PCA "Machine learning of frustrated classical spin models. I. Principal component analysis"~\cite{Wang2017} and the other examining kernel PCA "Machine learning of frustrated classical spin models II: Kernel principal component analysis"~\cite{Wang2018} applied to spin systems. In both articles, PCA is applied to Monte Carlo simulated data from the XY model in frustrated triangular and union jack lattices. The authors comment on the differences and the capabilities of kernel PCA to learn non-linear functions. Further, they point out that in addition to the principal component revealed by the linear PCA, the kernel PCA can find two more principal components using the data generated by Monte Carlo simulation for various temperatures as the input. One of them is related to the strength of the U(1) order parameter and the other directly manifests the chiral order parameter that characterizes the $\mathbb{Z}_2$ symmetry breaking. 

Another two-part series of articles consists of "Unsupervised machine learning for detection of phase transitions in off-lattice systems. I. Foundations"~\cite{Jadrich2018} and "...II. Applications"~\cite{Jadrich2018b}, where the authors demonstrate the utility of a PCA for the detection of phase transitions in off-lattice systems. In the first article, PCA is used to detect the freezing transitions of two-dimensional hard-disk and three-dimensional hard-sphere systems, as well as liquid-gas phase separation in a patchy colloid model. In these models, PCA is able to discover order-parameter-like quantities that report on phase transitions. In a second paper, the method is extended to explore the detection of phase transitions in various model systems controlled by compositional demixing, liquid crystalline ordering, and non-equilibrium active forces.

\subsubsection{SVM}
Support vector machines for classification are designed to find hyperplanes separating different classes \fig{fig:MLalgo}. These hyperplanes can either be in the original data space for linear support vector machines or in the case of kernel support vector machines in the kernel space allowing for non-linear decision functions. While they might not be as powerful as artificial neural networks, their advantage is that the decision function can be interpreted more easily. Further, by choosing a physics-inspired kernel function, it is possible to capture meaningful physical features.

The idea to use support vector machines for phase classification and the calculation of phase transitions was first proposed in "Kernel methods for interpretable machine learning of order parameters"~\cite{Ponte2017}. The authors apply SVMs to a set of two-dimensional spin models: the ferromagnetic Ising model, a conserved-order-parameter Ising model, and the Ising gauge theory. They find that SVMs can learn the mathematical form of physical discriminators, such as order parameters and Hamiltonian constraints.

The article "Interpretable machine learning study of the many-body localization transition in disordered quantum Ising spin chains"~\cite{Zhang2019} discusses the application of SVMs to a disordered quantum Ising chain with transverse external field to study the phase transition between many-body localized and thermal phases. By analyzing the decision function, the authors find meaningful physical descriptors that are analogous to the inverse participation ratio in configuration space.

In the following, we discuss a multi-part collection of articles that made very significant contributions to interpretable SVMs applied to the phase recognition problem.

In their first paper "Probing hidden spin order with interpretable machine learning"~\cite{Greitemann2019}, the authors construct a physically inspired kernel for SVMs and apply this algorithm to an artificially designed spin model that exhibits multipolar spin order. The algorithm is successful at identifying all phases and it is possible to extract the analytical form of nematic order parameter tensors.

The article "Identification of emergent constraints and hidden order in frustrated magnets using tensorial kernel methods of machine learning"~\cite{Greitemann2019b} describes the application of SVMs with a physically inspired tensorial kernel to uncover the phase diagram of a classical frustrated spin model. By interpreting the decision function, it is possible to reveal order parameter tensors of phases with broken symmetry and the local constraints that signal an emergent gauge structure and so characterize classical spin liquids. 

Further, in "Learning multiple order parameters with interpretable machines"~\cite{Liu2019}, the authors focus on addressing the multiclass classification problem of multiple phases of matter with their tensorial-kernel SVM.
 
Additionally, the article "Revealing the phase diagram of Kitaev materials by machine learning: Cooperation and competition between spin liquids"~\cite{Liu2021b} demonstrates the application of the tensorial-kernel SVM to the Kitaev model on the honeycomb lattice in a magnetic field. The authors point out the possibility of the existence and nature of a previously unknown phase.
Finally, the authors show in "Unsupervised interpretable learning of phases from many-qubit systems"~\cite{Sadoune2023} that the tensorial-kernel SVM can directly analyze positive-operator valued measure (POVM) measurements and identify local entanglement structures in quantum states.

A quantum-inspired version of SVMs is introduced in "Quantum kernels to learn the phases of quantum matter"~\cite{SanchoLorente2022}. The authors utilize SVM with a quantum kernel to predict and characterize second-order quantum phase transitions. This algorithm is tested on the transverse field Ising chain and further applied to classify the phases of matter in a quantum processor.

\subsubsection{Others}

"Interpretable and unsupervised phase classification"~\cite{Arnold2021} proposes an alternative, physically-motivated, data-driven scheme, which relies on the calculation of differences between mean input features. The method is demonstrated in the example of the physically rich ground-state phase diagram of the spinless Falicov-Kimball model.

Results by~\cite{Wetzel2017b} and~\cite{Suchsland2018} suggest neural networks learn concepts like the order parameter and energies to distinguish phases of the matter. Arnold et al.~\cite{Arnold2022,Arnold2023,Arnold2023b} devised a framework to verify this information in a general setting independently of the underlying machine learning algorithm as long as the model is converged to the minimum of the underlying objective function. By employing this scheme it is possible to prove that the energy is a sufficient quantity for learning the phases of the Ising model. More generally, they prove that the indicators of phase transitions of various machine learning schemes~\cite{Carrasquilla2017,vanNieuwenburg2017,Schaefer2019,Singh2021} approximate the system's Fisher information.

\section{Conclusions}
\label{chapter:conclusions}

In this review, we have provided an overview of current research directions in the emerging field of interpretable machine learning applied to physics. Interpretable Machine Learning serves as a transformative approach that bridges the gap between machine-driven insights and human understanding. As machine learning algorithms become more powerful and computational resources continue to expand, artificial intelligence will likely achieve scientific discoveries beyond human reach. This underscores the urgent need to develop a deeper understanding of machine learning models, ensuring that humans remain active participants in the era of AI-driven scientific discoveries.

\begin{acknowledgments}
We thank Julian Arnold, Anna Dawid, Javier Quetzalcoatl Toledo-Marin, Eric Raidl, Joachim Stein, Max Weinmann, and Oliver Buchholz for helpful discussions.

Research at Perimeter Institute is supported in part by the Government of Canada through the Department of Innovation, Science and Economic Development Canada and by the Province of Ontario through the Ministry of Economic Development, Job Creation and Trade. This work was supported by Mitacs through the Mitacs Accelerate program. Funded by Deutsche Forschungsgemeinschaft (DFG, German Research Foundation) under Germany's Excellence Strategy -- EXC 2075 -- 390740016 and by the WIN programme of the Heidelberg Academy of Sciences and Humanities, financed by the Ministry of Science, Research and the Arts of the State of Baden-W\"urttemberg. We acknowledge the further support of the Ministry of Science, Research and the Arts Baden-W\"urttemberg (Az. 33-7533-9-19/54/5) in "{K\"unstliche Intelligenz \& Gesellschaft: Reflecting Intelligent Systems for Diversity, Demography and Democracy}" (IRIS3D), and of the {Interchange Forum for Reflecting on Intelligent Systems} (IRIS) and the Stuttgart Center for Simulation Science (SimTech) at the University of Stuttgart. Z.L. is supported by IAIFI through NSF grant PHY-2019786 and the Google PhD Fellowship. S.W. is supported by generous grants from the Siegel Family Endowment and the Omidyar Network.

\end{acknowledgments}

\newpage
\sloppy
\bibliography{literature}
%\bibliographystyle{abbrvnat}

%\appendix
%\newpage
%\onecolumngrid
%\section{Do we need an appendix}\label{app:app}

\end{document}